\documentclass[12pt]{article}
\usepackage[pdftex]{graphicx} %%GRAPHICX
\usepackage{epsfig}
\usepackage{comment}
\usepackage{latexsym}
\usepackage{hyperref}
\usepackage{amsmath}
\usepackage{color}
\usepackage{amsbsy}

\newcommand{\mysquare}[0]{\raise-.2ex\hbox{{\Large$\Box$}}}
\def\lsim{\mathrel{\rlap {\raise.5ex\hbox{$ < $}}
{\lower.5ex\hbox{$\sim$}}}}
\def\gsim{\mathrel{\rlap {\raise.5ex\hbox{$ > $}}
{\lower.5ex\hbox{$\sim$}}}} \topmargin -1.5cm \textheight=22.5cm \textwidth=16.5cm
\catcode`\@=11
\newcount\hour
\newcount\minute
\newtoks\amorpm
\hour=\time\divide\hour by60 \minute=\time{\multiply\hour by60
\global\advance\minute by-\hour}
\edef\standardtime{{\ifnum\hour<12 \global\amorpm={am}%
        \else\global\amorpm={pm}\advance\hour by-12 \fi
        \ifnum\hour=0 \hour=12 \fi
        \number\hour:\ifnum\minute<10 0\fi\number\minute\the\amorpm}}
\edef\militarytime{\number\hour:\ifnum\minute<10 0\fi\number\minute}
\def\draftlabel#1{{\@bsphack\if@filesw {\let\thepage\relax
   \xdef\@gtempa{\write\@auxout{\string
      \newlabel{#1}{{\@currentlabel}{\thepage}}}}}\@gtempa
   \if@nobreak \ifvmode\nobreak\fi\fi\fi\@esphack}
        \gdef\@eqnlabel{#1}}
\def\@eqnlabel{}
\def\@vacuum{}
\def\draftmarginnote#1{\marginpar{\raggedright\scriptsize\tt#1}}
\def\draft{\oddsidemargin -.2truein
        \def\@oddfoot{\sl preliminary draft \hfil
        \rm\thepage\hfil\sl\today\quad\militarytime}
        \let\@evenfoot\@oddfoot \overfullrule 3pt
        \let\label=\draftlabel
        \let\marginnote=\draftmarginnote
   \def\@eqnnum{(\theequation)\rlap{\k

 ern\marginparsep\tt\@eqnlabel}%
\global\let\@eqnlabel\@vacuum}  }
\newcommand{\be}[0]{\begin{equation}}
\newcommand{\ee}[0]{\end{equation}}
\newcommand{\ba}[0]{\begin{eqnarray}}
\newcommand{\ea}[0]{\end{eqnarray}}

\def\bs{\begin{subequations}}
\def\es{\end{subequations}}

\def\AEF{A.E. Faraggi}

\def\np#1#2#3{Nucl.\ Phys.\ B {\bf{#1}} (#2) #3}
\def\pl#1#2#3{Phys.\ Lett.\ B {\bf{#1}} (#2) #3}

\def\pr#1#2#3{Phys.\ Rev.\ D {\bf{#1}} (#2) #3}
\def\thebibliography#1{%
\vskip 0.5cm \centerline{\bf \Large References}
\list{%
[\arabic{enumi}]}{\settowidth\labelwidth{[#1]} \leftmargin\labelwidth
\advance\leftmargin\labelsep
\usecounter{enumi}}
\def\newblock{\hskip .11em plus .33em minus .07em}
\sloppy\clubpenalty4000\widowpenalty4000 \sfcode`\.=1000\relax}

\renewcommand{\theequation}{\arabic{section}.\arabic{equation}}

\renewcommand{\section}{\setcounter{equation}{0}\@startsection
{section}{1}{0mm}{-\baselineskip}{0.5\baselineskip} {\normalfont\Large\bfseries}}

\renewcommand{\subsection}{\@startsection
{subsection}{2}{0mm}{-\baselineskip}{0.5\baselineskip} {\normalfont\large\bfseries}}

\renewcommand{\subsubsection}{\@startsection
{subsubsection}{3}{0mm}{-\baselineskip}{0.5\baselineskip}
{\normalfont\normalsize\slshape}}

\usepackage{amssymb}
\usepackage{amsthm}
\usepackage{amsmath}
\usepackage{amssymb,amsfonts}
\usepackage{graphicx}
\usepackage{cite}

\newcommand{\ie}{{\em i.e. }}
\newcommand{\eg}{{\em e.g. }}

\newcommand{\Real}{\mathbb{R}}
\newcommand{\Z}{\mathbb{Z}}
\newcommand{\sign}{\mbox{sign}}

\renewcommand{\O}{{\cal O}}
\newcommand{\abs}{|}

\renewcommand{\and}{\mbox{and}}

\newcommand{\F}{{\cal F}}
\newcommand{\N}{{\cal N}}
\renewcommand{\S}{{\cal S}}
\newcommand{\T}{{\cal T}}

\def\Im{\,{\rm Im}\, }
\def\Re{\,{\rm Re}\, }

\def\thefootnote{\fnsymbol{footnote}}
\def\be{\begin{equation}}
\def\ee{\end{equation}}
\def\ba{\begin{eqnarray}}
\def\ea{\end{eqnarray}}
\def\bs{\begin{subequations}}
\def\es{\end{subequations}}

\def\R{{\cal{R}}}

\def\thebibliography#1{%
\vskip 0.5cm \centerline{\bf References}
\list{%
[\arabic{enumi}]}{\settowidth\labelwidth{[#1]}
\leftmargin\labelwidth
\advance\leftmargin\labelsep
\usecounter{enumi}}
\def\newblock{\hskip .11em plus .33em minus .07em}
\sloppy\clubpenalty4000\widowpenalty4000
\sfcode`\.=1000\relax}

\begin{document}
%\verb|\usepackage{draftcopy}|\\
\begin{titlepage}
\begin{flushright}
LTH--1020,
LPTENS--14/11,
CPHT-RR049.0914,\;
October   2014
\vspace{-.0cm}
\end{flushright}
\begin{centering}
{\bf \Large Large volume susy breaking with \\ 
 a solution  to the decompactification problem}

\vspace{5mm}

 {\bf Alon E. Faraggi$^{1}$, Costas Kounnas$^{2}$ and Herv\'e Partouche$^3$}

 \vspace{1mm}

$^1$  Department of Mathematical Sciences, University of Liverpool,\\
Liverpool L69 7ZL, United Kingdom.\\
{\em alon.faraggi@liv.ac.uk}

$^2$ Laboratoire de Physique Th\'eorique,
Ecole Normale Sup\'erieure$^\dag$,\\
24 rue Lhomond, F--75231 Paris cedex 05, France\\
{\em  Costas.Kounnas@lpt.ens.fr}

$^3$  {Centre de Physique Th\'eorique, Ecole Polytechnique$^\ddag$,\\
F--91128 Palaiseau cedex, France\\
{\em herve.partouche@polytechnique.edu}}

\end{centering}
\vspace{0.1cm}
$~$\\
\centerline{\bf\Large Abstract}\\
\vspace{-0.6cm}

\begin{quote}

We study heterotic ground states in which supersymmetry is 
broken by coupling the momentum and winding charges of two  large 
extra dimensions to the R-charges of the supersymmetry generators. 
The large dimensions give rise to towers of heavy string thresholds
that contribute to the running of the gauge couplings. In the
general case, these contributions are proportional to the 
volume of the two large dimensions and invalidate the perturbative
string expansion. The problem is evaded if the susy breaking 
sectors arise as a spontaneously broken phase of 
$\N=4\rightarrow\N=2\rightarrow\N=0$
supersymmetry, provided that $\N=4$ supersymmetry is restored on the
boundary of the moduli space. We discuss the mechanism 
in the case of $\Z_2\times \Z_2$ orbifolds, which requires that 
the twisted sector that contains the large extra dimensions
has no fixed points. We analyze the full string partition function 
and show that the twisted sectors distribute themselves in non-aligned
$\N=2$ orbits, hence preserving the solution to the string 
decompactification problem. 
Remarkably, we find that the contribution to the vacuum 
energy from the $\N=2\to \N=0$ sectors is suppressed, 
and the only substantial contribution arises
from the breaking of the $\N=4$ sector to
$\N=0$.

\end{quote}

\vspace{3pt} \vfill \hrule width 6.7cm \vskip.1mm{\small \small \small
  \noindent
   $^\dag$\ Unit{\'e} mixte  du CNRS et de l'Ecole Normale Sup{\'e}rieure associ\'ee \`a
l'Universit\'e Pierre et Marie Curie (Paris 6), UMR 8549.\\
$^\ddag$\ Unit{\'e} mixte du CNRS et de l'Ecole Polytechnique,
UMR 7644.}\\

\end{titlepage}
\newpage
\setcounter{footnote}{0}
\renewcommand{\thefootnote}{\arabic{footnote}}
 \setlength{\baselineskip}{.7cm} \setlength{\parskip}{.2cm}

\setcounter{section}{0}

%%%%%%%%%%%%%%%%%%%%%%%%%%%%%%%%%%%%
%%%%%%%%%%%%%%%%%%%%%%%%%%%%%%%%%%%%

\section{Introduction}

String theory is the leading contender for a unified theory of 
all known interactions~\cite{SStringsIntr}, and numerous string models exhibiting rich 
phenomenological properties have been constructed. They  utilize  various
compactification techniques, like for instance the 
Calabi-Yau compactifications~\cite{SStringsIntr}, 
 the orbifold compactifications~\cite{orbifolds},  
the  2d-fermionic constructions~\cite{fermionic}, 
the self-dual lattice constructions~\cite{Luest},
 the asymmetric orbifold compactifications~\cite{AsOrbifolds}, 
the $\N=(2,2)$ 
superconformal constructions~\cite{Gepner}, 
as well as the $\N=(2,0)$ constructions~\cite{fermionic, AsOrbifolds}. 

However, all of the quasi-realistic string models that have been constructed to date, namely with the correct standard model spectrum,  
possess an $\N=1$ spacetime supersymmetry (susy), and the question of how this symmetry is broken is still an open problem. The mechanisms that have been proposed to address this point are either 
perturbative~\cite{AlvarezGaume,Rohm,SSstring,AntoniadisTeV} or non-perturbative~\cite{gc, KiritsisKounnas, StringDualities, Fluxes}. One can consider :
\begin{description}
\item[\footnotesize $\bullet$] A non-perturbative breaking {\em via} gaugino condensation~\cite{gc},
which up till now has to be discussed at the level of the effective
supergravity. Due to the non-perturbative nature of the mechanism, one looses 
the predictability associated to the underlying string model. One then has to resort 
to an effective parametrization of the susy breaking parameters. 
\item[\footnotesize $\bullet$] Perturbative and/or non-perturbative flux compactifications, where internal 
fluxes are introduced and break susy suitably. These models can be explored using the non-perturbative $S,T,U$-dualities between the heterotic, Type IIA, Type IIB and orientifold superstring 
vacua~\cite{StringDualities,RcorrectionsDuals,Fluxes}.
\item[\footnotesize $\bullet$] An interesting example of geometrical fluxes is the one associated with 
a Stringy Scherk-Schwarz (SSS) susy breaking compactification, which has the 
advantage to be implemented at the perturbative string level~\cite{SSstring}. Here, the symmetry breaking parameters are obtained directly from the perturbative string theory. 
\end{description}

In this last approach,  the Scherk-Schwarz mechanism~\cite{SS} defined in supergravity theories is promoted at the superstring level~\cite{Rohm, SSstring,AntoniadisTeV}. 
Denoting the string scale as $M_{\rm s}=1/\sqrt{\alpha'}$, the mechanism entails that some of the compactified dimensions of characteristic size $R/M_{\rm s}$ (measured in string frame) of the internal manifold  are large, {\it i.e.} of the order of the inverse of the supersymmetry breaking scale. In Einstein frame,  we have $m_{3\over 2}^{({\rm E})}=\O( M_{\rm Planck}/R)  = {\cal O}$(1--10)~TeV. This follows from the fact that supersymmetry is broken by coupling a $\Z_2$ freely acting shift in these compactified directions, with the R-charges of the supersymmetry generators. 
These large dimensions give rise to tower of states, charged under low-energy gauge groups, that populate the energy range between the susy breaking scale and the Planck scale. They induce thresholds, 
whose analysis was recently pioneered in~\cite{FlorakisN=0},
that contribute to the running of the gauge couplings, Yukawa couplings and soft
susy breaking parameters. 

However, a problem arises when the threshold corrections are proportional 
to the volume of the large dimensions. When the $\beta$-function coefficient is negative,  they drive the theory to strong coupling at energies lower than  the unification (or string) scale~\cite{solving}. 
This problem  is known as the  {\it decompactification problem} and some proposals exist 
on how to avoid it~\cite{SSstring, AntoniadisTeV, RcorrectionsDuals, solving}. 
A first idea supposes the existence  of models without $\N=2$ sectors, so that the
threshold corrections are independent of the volume moduli of the internal theory~\cite{AntoniadisTeV}. 
Alternatively, one can suppose the thresholds of different spin states cancel 
among each other at one-loop in the perturbative 
expansion~\cite{AntoniadisTeV}. However,
the stability of this mechanism against higher loop corrections 
has not been demonstrated. Moreover, no quasi-realistic model realizing  one of the above two proposals has  been constructed so far. 

In this paper, we examine a different possibility, which was introduced in  Ref.~\cite{solving} in the context of $\N=2$ supersymmetric models. Due to  the properties of the $\N=4\to \N=2$ spontaneous breaking 
{\em via} freely acting orbifolds, the behavior of thresholds as functions of the moduli of the
internal manifold is radically different from that of the generic orbifold models, where 
the breaking from $\N=4$ to $\N=2$ is not spontaneous~\cite{solving}. The reason for this distinction is that $\N=4$ supersymmetry  is restored on the boundary of the
moduli space. In this case, for large values of the relevant moduli, the thresholds vanish (up to logarithmic corrections).

In order to extend the above idea to non-supersymmetric models, we first present in Sect.~\ref{2s} the class of string theories we consider, namely the heterotic $\Z_2\times \Z_2$ non-left/right-symmetric orbifold models realized in ``moduli-deformed fermionic constructions", where the $\N=1$ supersymmetry is further spontaneously broken to $\N=0$ by a SSS mechanism. In Sect. \ref{N=40}, we provide some preliminary introduction on  
how the gauge coupling threshold corrections in simple $\N=4$ models spontaneously broken to $\N=0$ do not develop dangerous linear dependences on volume moduli. We turn back to the $\Z_2\times \Z_2$ models from Sect.~\ref{twistshift} to the end of the article.  For simplicity, we specialize to the case where only one of the $\Z_2$ actions is freely acting. The second {\em together with} the diagonal action of the  two are supposed to have fixed points. As we will see, this restriction forces the spontaneous breaking of the supersymmetries to involve only one of the three internal 2-tori, for the decompactification problem not to arise.   

In Sect.~\ref{twistshift}, we evaluate the threshold corrections and effective potential generated at one-loop in the sectors arising from the action of a single $\Z_2$, namely the $\N=4$ sector and the so-called $\N=2$ $1^{\rm st}$ complex plane. For the associated $\N=4\to \N=2$ susy breaking to be spontaneous, the $\Z_2$ twist acts simultaneously as a shift along some of the two untwisted internal directions. The SSS mechanism responsible of the final spontaneous susy breaking to $\N=0$ is implemented by an additional $\Z_2^{\rm shift}$. The action of the latter on the above two untwisted internal directions introduces sub-sectors we analyze carefully. We find that only the $\N=4\to \N=0$ sub-sector (denoted as $B$), together with two sub-sectors (denoted as $C$ and $D$) preserving distinct $\N=2$ supersymmetries  contribute substantially. 

Sect.~\ref{5} discusses physically the formal results obtained in the sub-sectors $B,C,D$.  Three moduli-dependent mass scales $M^{({\rm E})}_{B,C,D}$ are introduced, the lowest of which being in the TeV region in realistic models. These scales, which  are different from the gravitini masses  present in each sector, control the contributions of the whole towers of Kaluza-Klein states that contribute to the running effective gauge couplings. Some examples are also presented. 

Sect.~\ref{6} completes the sector by sector analysis of the $\Z_2\times \Z_2$ models, by considering the additional contributions arising from the action of the second $\Z_2$, namely the $2^{\rm nd}$ and  $3^{\rm rd}$ complex planes, together with the $\N=1$ sector.  Under our hypothesis (only the $1^{\rm st}$ $\Z_2$ action is freely acting), the above  two planes have fixed points and  the SSS susy breaking to $\N=0$ must only involve the $1^{\rm st}$ plane moduli. This has two consequences. First, the gravitino mass $m_{3\over 2}$ of the $\N=1\to \N=0$ model is of order $1/\sqrt{\Im T_1}$, the inverse of the volume of the internal $1^{\rm st}$ plane. Moreover, the $2^{\rm nd}$ plane, $3^{\rm rd}$ plane and $\N=1$ sectors preserve exact supersymmetries  at tree level  and the threshold scales $M_I^{({\rm E})}$ associated to the complex planes $I=2,3$ must be of the order of the Planck scale.  We also collect  our results in order to write the expression of the effective coupling constants in the $\N=1\to \N=0$ models  we consider.  Moreover,  it is remarkable that the effective potential arises only from the $\N=4\to \N=0$ sector $B$, the other sectors being either supersymmetric or exponentially suppressed, when $m_{3\over 2}^{({\rm E})}$ is lower than the Planck scale. 

 Finally, our conclusions can be found in  Sect.~\ref{8}, while Appendix~\ref{deformed} is a review of the moduli-deformed fermionic construction. 

%%%%%%%%%%%%%%%%%%%%%%%%%%%%%%%%%%%%%%%%%%
\section{The $\boldsymbol{\Z_2\times \Z_2}$ models with spontaneously broken susy}
\label{2s}

The context in which we will propose a solution to the decompactification problem consists in $\Z_2\times \Z_2$ non-symmetric orbifolds obtained {\em via} the ``moduli-deformed fermionic construction" defined in 
Appendix~\ref{deformed}, and describing a spontaneous $\N=1\to \N=0$ susy breaking. As we will see in Sect.~\ref{twistshift}, the relevant models rely on an underlying $\N=4$ structure. Specifically,  at least  one of the two  $\Z_2$'s must act freely, so that an $\N=2$ sector will have the desired  properties of spontaneously broken $\N=4 \to \N=2$~\cite{solving}.  It is however important to note that this condition is incompatible with the existence of a chiral spectrum, as explained in Sect.~\ref{6}.  The final implementation of the   $\N=1 \to \N=0$  spontaneous breaking is done by
coupling another $\Z_2$ freely acting shift in the large internal directions, with the 
supersymmetric R-symmetry charges (\eg  the four $SO(1,9)$ helicity charges of the ten dimensional mother theory).   In the present section, our goal is to review the expression of the gauge  threshold corrections in heterotic string and to present the structure of the partition function in the {\em most general}  $\Z_2\times \Z_2$ non-symmetric orbifold models arising from deformed fermionic construction.  

For a gauge group factor $G^i$ at Kac-Moody level $k^i$, the running effective field theory coupling constant of a string model is~\cite{thresholds, IRregular, universality, solving, RcorrectionsDuals}
\be
{16\, \pi^2\over g_i^2(\mu)} = k^i{16\, \pi^2\over g_{\rm s}^2} +
 b^i\log {M_{\rm s}^2\over \mu^2} +\Delta^i\, ,
\label{D}
\ee
where $b^{i}$ is the $\beta$-function coefficient, $g_{\rm s}$ is the string coupling and $\mu$ plays the role of renormalization scale in the effective field theory. In string calculations, a mass gap $\mu$ is introduced to regularize the infrared~\cite{IRregular}. The analytic expression of the threshold corrections takes the form  
\be
\Delta^i=\int_{\cal F}{d^2\tau \over \tau_2}\!
\left(
{1\over 2}\sum_{a,b}{\cal Q}[^a_b](2v)
\left({\cal P}_i^2(2\bar w) - {k^{i}\over 4\pi{\tau_2}}\right)\! \tau_2 \, {Z}[{^a_b}](2v,2\bar w)-b^i
\right)\!\!\Bigg\abs_{v=\bar w=0}
+b^i\log {2\,e^{1-\gamma}\over \pi\sqrt{27}} \, ,
\label{2}
\ee 
where ${Z}[^a_b](2v,2\bar w)$ is the partition function for given spin structures $(a,b)$ of the worldsheet  fermionic supercoordinates. $(a,b)$ are integer modulo 2 : Spacetime bosons have $a=0$, while spacetime fermions have $a=1$. As indicated by the presence of the variables $v$ and $\bar w$, $Z[{^a_b}](2v,2\bar w)$ is actually a refined partition function, on which  the helicity operator ${\cal Q}[^a_b](2v)$ acts on the left-moving part,
\be
\label{Qop}
{\cal Q}[^a_b](2v) ={i\over \pi}\, \partial_{\tau}\!\left(\log {\theta[^a_b](2v)\over\eta} \right)\!\equiv {1\over 16\pi^2}\, {\partial_v^2(\theta[^a_b](2v))\over \theta[^a_b](2v)}-{i\over\pi}\, \partial_\tau\log\eta\, .
\ee
Our conventions for the $\theta[^\alpha_\beta](v\abs \tau)$-functions can be found in Eq.~(\ref{th}) or in Appendix~C of Ref.~\cite{KiritsisBook} and it is understood that $\theta[^\alpha_\beta](v)$ denotes $\theta[^\alpha_\beta](v\abs\tau)$, while  $\theta[^\alpha_\beta]$ stands for $\theta[^\alpha_\beta](0\abs \tau)$. 
On the contrary, ${\cal P}_i(2\bar w)$ is the charge operator of the gauge group factor $G^i$, thus acting  on the right-moving sector of the heterotic string  as a derivative operator.   
Finally, no infrared divergence occurs in the expression of $\Delta^i$, due to the relation  
\be
b^i=
\lim_{\tau_2 \to\infty}\,
{1\over 2}\sum_{a,b}
\! \left.{\cal Q}[^a_b](2v)\, {\cal P}_i^2(2\bar w)  \,\tau_2 \,  {Z}[^a_b](2v,2\bar w)\!\right\abs_{v=\bar w=0} \, .
\label{2a}
\ee

In all orbifold models  that preserve $\N=1$ supersymmetry,  the $\N=4$ sector gives vanishing contribution and only the $\N=2$ sectors contribute. Thus, in the $\Z_2\times \Z_2$ non-symmetric case, one has
\be
\Delta^i=\sum_{I=1}^3\Delta_I^i(T_I,U_I)\, ,
\ee
where the threshold corrections $\Delta_I^i(T_I,U_I)$ come from the three different $\N=2$ planes. In this expression, $T_I, U_I$, $I=1,2,3$, are the moduli of the three $\Gamma_{2,2}$-lattices associated to  the six internal dimensions. Notice that in all $\Z_2\times \Z_2$ non-symmetric orbifold models, there are no $\N=1$ sectors. The full $\beta$-function coefficient in these $\N=1$ theories is thus
\be
b^i=\sum_{I=1}^3 b_I^i\, ,\qquad b^i_I={1\over 2}\sum_{a,b}
 {\cal Q}[^a_b]\, {\cal P}_i^2  \, \tau_2\, {Z}_I[^a_b]\abs_{v=\bar w=0}\, , 
\ee
where ${Z}_I[^a_b]$ is the contribution from the plane $I$, and the modular
covariant helicity operator ${\cal Q}[^a_b]$ can be replaced by ${i\over \pi}\partial_{\tau} \!\log {\theta[^a_b]}$,
since the $-{i\over \pi}\partial_{\tau} \!\log \eta$ contribution is proportional
to zero, due to the preservation of supersymmetry.

Our goal is to derive the analogous structure of the threshold corrections to the couplings and to the effective potential in $\Z_2\times \Z_2$ non-symmetric orbifold models,  where $\N=1$ supersymmetry is spontaneously broken ``\`a la Stringy Scherk-Schwarz". This is done in the context of the moduli-deformed  fermionic construction, where the dependence in the moduli $T_I,U_ I$, $I=1,2,3$, of the three $\Gamma_{2,2}$-lattices are implemented. For this  purpose,  we need the  generic form of the associated partition functions, which is found by first following the rules  of the fermionic construction and then implementing the moduli deformations, as explained  in Appendix~\ref{deformed}. We obtain in this way not only the generic form of the partition function in symmetric  $\Z_2\times \Z_2$ orbifolds, but also in 
non-left/right-symmetric ones. 

Limiting ourselves to the continuous deformations parameterized by $T_I,U_ I$, but including however all possible $1\over 2$-discrete  Wilson lines, the generic modular invariant partition function turns out to be 
\begin{align}
  Z(2v,2\bar w)=& \;{1\over \tau_2(\eta \bar\eta)^2}\,{1\over 2} \sum_{a,b} {1\over 4}  
\sum_{H_I,G_I} {1\over 2^N}\sum_{h_I^i,\hat h_I^i,g_I^i, \hat g_I^i} e^{i\pi (a+b+ab)}\, {\theta[^a_b ](2v) \over \eta}\, 
{\theta[^{a+H_2}_{b+G_2}] \over \eta}\,{\theta[^{a+H_1}_{b+G_1}] \over \eta}\,
{\theta[^{a+H_3}_{b+G_3}] \over \eta}
\nonumber\\
&\; \times \,  S\Big[{}^{a,\, h^i_I,\, \hat h^i_I,\,  H_I} _{b,\;  g^i_I,\;  \hat  g^i_I,\; G_I}  \Big]  \, \, 
Z_{2,2}\Big[{}^{h_1^i,  \,  \hat h_1^i}_{g_1^i, \; \hat g_1^i} 
 \Big \abs {}^{H_2}_{G_2} \Big]\, 
 Z_{2,2}\Big[{}^{h_2^i, \, \hat h_2^i}_{g_2^i ,\; \hat g_2^i} \Big \abs{}^{H_1}_{G_1} \Big]\,
 Z_{2,2}\Big[{}^{h_3^i  ,\,  \hat h_3^i}_{g_3^i ,\; \hat g_3^i} \Big \abs{}^{H_3}_{G_3} \Big] \,
 Z_{0,16}\Big[{}^{h_I^i , \, \hat h^i_I, \,  H_I}_{g_I^i , \;\hat g_I^i,\; G_I}  \Big] \;\!\!(2\bar w)\,  ,
\label{partition}
\end{align}
in terms of which the effective potential can be expressed as
\be
V_{\rm eff}= -{1\over (2\pi)^4}\int_\F {d^2\tau\over 2\tau_2^2}\, Z\abs_{v=\bar w=0}\, .
\ee
 In Eq. (\ref{partition}), the variable $\bar w$ refers to a gauge group factor realized by the $Z_{0,16}$ block but may have been implemented in one of the $Z_{2,2}$'s (see the following).
Our notations are as follows :  
\begin{description}
\item[\footnotesize $\bullet$] $(H_1,G_1)$ and  $(H_2,G_2)$ are integer modulo 2, associated to the $\Z_2\times \Z_2$ action, whose generators twist the internal coordinates $X^{6,7,8,9}$ and  $X^{4,5,8,9}$, respectively. We denote $(H_3,G_3)\equiv -(H_1+H_2, G_1+G_2)$, which is associated to the diagonal action. It is then natural to separate the contributions of the partition function in the following sectors\;: 
\begin{description}
\item[-]The $\N=4$ sector, which corresponds to $(H_1,G_1)=(H_2,G_2)=(H_3, G_3)=(0,0)$.
\item[-]Three $\N=2$ twisted sectors, \ie the so-called complex planes :
\begin{description}
\item[]Complex plane $I=1$ : $(H_1,G_1)=-(H_3,G_3) \ne (0,0)$ with $(H_2,G_2)= (0,0)$.
\item[]Complex plane $I=2$ : $(H_2,G_2)=-(H_3,G_3) \ne(0,0)$ with $(H_1,G_1)= (0,0)$.
\item[]Complex plane $I=3$ : $(H_1, G_1)=-(H_2, G_2)\ne  (0,0)$ with  $(H_3,G_3)=(0,0)$.
\end{description}
\item[-] The $\N=1$ twisted sector :  $(H_1,G_1)\neq (0,0)$,  $(H_2,G_2)\neq (0,0)$,  $(H_3,G_3)\neq (0,0)$.
\end{description}
As we  will explain later in more details, $\N=4,2,1$ denotes in the above list the number of fermionic zero modes present in each sector, when no spontaneous breaking of supersymmetry  to  $\N=0$ is implemented. Indeed, the (extended) supersymmetry  of each sector may or may not be in a spontaneously broken phase, $\N=4,2,1\to \N=0$, depending on the choice of $S$ introduced below.  
\item[\footnotesize  $\bullet$] $(h^i_I,g^i_I)$, $(\hat h^i_I, \hat g^i_I)$, $i=1,2$,   
 $I=1,2,3$, are integer modulo 2. $(h^i_I,g^i_I)$ are shifts and $(\hat h^i_I, \hat g^i_I)$ are ``dual shifts" of the three untwisted $\Gamma_{2,2}$-lattices, which are given as  sums over two momenta $m_I^i$ and two winding numbers $n_I^i$ associated to each complex  plane $I$ (see 
 Appendix~\ref{deformed}). 
\item[\footnotesize $\bullet$] The contribution of the six internal coordinates (shifted by $(h_I^i,g_I^i)$, dual shifted by $(\hat h_I^i,\hat g_I^i)$ and twisted by $(H_I,G_I)$),  is given in the second line of 
Eq.~(\ref{partition}), in terms of the $(2,2)$-conformal blocks $Z_{2,2}\Big[{}^{h_I^i,\, \hat h_I^i } _{g_I^i,\; \hat g_I^i} \Big \abs{}^{H_I}_{G_I} \Big]$, $I=1,2,3$.
\item[\footnotesize $\bullet$] The fact that the shifts $(h_I^i,g_I^i)$,  the dual shifts $(\hat h_I^i,\hat g_I^i)$ and the twists $(H_I,G_I)$ 
are not  in general independent leads to  an effective normalization factor $1/2^N$ in the partition function, with $N$ the number of independent  pairs  $(h_I^i,g_I^i)$ and $(\hat h_I^i,\hat g_I^i)$.
\item[\footnotesize $\bullet$] $S$ is a phase that  can implement the breaking of  $\N=1$ spacetime supersymmetry to $\N=0$. 
When $S\Big[{}^{a,\,h^i_I,\, \hat h^i_I,\, H_I} _{b,\; g^i_I,\; \hat g^i_I,\;G_I}  \Big]\equiv 1$, the theory is $\N=1$ supersymmetric. 
The latter can be broken spontaneously ``\`a la Stringy Scherk-Schwarz"  
once some of the 10-dimensional helicity  characters (R-parity charges) 
\be
\left(^a_b \right),\quad \left(^{a+H_1}_{b+G_1} \right) , 
\quad \left(^{a+H_2}_{b+G_2} \right),\quad\left(^{a+H_3}_{b+G_3}\right)
\ee
are coupled with the lattice charges, \ie with some shifts $(h_I^i,g_I^i)$ and/or  dual shifts $(\hat h_I^i,\hat g_I^i)$.
\item[\footnotesize $\bullet$] Finally, the contribution of the 32 extra right-moving worldsheet fermions is denoted 
$Z_{0,16}\Big[{}^{h_I^i,\, \hat h_I^i,\, H_I} _{g_I^i, \; \hat g_I^i,\;  G_I}  \Big]$. In the absence of shifts, dual shifts and twists, $Z_{0,16}$ 
is  the partition function associated to the $E_8\times E_8$ or $SO(32)$ root lattice. 
When shifts, dual shifts or twists are non-trivial, the initial gauge group is broken to a product 
of lower dimensional subgroups (modulo some stringy extended symmetry points). Therefore, the 
role of the non-trivial (dual) shifts and twists is to generate non-zero discrete and
continuous Wilson lines. According to the fermionic construction rules, the choice of (dual) shifts and twists in  realistic  models is such that  the  right-moving gauge group contains an  $SO(10)$  factor,  which is further broken 
to a subgroup that includes the desired standard model gauge group, coupled to acceptable particle content, with three generations (see for instance Ref.~\cite{classi}).  
\end{description}

 If no particular attention is devoted to the choice of shifts $(h^i_I,g^i_I)$ and  dual shifts $(\hat h^i_I, \hat g^i_I)$, when   supersymmetry is broken to $\N=0$  ``\`a la Stringy Scherk-Schwarz",   the resulting  $\N=0$ model  may suffer  from the so-called decompactification problem. The reason for this  is related to the supersymmetry breaking  scale,  which is fixed by the inverse of the characteristic size $R$ of the internal compactified dimensions involved in the breaking, $m_{3\over 2} =\O(M_{\rm s}/R)$. Indeed,  in order to have a small supersymmetry breaking scale compared to the string scale,  $m_{3\over 2}= 10^{-14} M_{\rm s}$,  $R$  must be enormous. 
Consequently, when the  threshold corrections due to the tower of Kaluza-Klein states are proportional to the volume of the large extra dimensions and dressed with a  negative $\beta$-function coefficient, the perturbative expansion is invalidated~\cite{solving,RcorrectionsDuals}. However, this is not always the case. The next section is devoted to the presentation of the simplest example, where such a volume term is not generated.

\section{The $\boldsymbol{\N=4 \rightarrow \N=0}$ sector} 
\label{N=40}

The partition function (\ref{partition}) can be separated in sectors according to the $\Z_2 \times \Z_2$ action. In this section, we focus on the $\N=4$ sector  $(H_1,G_1)=(H_2,G_2)=(0,0)$, which can be spontaneously broken to $\N=0$, when the SSS phase $S$ is non-trivial. In this case, the induced contribution to the thresholds yields a  logarithmic dependence on the volume of the internal directions involved in the susy breaking. 
Actually, the threshold corrections of the  $\N=4\to \N=0$ sector appearing 
in the $\Z_2\times \Z_2$ non-symmetric orbifold models are smaller by a factor 4, compared  to those of the full ``mother" $\N=4\to \N=0$ theory. As a first step, we compute here the threshold corrections in an $\N=4\to \N=0$ theory and will remind that in the final result a factor of ${1\over 4}$ arising from a $\Z_2\times \Z_2$ projection must be included. We will present in detail the simple case, where a single factorized circle is involved in the process of supersymmetry breaking. This example can be considered as an introduction, since Sects~\ref{twistshift}--\ref{6} will present the analysis valid in  $\Z_2\times \Z_2$ models obtained by  moduli-deformed  fermionic constructions and where only the $\Z_2$ action parameterized by $(H_1,G_1)$ is freely acting. 

In an $\N=4$ model, two possibilities may arise once a phase $S$ is introduced. 
If $S$ is independent of $(a,b)$, then the $\N=4$ supersymmetry is unbroken. In this case, the contribution of the worldsheet fermions to the partition function yields
\be
{1\over 2}\sum_{a,b}(-)^{a+b+ab}\,  \theta[^a_b ] (2v)\,\theta[^a_b ]^3=\theta[^1_1 ]^4 ({v}) =\O(v^4)\, ,
\ee
where we use the Jacobi $\theta$-function  identity and the relation $\theta[^1_1](v\abs \tau)= 2\pi\eta^3(\tau)\, v+ {\cal O} (v^3)$. Therefore, the partition function (and effective potential) vanish. Similarly, the helicity insertion, which defines the corrections to the coupling constants, gives
\be
\label{Qsusy}
{1\over 2}\sum_{a,b}(-)^{a+b+ab} \, {\cal Q}[^a_b ](2v) \,\theta[^a_b ] (2v)\,\theta[^a_b ]^3=
{1\over 16\pi^2}\, \partial_{v}^2 \!\left(\theta[^1_1 ]^4 ({v})\right)={\cal O} (v^2) \, ,
\ee
which shows that the gauge coupling thresholds vanish as well.

The second possibility is when the phase $S$ couples non-trivially the helicity charges
$(a,b)$, with the shifts and/or dual shifts of the internal lattice. This will break spontaneously the $\N=4$ supersymmetry to $\N=0$. In order to simplify our discussion in this section, we restrict ourselves to the case where only one $S^1$ cycle  is involved in the susy breaking, and is very large. In this direction, we also consider shifts only, $(h^1_1,g^1_1)$ we denote as $(h,g)$, and take
 \be
 S=e^{i\pi(ah+bg+hg)} \, .
 \ee
Moreover, we specialize to the case where the $S^1$ shifted lattice is factorized,
\be
\Gamma_{6,6+16}[^h_g]=\Gamma_{1,1}[^h_g](R_1)\, \Gamma_{5,21}[^h_g]\, ,
\ee
where $\Gamma_{5,21}[^h_g]$ is a  shifted lattice associated to the remaining 5 internal coordinates and the 32 right-moving worldsheet fermions of the heterotic string.\footnote{In $\Z_2\times \Z_2$ models, $\Gamma_{5,21}$ is further factorized as in Eq.~(\ref{partition}).} For instance, the dependence of the $\Gamma_{5,21}$-lattice on $(h,g)$ may induce a Higgs mechanism by acting on the right-moving worldsheet degrees of freedom. In any case, due to our assumptions, this dependence must not imply a participation of the $\Gamma_{5,21}$ moduli in the super-Higgs mechanism, which would otherwise induce a very large gravitino mass. 
The $S^1$ shifted lattice, $\Gamma_{1,1}[^h_g]$, admits two representations, Hamiltonian or Lagrangian, which are related to one another by Poisson resummation on the momentum quantum number 
$m$~\cite{solving,RcorrectionsDuals} :
\begin{align}
\label{Gshift}
\Gamma_{1,1}[^h_g](R_1)&=\sum_{m,n}(-)^{mg}\, 
q^{{1\over 2}p_L^2}\bar q^{{1\over 2}p_R^2}\,  ,\quad \mbox{where}\quad 
p_{\underset{\scriptstyle R}{L}}=
{1\over \sqrt{2}}\left[{m\over R_1}\pm \Big(n+{h\over 2}\Big)R_1\right] \nonumber\\
&={R_1\over \sqrt{\tau_2}}
\sum_{n,\tilde m}e^{-{\pi R_1^2\over \tau_2}\left 
\abs (\tilde m+{g\over 2})+(n+{h\over 2})\tau\right\abs^2}\, .
\end{align}
In fact, restricting the internal lattice to the above factorized form will not affect the asymptotic behavior of the threshold corrections for large $R_1$. 

Because of the non-trivial correlation of the helicity and lattice charges through the SSS susy breaking phase, both the partition function and the coupling constant corrections are not zero. Indeed, in the partition function, the worldsheet fermions and SSS phase give 
\be
{1\over 2}\sum_{a,b}(-)^{a+b+ab} \, e^{i\pi(ag+bh+hg)} \theta[^a_b ]^4= 
{1\over 2}\sum_{A,B}e^{i\pi (A+B +AB+h+g)}  \theta\!\left[^{A+h}_{B+g} \right]^4  =
e^{i\pi (h+g+1)}\theta\!\left[^{1-h}_{1-g} \right]^4 ,
\ee
which contribute to the effective potential when 
$(h,g)\neq (0,0)$~\cite{CosmologicalTerm}. 
Moreover, using the above equation, the integrand involved in the gauge threshold corrections becomes
\begin{align}
{1\over 2}\sum_{a,b} {\cal Q}[^a_b] \!\left({\cal P}_i^2 - {k^i\over 4\pi{\tau_2}}\right) \!{\tau_2 \,Z}[^a_b]\bigg\abs_{v=\bar w=0}\!\!= \; &{1\over 2}\sum_{h,g}e^{i\pi(h+g+1)}\, {i\over \pi}\!\left({1\over 4}\partial_\tau \theta[^{1-h}_{1-g}]^4-(\partial_\tau\log \eta)\theta[^{1-h}_{1-g}]^4\right)\times\nonumber \\
&{1\over \eta^{12}\bar \eta^{24}}\, \Gamma_{1,1}[^h_g](R_1) \!\left({\cal P}_i^2(2\bar w) - {k^i\over 4\pi{\tau_2}}\right)\! \Gamma_{5,21}[^h_g](2\bar w)\Big\abs_{\bar w=0} .
\end{align}
The second part of the helicity operator ${\cal Q}[^a_b]$ proportional to $\partial_{\tau}\log\eta$ gives non-trivial contribution, when supersymmetry is broken to $\N=0$ \ie when $(h,g)\neq (0,0)$. 

To perform the integral over the fundamental domain, one can use the unfolding method introduced in Ref.~\cite{McClain:1986id} and used
 in~\cite{thresholds,solving,RcorrectionsDuals,FlorakisIR}. Defining $N=2n+h$ and $\tilde M=2\tilde m+g$, when $R_1$ is sufficiently large to guaranty the absolute convergences of the series, one can map the integral over the fundamental domain $\F$ into an integral over $\F$ restricted to the pair $(N,\tilde M)=(0,0)$, plus an integral over the ``upper half strip'' ($-{1\over 2}<\tau_2<{1\over 2}$, $\tau_2>0$) restricted to $N=0$, $\tilde M\neq 0$. In the strip representation, the winding contributions to the fundamental domain integral are mapped to the momentum contributions in the ultraviolet region of the strip, $\tau_2 < 1$. In our case, all integrands with $N=0$ (\ie $n=h=0$) and $\tilde M$ even (\ie $g=0$) preserve $\N=4$ and therefore vanish, as shown in Eq.~(\ref{Qsusy}). This is fundamental, since the key point to not have a contribution to the thresholds proportional to a large volume ($R_1$ in the present case) is that the integrand with $(N,\tilde M)=(0,0)$ vanishes.   Thus, we are left with an integral over the strip, with $(h,g)=(0,1)$,
\begin{align}
\label{N4delta}
\Delta^i=\lim_{\mu\to 0}\bigg[&\int_{\abs\abs}{d^2\tau\over \tau_2} \, 
{1\over 2}\, {i\over \pi}\!\left({1\over 4}{\partial_\tau \theta[^1_0]^4\over \eta^{12}}-(\partial_\tau\log \eta){\theta[^1_0]^4\over \eta^{12}}\right)\!
{R_1\over \sqrt{\tau_2}}\sum_{\tilde m}
e^{-{\pi R_1^2\over 4\tau_2}\left(2\tilde m+1\right)^2 -\pi\mu^2\tau_2}\,\times\nonumber \\
&\left({\cal P}_i^2(2\bar w) - {k_{i}\over 4\pi{\tau_2}}\right)\!
{\Gamma_{5,21}[^{0}_{1} ](2\bar w)\over \bar \eta^{24}}
-b^i\,\int_\F {d^2\tau \over \tau_2}\, e^{-\pi\mu^2\tau_2}\left.\!\bigg] \!\right\abs_{\bar w=0}  +
b^i \ln {2\,e^{1-\gamma}\over \pi\sqrt{27}}\, .
\end{align}
In Eq.~(\ref{N4delta}), we introduced a small mass $\mu$ in order to regulate the infrared 
divergences in the large $\tau_2$ limit~\cite{IRregular}. 
Other ways to regularize the infrared regime have been proposed 
recently~\cite{FlorakisIR} and have the advantage of preserving in a very elegant 
way both worldsheet and target space dualities. 
Our results, however, do not depend of the regularization scheme. 

The would be tachyonic level appearing 
in the right-moving sector is projected out by the level matching condition 
induced {\em via} $\tau_1$-integration over the strip. In the large $R_1$ limit, 
the massive string states give exponentially 
suppressed contributions to the integral over $\tau_2$ and can be consistently neglected.
The dominant contribution comes from the massless level and even if supersymmetry is broken, there are no-tachyons arising from  the left-moving sector. More specifically, we have    
\be
 \left( {i\over \pi}\partial_{\tau}\log\theta[^{1}_{0} ] - 
{i\over \pi} \, \partial_{\tau}\log\eta \right)\!{\theta[^{1}_{0} ]^4 \over \eta^{12}}
= \left(-{1\over 4}+{1\over 12}\right) \!16+ {\cal O}(q)=-{8\over 3} +{\cal O}(q)\, , 
\ee
which is an expected result, since the constant term in the above $q$-expansion must be proportional to  the $\beta$-function contribution  of the bosons of the $\N=4$ vector multiplets. On the contrary, the gauge group contribution comes from the 
${\cal P}_i^2$ charge operator, which acts on the right-moving sector.  
Actually, in our conventions, the $\beta$-function contributions of massless degrees of freedom are :
 \be
 \label{coeff}
b(\mbox{gauge boson}) =-{11\over 3} \, C({\cal R})\, , \;  b(\mbox{real scalar}) ={1\over 6} \,  C({\cal R})\, , \; b(\mbox{Majorana fermion}) = {2\over 3}\, C({\cal R})\, ,
\ee
where $C({\cal R})\delta^{ab}={\rm Tr} (T^aT^b)$ is the group factor coefficient associated to the  generators $T^a$ in the representation ${\cal R}$ of $G^i$.  In an $\N=4$ vector multiplet, ${\cal R}$ is the adjoint representation, and there are 6 real scalars and 4 Majorana gauginos per gauge boson,  
leading to $b({\rm bosons}) =-{8\over 3}\, C({\cal R})$ and 
$b({\rm fermions})={8\over 3}\, C({\cal R})$. When supersymmetry is unbroken, the 
$\N=4$ $\beta$-functions vanish. However, in our case, supersymmetry is spontaneously  broken {\em via} the SSS mechanism. The gravitinos and gauginos are getting masses that can be read in the Hamiltonian form of the  $\Gamma_{1,1}[^0_1]$-lattice in Eq.~(\ref{Gshift}) and are  proportional to the inverse of the internal radius,
\be
m^2_{3\over 2}=m^2_{1\over 2}={M_{\rm s}^2\over R_1^2}\, ,
\ee  
while the gauge bosons and scalars remain massless,
 \be
m^2_{1}=m^2_{0}=0\, .
\ee  
Thus, the logarithmic behavior of the $\beta$-function is fully controlled by the massless 
bosons, while the main corrections in the thresholds come from the tower of states
organized by the shifted $\Gamma_{1,1}[^0_1](R_1)$-lattice. 

Neglecting in Eq.~(\ref{N4delta}) the exponentially suppressed  
contributions  for large radius, $\Delta^i$ gets simplified enormously,
\be
\Delta^{i}=b^i\Delta-k^iY\, ,
\ee
where $b^i \Delta$ comes from the ${\cal P}_i^2$ action and 
$k^iY$ is the universal contribution arising from its modular covariant term  ${k^i\over 4\pi{\tau_2}}$. The former is
\begin{align}
 \Delta&=\lim_{\mu\to 0}\left[R_1\sum_{\tilde m} \int_0^{+\infty}{d\tau_2\over \tau_2^{3/2}}\,  e^{-{\pi R_1^2\over 4\tau_2}(2\tilde m+1)^2}e^{-\pi \tau_2 \mu^2}-\int_1^{+\infty}{d\tau_2\over \tau_2}\, e^{-\pi \tau_2 \mu^2}\right]\!-\ln\pi-\gamma+\cdots\nonumber\\
 & =\lim_{\mu\to 0}\left[2\sum_{\tilde m} {1\over \abs 2\tilde m +1\abs}\, e^{-\pi R_1\abs2\tilde m +1\abs \mu}-\Gamma(0,\pi \mu^2)\right]\!-\ln\pi-\gamma+\cdots ,
 \end{align}
 where the dots stand for ${\cal O}(e^{-cR_1})$ corrections, with $c$ positive and of the order of the lowest mass $M_0$ of the massive spectrum divided by $M_{\rm s}$.\footnote{$M_0$ depends on the moduli appearing in the $\Gamma_{5,21}[^h_g]$-lattice and is at most equal to $M_{\rm s}$.}
In the above expression, $\Gamma(s,x)$ is the upper incomplete $\Gamma$-function. Using the fact that $\Gamma(0,x)=-\ln(x)-\gamma+\O(x)$, one finally finds
 \be
 \Delta  =\lim_{\mu\to 0}\left[2 \ln \left({1+ e^{-\pi R_1 \mu}\over 1-e^{-\pi R_1 \mu}}\right)+\ln \mu^2 \right]\!+\cdots= - \log\!\left({\pi^2\over 4} R^2_1\right)+\cdots.
 \ee
For the determination of $Y$, the infrared regulator $\mu$ is not needed since 
the integral is infrared convergent,
\be
Y={C_0\over 4\pi}\, \sum_{\tilde m}\int_0^{\infty}{d\tau_2\over \tau_2^{5/2}}\, 
R_1 e^{-{\pi R_1^2\over 4\tau_2}(2\tilde m+1)^2}+\cdots=
{7\zeta(3)\over 4\pi^2}\, {C_0\over R_1^2} +\cdots  .
\label{c0}
\ee
 In (\ref{c0}), $C_0$ is the product of the contribution of the helicity operator 
${\cal Q}[^a_b]$ acting on the left-moving sector, $-{8\over 3}$, with a coefficient $2+d_G-n_{\rm F}$  associated to the right-moving sector,  
\be
C_0={1\over 2}\sum_{a,b}
 {\cal Q}[^a_b] \,{\tau_2 Z}[^a_b]\Big\abs_{q^0\bar q^0, v=0}=-{8\over 3} (2+d_{G}-n_{\rm F}) \, .
\ee 
$d_{G}$ is the number of vector bosons in the $\N=4$ vector multiplets of the parent $\N=4$ theory that remain massless after spontaneous breaking to $\N=0$. In other words, $d_{G}$ is the dimension of the gauge group. Similarly, $4n_{\rm F}$ is the number of Majorana fermions in the $\N=4$ vector multiplets of the parent $\N=4$ theory that remain massless after spontaneous breaking to $\N=0$. When the shifts $(h,g)$ are not acting on the right-moving sector, then $n_F=0$. However, in the generic case, $n_F$ is non-trivial, as is the case in the examples presented in Sects~\ref{eg2} and \ref{eg3}.
Therefore, the corrections to the coupling constants in this $\N=4\to \N=0$ model are 
\be
\label{n=4}
\Delta^i=b^i\Delta-k^iY=-b^i\log\!\left({\pi^2\over 4} R^2_1\right)\!+k^i\,  {14\zeta(3)\over 3\pi^2}\, {2+d_{G}-n_{\rm F}\over R_1^2}+{\cal O}\!\left(e^{-cR_1}\right).
\ee
The dangerous volume dependence (linear term in $R_1$) is absent, and 
the reason for this is the restoration of the $\N=4$ supersymmetry in the $R_1\to \infty$
limit. Since the universal contribution $Y$ scales like $m_{3\over2}^2/M^2_{\rm s}$, it is a tiny correction to the logarithmic term and may be neglected.

As said at the beginning of this section, the contribution of the $\N=4\to \N=0$ sector in a $\Z_2\times \Z_2$ model is obtained from  Eq.~(\ref{n=4}) by changing $b^i\to b^i/4$ and $C_0\to C_0/4$, where the $\beta$-function $b^i$ and $C_0$ refer to the $\N=4\to \N=0$ parent theory. However, the presence of $\N=2$ sectors requires more attention in the choice of susy breaking (dual) shifts. For instance, an $\N=2\to \N=0$ model containing a sector of the form 
\be
{S^1\over \Z^{\rm shift}_2}\times {T^4\over \Z_2}\, , 
\ee
where the circle of radius $R_1$ is shifted as before to break susy spontaneously to $\N=0$, will contain a contribution to the thresholds arising from the integration over $\F$ of the lattice term with $(N,\tilde M)=(0,0)$, which is proportional to the large radius $R_1$. This contribution arises from an $\N=2$ preserving sector, which therefore {\em does not vanish} as is the case when $\N=4$ is preserved. On the contrary, an $\N=2\to \N=0$ model based on an internal space containing a factor 
\be
{\displaystyle {S^1/ \Z^{\rm shift}_2}\times T^3\over \Z_2}
\ee
is safe. The reason for this is that the only $R_1$-dependent contribution to the partition function  arises from the untwisted sector, which realizes an $\N=4\to \N=0$ spontaneous breaking. Unfortunately, there is no model based on a single large $S^1$ shifted direction that  realizes a SSS spontaneous breaking of $\N=1$ supersymmetry to $\N=0$ and solves the decompactification problem. Therefore, we proceed in the next section with the more sophisticated case where two internal shifted directions involved in the breaking are large.

%%%%%%%%%%%%%%%%%%%%%%%%%%%%%%%%%%
\section{$\boldsymbol{\N=4}$ and $\boldsymbol{1^{\rm st}}$  plane contributions : 
$\boldsymbol{(H_2 , G_2)=(0,0)}$}
\label{twistshift}

From now on, we come back to   $\Z_2\times \Z_2$ models defined in Eq.~(\ref{partition}).   In this section and the following, we develop  a sector by sector analysis of the contributions to the gauge threshold corrections and effective potential. The susy breaking  is defined by the SSS phase $S\Big[{}^{a,\,h^i_I,\, \hat h^i_I,\, H_I} _{b,\; g^i_I,\; \hat g^i_I,\;G_I}  \Big]$ that correlates non-trivially the (dual) shifts and the twists charges with the helicity and R-symmetry 
charges. However, $S$ being sector-dependent, it can be trivial ($S=1$) in
some sectors, thus preserving supersymmetry, and non-trivial ($S\ne 1$) in some others, thus inducing a spontaneous breaking of supersymmetry. 

In the present section, we focus on the $\N=4$ sector $(H_2,G_2)=(H_1,G_1)=(0,0)$, together with the $1^{\rm st}$ $\N=2$ plane $(H_2,G_2)=(0,0)$, $(H_1,G_1)\neq (0,0)$. We derive here the formal results, and will comment on them physically  in Sect.~\ref{5}. Both sectors contain sub-sectors, which preserve or break supersymmetry. The contribution of the untwisted internal coordinates ($(H_2,G_2)=(0,0)$) in the partition function (\ref{partition}) involves  shifts $(h^i_1,g^i_1)$, and we restrict ourselves to the case where no dual shifts are introduced, $(\hat h_1^i, \hat g_1^i)\equiv(0,0)$. In this class of models, we have 
\be
 Z_{2,2}\Big[{}^{h_1^i} _{g_1^i} \Big \abs {}^{0}_{0} \Big]\!= {\Gamma_{2,2}\Big[{}^{h_1^1,\, h_1^2} _{g_1^1,\; g_1^2} \Big]\over (\eta\bar\eta)^2}\, ,
\ee
which depends on the $T_1,U_1$ moduli implemented in the moduli-deformed fermionic model, as explained in Appendix~\ref{deformed}. The shifted lattice dependence on $T_1,U_1$ (denoted $T, U$ in this section and Sect.~\ref{5}) is 
\begin{align}
\Gamma_{2,2}\Big[{}^{h^1,\, h^2} _{g^1,\;g^2} \Big]&=\sum_{m^i,n^i}(-)^{m_1g^1+m_2g^2}\, e^{2i\pi\tau\left[m_1\left(n^1+{h^1\over 2}\right)+m_2\left(n^2+{h^2\over 2}\right)\right]}\,\times\nonumber\\
&\qquad\quad\;\; e^{-{\pi\tau_2\over \Im T\Im  U}\left\abs T\left(n^1+{h^1\over 2}\right)+TU\left(n^2+{h^2\over 2}\right)-Um^1+m^2\right\abs^2}\phantom{\underset{\cdot}{\cdot}}\nonumber\\
&={\sqrt{\det G}\over \tau_2}\sum_{\tilde m^i,n^i}e^{-{\pi\over \tau_2}\big[\tilde m^i+{g^i\over 2}+\left(n^i+{h^i\over 2}\right)\tau\big](G_{ij}+B_{ij})\big[\tilde m^j+{g^j\over 2}+\left(n^j+{h^j\over 2}\right)\bar \tau\big]}, 
\label{latt}
\end{align}
where the dictionary between $T,U$ and the internal metric and antisymmetric tensor in the two associated compact directions is  
\be
G_{ij}={\Im T\over \Im U}\left(\!\!\!\begin{array}{cc} 1 &\Re U\\ \Re U & \abs U\abs^2\end{array}\!\!\right)\; , \qquad B_{ij}=\Re T \left(\!\!\begin{array}{cc} 0 &1\\ -1 & 0\end{array}\!\!\right).
\ee
As explained before, our solution to the decompactification problem requires the breaking of $\N=4\to \N=2$ to be spontaneous. This is implemented by imposing the twist action labeled by $(H_1,G_1)\equiv (H,G)$ to act simultaneously as a shift in the above $\Gamma_{2,2}$-lattice. As in the previous section,  independent  charges $(h,g)$ that are integer modulo 2 must be used to define the $\N=2\to \N=0$ SSS  susy breaking phase.  In the sectors we consider here, two options parameterized by $\zeta'=0$ or 1 can be chosen for the phase $S$ :
\be
\label{S}
\mbox{In the sectors $(H_2,G_2)=(0,0)$, } \quad S=e^{i\pi[ag+bh+hg+\zeta' (aG+bH+HG)]}\, .
\ee
 Anticipating the arguments  of Sects \ref{5} and  \ref{6}, when neither of the  $\N=2$ sectors associated to the $2^{\rm nd}$ and $3^{\rm rd}$ planes are realized as a spontaneous breaking of $\N=4$ supersymmetry (a fact that we suppose from now on), the moduli $T_I,U_I$ involved in these planes must not be too far from 1, for the decompactification problem no to occur. In this case, $(h,g)$ must be associated to the $\Gamma_{2,2}$-lattice of the $1^{\rm st}$ internal 2-torus, for the gravitino masses to be low. Therefore, both shifts $(h_1^i,g_1^i)$, $i=1,2$,  are involved and three classes of two models (labeled by $\zeta=0$ or 1)  can be analyzed\footnote{The {\em a priori} remaining cases $\Gamma_{2,2}\Big[{}^{h+H,\, 0} _{g+G,\;0} \Big]$, $\Gamma_{2,2}\Big[{}^{0,\, h+H} _{0,\;g+G} \Big]$ and $\Gamma_{2,2}\Big[{}^{h+H,\, h+H} _{g+G,\; g+G} \Big]$ lead to a volume dependence in the gauge thresholds, arising from the sub-sector $(h,g)=(H,G)\neq (0,0)$, which preserves $\N=2$ supersymmetry.} :
\begin{align}
a)& \;\;   \Gamma_{2,2}\Big[{}^{h+\zeta H,\, H} _{g+\zeta G,\;G} \Big] \;\; i.e.\;\; (h_1^1,g_1^1)\equiv (h,g)+\zeta(H,G), \; (h_1^2,g_1^2)\equiv(H,G)\nonumber\\
b)& \;\;  \Gamma_{2,2}\Big[{}^{H,\,h+\zeta H} _{G,\;g+\zeta G} \Big] \;\; i.e.\;\; (h_1^1,g_1^1)\equiv (H,G), \; (h_1^2,g_1^2)\equiv(h,g)+\zeta (H,G)\nonumber\\
c)& \;\;  \Gamma_{2,2}\Big[{}^{h+\zeta H,\,h+(1-\zeta)H} _{g+\zeta G,\;g+(1-\zeta)G} \Big] \;\; i.e.\;\; (h_1^1,g_1^1)\equiv (h,g)+\zeta (H,G), \; (h_1^2,g_1^2)\equiv(h,g)+(1-\zeta)(H,G)\, .
\label{lattt}
\end{align}
In the absence of SSS phase and $\Z_2^{\rm shift}$ action parameterized by $(h,g)$,  the models would describe the partial spontaneous 
breaking of supersymmetry from $\N=4 \to \N=2 $, which was considered 
in~\cite{solving}.
In this reference, it was shown 
that the pathological volume behaviors of the gauge couplings are  absent,  thanks
to the restoration of $\N=4$ supersymmetry in the large volume limit. 
In the  presence of non-trivial SSS phase, the $\Z_2^{\rm shift}$ action parameterized by $(h,g)$ breaks further the supersymmetry to $\N=0$. In this case, the decompactification problem becomes more involved, due to extra contributions coming from the sectors with non-trivial charges $(h,g)$. 
 
The separation of the $(H_2,G_2)=(0,0)$ sector of the partition function (\ref{partition}) in sub-sectors is more  transparent once we perform the summation  
over the helicity charges  $(a,b)$, keeping the non-trivial characters $(h,g)$ and $(H,G)$ fixed\footnote{The factor $1\over 2$ in the l.h.s. refers to the $\Z_2^{\rm shift}$ projection obtained once the sum over $h$ and $g$ is performed. The analogous $1\over 4$ factor associated to the $\Z_2\times \Z_2$ twist (or ${1\over 2}$  for a single $\Z_2$ twist) will be included later.} :
\begin{align}
{1\over 2}Z\Big[{}^{h,\, H}_{g, \; G} \Big](2v,2\bar w)&={1\over 4 \eta^8}
\sum_{a,b}e^{i\pi (a+b+ab)}\, e^{i\pi[ag+bh+hg+\zeta' (aG+bH+HG)]}\,\times \nonumber \\
&\qquad\qquad\quad \theta[^a_b ] (2v)\, \theta [^a_b ]\, \theta[^{a+H}_{b+G} ]\, \theta[^{a-H}_{b-G} ] \, 
\Gamma_{2,2}\Big[{}^{h_1^1,\, h_1^2}_{g_1^1, \; g_1^2} \Big]{1\over \bar\eta^4}  \, 
Z_{4,20}\Big[{}^{h,\, H}_{g, \; G} \Big]\!\;\!(2\bar w)\nonumber\\
& ={1\over  2\eta^8}\, e^{i\pi [hg+G(1+h+H)]}\, \theta[^{1-h}_{1-g} ]^2(v) \, 
 \theta[^{1-h+H}_{1-g+G} ]^2(v)  \, 
\Gamma_{2,2}\Big[{}^{h_1^1,\, h_1^2}_{g_1^1, \; g_1^2} \Big]{1\over \bar\eta^4}  \, 
Z_{4,20}\Big[{}^{h,\, H}_{g, \; G} \Big]\!\;\!(2\bar w) .
\label{theta}
\end{align}
The above result is obtained by redefining $a=A-h-\zeta' H$, $b=B-g-\zeta' G$ and summing over $A,B$ equal to 0 or 1. Note that $\zeta'$ has disappeared, which shows that the two SSS phases $S$ in Eq.~(\ref{S}) are actually equivalent, the different sectors of the theory being simply reshuffled.   
In Eq.~(\ref{theta}), the conformal block $Z_{4,20}\Big[{}^{h,\, H}_{g, \; G} \Big]$ for $(H,G)=(0,0)$ involves an untwisted lattice $\Gamma_{4,20}[^h_g]$, which depends on  the moduli  $T_I,U_I$, $I=2,3$. As said before, the latter are close to 1 and therefore must not participate in the super-Higgs mechanism that breaks susy to $\N=0$. Otherwise, a gravitino mass close to $M_{\rm Planck}$ would be generated in the sub-sector $(h,g)\neq (0,0)$, $(H,G)=(0,0)$ \ie far above the acceptable 1--10 TeV region.  However, the dependence of the $\Gamma_{4,20}$-lattice on $(h,g)$ may  induce a Higgs mechanism arising from an action on the right-moving worldsheet degrees of freedom. Several examples will be given in Sect.~\ref{5}.

In Eq.~(\ref{theta}), the number of odd $\theta$-functions $\theta[^{1+X}_{1+Y}](v)$, with $(X,Y)=(0,0)$,  counts the preserved supersymmetries, according to the number of fermionic zero modes in each sub-sector.  In the following, we use this number  of preserved supersymmetries to classify the sub-sectors and derive the effective potential and gauge couplings corrections in each case.  

\subsection{$\boldsymbol{A}$ : The exact  $\boldsymbol{\N=4}$ sector $\boldsymbol{(h,g)=(0,0)}$, $\boldsymbol{(H,G)=(0,0)}$}
 In this sector we denote $A$, $\N=4$ supersymmetry is unbroken. Therefore,  the contributions ${V_{\rm eff}}_A$ and $\Delta_A^i$ to the partition function (or effective potential) and to the gauge 
couplings vanish. This is due to the fact that the partition function (\ref{theta}) is   in this case proportional to  
$\theta[^{1}_{1} ]^4(v) ={\cal O}{(v^4)}$ and the $\beta$-functions  are of order ${\cal O}(v^2)$,
\be
 \Delta_A^i=0\, ,\quad {V_{\rm eff}}_A=0\, .
 \ee 
  The four gravitini in this sector are massless,
  \be
  m^i_{3\over 2}=0\, , \quad i=1,2,3,4 \, .
  \ee

 \subsection{$\boldsymbol{B}$ : The $\boldsymbol{\N=4 \rightarrow \N=0}$ sector $\boldsymbol{(h,g)\ne (0,0)}$, $\boldsymbol{(H,G)=(0,0)}$} 
In this sector we denote $B$, all arguments of the $\theta$-functions in 
Eq.~(\ref{theta}) are identical but not equal to $[^1_1]$. The partition function being  proportional to $\theta[^{1+h}_{1+g} ]^4(v)$, both  corrections ${V_{\rm eff}}_B$ and $\Delta_B^i$ to the effective potential and to the 
$\beta$-functions are non-vanishing. 
The four gravitini have  equal non-zero masses, which can be read from the Hamiltonian form  of the lattice (\ref{latt}) (the first equality),\footnote{We display the masses for $\Re(U)$ in the range $(-1,1]$.}$^,$\footnote{We define $\sign(0)=+1.$}  
\be
\label{m32}
m^i_{3\over 2}\!\equiv \!m_B=\!{\abs \alpha_B U-\sign(\!\Re U)\beta_B\abs\over \sqrt{\Im T\Im U}}\, M_{\rm s}=\!{\sqrt{(\alpha_B \Im U)^2+(\alpha_B\abs\!\Re U\abs-\beta_B)^2} \over \sqrt{\Im T\Im U}}\, M_{\rm s}\, , \,i=1,2,3,4,
\ee
where we define 
\be 
\label{alphaB}
(\alpha_B,\beta_B)=\left\{\!\!\begin{array}{ll}(1,0) & \mbox{in case }a) \\(0,1)& \mbox{in case }b) \\(1,1) &\mbox{in case }c)\, . \end{array}\right.
\ee
In Sect.~\ref{N=40}, we evaluated the coupling constant correction in case $a)$, when only one radius denoted by $R_1$ was very large. In this regime,  the  contribution of the remaining  $\Gamma_{5,21}$-lattice was trivial. However, there are extra  
contributions when both compact directions in the $1^{\rm st}$ plane are large. In the following, utilizing the techniques of 
Ref.~\cite{solving}, we compute the thresholds in cases $a)$, $b)$ and $c)$ in the regime where  the complex moduli $T$ and $U$ satisfy $\!\Im T\gg 1$, $U$ finite, which guaranties $m_B\ll M_{\rm s}$. 

Thanks to the Lagrangian expression  of the lattice (\ref{latt}) (the second equality), the sector 
$h=1$ is exponentially suppressed. Keeping explicitly the sector $(h,g)=(0,1)$, the threshold corrections  in sector $B$ are
\begin{align}
\Delta^i_B=&\int_\F{d^2\tau\over \tau_2}\left\{{1\over  \eta^4\bar\eta^{4}}\, {i\over 4\pi}\partial_\tau\!\!\left({\theta[^1_0]^4\over \eta^4}\right){1\over 2}\Gamma_{2,2}\big[{}^{\;0\;\;,\;\; 0}_{\alpha_B,\,\beta_B}\big]\!\left({\cal P}_i^2(2\bar w)-{k_i\over 4\pi\tau_2}\right)\!Z_{4,20}\big[{}^{0,\, 0}_{1, \, 0} \big]\!\!\;(2\bar w)-b^i_B\right\}\!\bigg\abs_{\bar w=0}\nonumber \\
&+b^i_B\log{2e^{1-\gamma}\over \pi\sqrt{27}}
+\cdots,
\end{align}
where the coefficient $b^i_B$ is introduced to cancel the infrared divergence and the dots stand for exponentially small contributions for large $\!\Im T$ and finite $U$. Similarly, the effective potential based on the partition function (\ref{theta}) with $(H,G)=(0,0)$ is
\be
{V_{\rm eff}}_B= -{1\over (2\pi)^4}\int_\F {d^2\tau\over 2\tau_2^3}\, {\theta[^1_0]^4\over \eta^{8}\bar\eta^{4}}\, {1\over 2}\Gamma_{2,2}\big[{}^{\;0\;\;,\;\; 0}_{\alpha_B,\,\beta_B}\big] Z_{4,20}[^{0,\, 0}_{1,\, 0}]\big\abs_{\bar w=0}+\cdots.
\ee
In the above two expressions, the dressing with the Lagrangian form of the $\Gamma_{2,2}$-lattice implies the non-level matched modes as well as the massive (level-matched) physical states to yield exponentially suppressed contributions. As a result, the universal form of the thresholds in sector $B$, 
\be
\label{deltaB}
\Delta^i_B=b^i_B\Delta_B-k^iY_B\, , 
\ee
as well as the effective potential take the simple forms obtained from the massless states and associated Kaluza-Klein modes :
\begin{align}
\Delta_B&=\int_\F{d^2\tau\over \tau_2}\left(\Gamma_{2,2}\big[{}^{\;0\;\;,\;\; 0}_{\alpha_B,\,\beta_B}\big]-1\right)+\log{2e^{1-\gamma}\over \pi\sqrt{27}}+\cdots,\phantom{\underset{\underset{a}{\cdot}}{\abs}}\nonumber\\
Y_B&={C_B\over 8\pi}\int_\F{d^2\tau\over \tau^2_2}\, \Gamma_{2,2}\big[{}^{\;0\;\;,\;\; 0}_{\alpha_B,\,\beta_B}\big]+\cdots,\phantom{\underset{\underset{a}{\cdot}}{\abs}} \nonumber \\
{V_{\rm eff}}_B&=-{C_V\over 2(2\pi)^4}\int_\F{d^2\tau\over \tau^3_2}\, \Gamma_{2,2}\big[{}^{\;0\;\;,\;\; 0}_{\alpha_B,\,\beta_B}\big]+\cdots,
\label{deltaY}
\end{align}
where $C_B= -{8\over 3}(2+d_{G_B}-n_{{\rm F}_B})$ and $C_V=8(2+d_{G_B}-n_{{\rm F}_B})$. In these coefficients, $d_{G_B}$ is the number of vector bosons in the $\N=4$ vector multiplets of the parent $\N=4$ theory that remain massless after spontaneous breaking to $\N=0$, \ie the dimension of the gauge group realized in the sector $B$. Similarly, $4n_{\rm F}$ is the number of Majorana fermions in the $\N=4$ vector multiplets of the parent $\N=4$ theory that remain massless after spontaneous breaking to $\N=0$. In other words, $C_V$ is the index that counts the number of massless bosonic degrees of freedom minus the number of massless fermionic degrees of freedom in the $\N=0$ sector $B$,
\be
C_V=8(2+d_{G_B}-n_{{\rm F}_B})\equiv \mbox{massless Bosons} - \mbox{massless Fermions}\quad \mbox{in the sector $B$} \, .
\ee

A simple way to evaluate $\Delta_B$ is based on the relation between the shifted lattices $\Gamma_{2,2}\big[{}^{\;0\;\;,\;\; 0}_{\alpha_B,\,\beta_B}\big]$ and the unshifted one, $\Gamma_{2,2}(T,U)$. For the cases $a)$, $b)$ and $c)$, we use respectively
\begin{align}
\Gamma_{2,2}\big[{}^{0,\,0}_{1,\,0}\big](T,U)&={\sum_{h,g}}'\Gamma_{2,2}\big[{}^{h,\,0}_{g,\;0}\big](T,U)+\cdots=2\, \Gamma_{2,2}\Big({T\over 2},2U\Big)\!-\Gamma_{2,2}(T,U)+\cdots,\nonumber\\
\Gamma_{2,2}\big[{}^{0,\,0}_{0,\,1}\big](T,U)&={\sum_{h,g}}'\Gamma_{2,2}\big[{}^{0,\,h}_{0,\,g}\big](T,U)+\cdots=2\, \Gamma_{2,2}\Big({T\over 2},{U\over 2}\Big)\!-\Gamma_{2,2}(T,U)+\cdots,\nonumber\\
\Gamma_{2,2}\big[{}^{0,\,0}_{1,\,1}\big](T,U)&={\sum_{h,g}}'\Gamma_{2,2}\big[{}^{h,\,h}_{g,\;g}\big](T,U)+\cdots=2\, \Gamma_{2,2}\Big({T\over 2},{1+U\over 1-U}\Big)\!-\Gamma_{2,2}(T,U)+\cdots,
\label{dlatt}
\end{align}
where  the primes indicate the sums are  over $(h,g)\neq (0,0)$. Using the well know integral~\cite{thresholds, ek}
\be
\int_\F{d^2\tau\over \tau_2}\left(\Gamma_{2,2}(T,U)-1\right)+\log{2e^{1-\gamma}\over \pi\sqrt{27}}=-\log\!\Big(4\pi^2\, \abs\eta(T)\abs^4\, \abs\eta(U)\abs^4 \Im T\Im U\Big) ,
\label{416}
\ee
one obtains
\be
\label{DDelta}
\Delta_B=-\log\!\left({\pi^2\over 4}\, \big|\theta[^0_1](T) \big|^4\,
\big|\theta\big[{}^{1-\beta_B}_{1-\alpha_B}\big](U) \big|^4\Im T\Im U  \right)\!+{\cal O}\!\left(e^{-c\,\sqrt{\Im T}}\right) ,
\ee
where $c$ is positive and of the order of the lowest mass of the massive spectrum divided by $M_{\rm s}$. This lowest non-vanishing mass depends on the modulus $U$, together with the moduli of the $\Gamma_{4,20}[^h_g]$-lattice present in the sector $B$ and introduced below Eq.~(\ref{theta}). Supposing that the order of magnitude of $U$ is not too far from 1, a fact that will be justified in Sect.~\ref{5}, and given the fact  that the $\Gamma_{4,20}[^h_g]$-lattice moduli are also not too far from 1, we have $c=\O(1)$. Moreover, since 
\be
\log \big\abs\theta[^0_1](T)\big\abs^4= \O\!\left(e^{-\pi\Im T}\right),
\ee
this contribution can be omitted in Eq.~(\ref{DDelta}). Thus,  the $\!\Im T$ volume dependence of $\Delta_B$ is only logarithmic. The key point for this is the following. In the integral (\ref{416}), the contribution $\tilde m^i=n^i=0$ in the unshifted lattice (\ref{latt}) is proportional to $\sqrt{\det G}=\Im T$, which is responsible for a ${\pi\over 3}\Im T$ dominant contribution in the result. On the contrary, the shifted lattice in $\Delta_B$ is expressed in 
Eq.~(\ref{dlatt}) as a difference of two unshifted lattices, where the contribution $\tilde m^i=n^i=0$ cancels out. 

For the second part of the thresholds, $Y_B$, and the effective potential, we use the fact that the contributions with non-trivial winding numbers $n^i$ in the lattice (\ref{latt}) are exponentially suppressed, 
\be
\Gamma_{2,2}\big[{}^{\; 0\;\;,\;\; 0}_{\alpha_B,\,\beta_B}\big]\!={\Im T\over \tau_2}\sum_{\tilde m_1,\tilde m_2}e^{-{\pi\over \tau_2}\!{\Im T\over\Im U}\abs\tilde m_1+{\alpha_B\over 2}+(\tilde m_2+{\beta_B\over 2})U\abs^2}+\cdots.
\ee
This expression also justifies that, at our level of approximation, we are free to extend the integration domain from $\F$ to the full upper half strip. This leads 
\begin{align}
Y_B&=-{2+d_{G_B}-n_{{\rm F}_B}\over 3\pi^3}\, {1\over  \Im T}\, E_{(\alpha_B,\beta_B)}(U\abs\, 2)+{\cal O}\!\left(e^{-c\,\sqrt{\Im T}}\right),\phantom{\underset{\underset{a}{\cdot}}{\abs}}\nonumber\\
{V_{\rm eff}}_B&=  -{2+d_{G_B}-n_{{\rm F}_B}\over 2\pi^7}\, {1\over (\Im T)^2}\, E_{(\alpha_B,\beta_B)}(U\abs \, 3)+{\cal O}\!\left(e^{-c\,\sqrt{\Im T}}\right),
\label{YY}
\end{align}
where we have defined ``shifted real analytic  Eisenstein series" as
\be
E_{(g_1,g_2)}(U\abs \, s)={\sum_{\tilde m_1,\tilde m_2}}^{\!\!\!\prime}{(\!\Im U)^s\over \abs\tilde m_1+{g_1\over 2}+(\tilde m_2+{g_2\over 2})U\abs^{2s}}\, .
\ee
In these functions, $g_1$ and $g_2$ are integer modulo 2 and the prime means $\tilde m_1=\tilde m_2=0$ is excluded from the sum when $g_1=g_2=0$. They satisfy modular properties as follows :
\be
E_{(g_1,g_2)}(M(U)\abs \, s)=E_{(g_1,g_2)M^T}(U\abs\,  s)\, , \;\; \mbox{where}\;\; M(U)={aU+b\over cU+d}\, , \;\; M=\left(\!\!\begin{array}{cc}a&b\\c&d\end{array}\!\!\right)\in SL(2,\Z)\, .
\ee
Note that the sign of the index $C_V=\mbox{massless Bosons}-\mbox{massless Fermions}$ in the sector $B$ is essential to discuss questions about moduli 
stabilization~\cite{CosmologicalTerm}.   

\subsection{$\boldsymbol{C}$ : The exact $\boldsymbol{\N=2}$ sector with $\boldsymbol{(h,g)\!=\!(0,0)}$, $\boldsymbol{(H,G)\!\ne\!(0,0)}$}
The partition function (\ref{theta}) associated to this sector, which we will denote by $C$, is proportional to 
$ \theta[^{1}_{1} ]^2(v) \, \theta[^{1+H}_{1+G} ]^2(v)={\cal O}{(v^2)}$. 
Thus, the contribution ${V_{\rm eff}}_C$ to the effective potential is zero, while the threshold correction $\Delta_C^i$ is not vanishing and  proportional to an  $\N=2$ $\beta$-function coefficient $b_C^i$.  
Two of the four gravitini are massless, while the masses of the other two are 
given in terms of the $T$ and $U$ moduli,
\be
m_{3\over 2}^{1,2}=0\, , \qquad m^{3,4}_{3\over 2}\equiv m_C={\sqrt{(\alpha_C \Im U)^2+(\alpha_C\abs\!\Re U\abs-\beta_C)^2} \over \sqrt{\Im T\Im U}}\, M_{\rm s}\, , 
\label{secC}
\ee
where we have 
\be 
(\alpha_C,\beta_C)=\left\{\!\!\begin{array}{ll}(\zeta,1) & \mbox{in case }a) \\(1,\zeta)& \mbox{in case }b) \\(\zeta,1-\zeta) &\mbox{in case }c)\, . \end{array}\right.
\ee
The threshold corrections in this sector are 
those of $\N=2$ theories that are obtained by an $\N=4\rightarrow  \N=2$  spontaneous susy breaking {\em via} a free $\Z_2$ orbifold action. They have been computed in Ref.~\cite{solving} but we briefly rederive the results we need here. 

The Lagrangian form of the lattice (\ref{latt}) implies the sector $H=1$ to be exponentially suppressed, when $\!\Im T\gg1$ and $U$ is finite. Keeping explicitly the sector $(H,G)=(0,1)$, one obtains using again $\theta[^1_1](v\abs\tau)=2\pi\eta^3(\tau)\, v+\O(v^3)$,
\begin{align}
\Delta^i_C=&\int_\F{d^2\tau\over \tau_2}\left\{\Gamma_{2,2}\big[{}^{\;0\;\;,\;\; 0}_{\alpha_C,\,\beta_C}\big]\!\left({\cal P}_i^2(2\bar w)-{k_i\over 4\pi\tau_2}\right)\bar \Omega(2\bar w)-b^i_C\right\}\!\bigg\abs_{\bar w=0}+b^i_C\log{2e^{1-\gamma}\over \pi\sqrt{27}}+\cdots,\nonumber \\
&\mbox{ where}\quad \bar \Omega(2\bar w)= {\theta[^1_0]^2\over 4\eta^2\bar\eta^4}Z_{4,20}\big[{}^{0,\, 0}_{0, \, 1} \big]\!\!\;(2\bar w) .
\end{align}
In fact, since the 4 directions associated to the $2^{\rm nd}$ and $3^{\rm rd}$ planes are twisted, $Z_{4,20}\big[{}^{0,\, 0}_{0, \, 1} \big]$ contains an overall  factor $\eta^2/\theta[^1_0]^2$ making $\bar \Omega$ an antiholomorphic function. The contribution $b_C^i$  to the full $\beta$-function coefficient  subtracts the infrared divergence.  Proceeding as in the  sector $B$, only the massless contributions dressed by the $\Gamma_{2,2}\big[{}^{\;0\;\;,\;\; 0}_{\alpha_C,\,\beta_C}\big]$-lattice are non-negligible, leading to formally identical results :
\be
\label{deltaC}
\Delta^i_C=b^i_C\Delta_C-k^iY_C\, , 
\ee
where
\begin{align}
\Delta_C&=-\log\!\left({\pi^2\over 4}\, \big|\theta[^0_1](T) \big|^4\,
\big|\theta\big[{}^{1-\beta_C}_{1-\alpha_C}\big](U) \big|^4\Im T\Im U  \right)\!+{\cal O}\!\left(e^{-c\,\sqrt{\Im T}}\right) ,
\phantom{\underset{\underset{a}{\cdot}}{\abs}}\nonumber \\
Y_C&=-{2+n_{{\rm V}_C}-n_{{\rm H}_C}\over 3\pi^3}\, {1\over  \Im T}\, E_{(\alpha_C,\beta_C)}(U\abs \, 2)+{\cal O}\!\left(e^{-c\,\sqrt{\Im T}}\right).
\label{deltaYC}
\end{align}
In the above expression, $n_{{\rm V}_C}$ and $n_{{\rm H}_C}$ are the numbers of massless vector multiplets and hypermultiplets in the sector $C$. Thus $n_{{\rm V}_C}$ is the dimension of the gauge group $G_C$ realized in this sector, while
\be
{\cal I}_C=n_{{\rm V}_C}-n_{{\rm H}_C}
\ee
is an index arising naturally from the extended supersymmetry we will denote $\N_C=2$. 
 As in sector $B$, the $\big\abs\theta[^0_1](T)\big\abs^4$-term can be omitted and the thresholds are only logarithmic in $\!\Im T$. 
As said before, it is interesting enough that in this sector the cosmological term vanishes, ${V_{\rm eff}}_C= 0$,  
thanks to the exact $\N_C=2$ supersymmetry.

\subsection{\!$\boldsymbol{D}$ : \!The  exact $\boldsymbol{\N=2}$ sector with 
$\boldsymbol{(h,g)\!=\!(H,G)}$, $\boldsymbol{(G,H)\!\ne\!(0,0)}$} 

We denote this sector as $D$. As in sector $C$, the partition function (\ref{theta})  vanishes, 
since it is proportional to $\theta[^{1+H}_{1+G} ]^2(v)\, \theta[^{1}_{1} ]^2(v) ={\cal O}{(v^2)}$. 
There is an exact $\N=2$ supersymmetry, which is not that of the sector $C$,  the two $\N=2$ susymmetries being {\it not aligned}. The two massless and two massive gravitini are not the same,
\be
m_{3\over 2}^{1,2}\equiv m_D={\sqrt{(\alpha_D \Im U)^2+(\alpha_D\abs\!\Re U\abs-\beta_D)^2} \over \sqrt{\Im T\Im U}}\, M_{\rm s}\, , \qquad m^{3,4}_{3\over 2}=0
\ee
and the non-vanishing masses are even different to those in sector $C$. This is due to the fact that the pairs $(\alpha_D, \beta_D)$ and  $(\alpha_C,\beta_C)$ are not equal,
\be 
(\alpha_D,\beta_D)=\left\{\!\!\begin{array}{ll}(1-\zeta,1) & \mbox{in case }a) \\(1,1-\zeta)& \mbox{in case }b) \\(1-\zeta,\zeta) &\mbox{in case }c)\, . \end{array}\right.
\label{secD}
\ee
Actually, we see that the sectors $C$ and $D$ are replaced by one another under the change $\zeta\to 1-\zeta$,
\be
\mbox{sector $C$} \leftrightarrow \mbox{sector $D$} \quad \Longleftrightarrow \quad \zeta\to 1-\zeta\, .
\ee
As a result, the threshold corrections to the gauge couplings are 
\be
\label{deltaD}
\Delta^i_D=b^i_D\Delta_D-k^iY_D\, , 
\ee
where
\begin{align}
\Delta_D&=-\log\!\left({\pi^2\over 4}\, \big|\theta[^0_1](T) \big|^4\,
\big|\theta\big[{}^{1-\beta_D}_{1-\alpha_D}\big](U) \big|^4\Im T\Im U  \right)+{\cal O}\!\left(e^{-c\,\sqrt{\Im T}}\right) ,\phantom{\underset{\underset{a}{\cdot}}{\abs}}
\nonumber \\
Y_D&=-{2+n_{{\rm V}_D}-n_{{\rm H}_D}\over 3\pi^3}\, {1\over  \Im T}\, E_{(\alpha_D,\beta_D)}(U\abs \,2)+{\cal O}\!\left(e^{-c\, \sqrt{\Im T}}\right).
\label{deltaYD}
\end{align}
$n_{{\rm V}_D}$ and $n_{{\rm H}_D}$ count the massless vector multiplets and hypermultiplets in the sector $D$, while
\be
{\cal I}_C=n_{{\rm V}_D}-n_{{\rm H}_D}
\ee
is the index arising from the second non-aligned extended supersymmetry we will denote $\N_D=2$.  Of course,  $n_{{\rm V}_D}$ is nothing but the dimension of the gauge group $G_D$ realized in this sector. As before, the $\big\abs\theta[^0_1](T)\big\abs^4$-term in $\Delta_D$ can be omitted and the contribution to the cosmological term vanishes : ${V_{\rm eff}}_D= 0$.

\subsection{$\boldsymbol{E \,\&\, F}$ : The $\boldsymbol{\N_{C,D}=2 \rightarrow \N_{C,D}=0}$ sectors $\boldsymbol{hG-gH\neq 0}$}  

The previous sectors $A,B,C,D$ have $(H,G)$ or $(h,g)$ equal to $(0,0)$, or  $(H,G)=(h,g)$. All these conditions are equivalent to saying that the determinant  $\big\abs{}^{h\;H}_{g\; \,G}\big\abs$ vanishes. In the remaining sectors, namely $E$ and $F$, one has $\big\abs{}^{h\;H}_{g\; \,G}\big\abs\neq 0$, which implies not only that $(H,G)\ne (0,0)$, but also that $(h,g)\ne (0,0)$ and $(h,g)\ne (H,G)$. In other words, the supersymmetries $\N_C=2$ of sector $C$ and $\N_D=2$ of sector $D$ are both broken to $\N_C=0$ and $\N_D=0$. Indeed, one finds that in the partition function (\ref{theta}), the left-moving part (including the four twisted left-moving internal coordinates) is not vanishing and universal, modulo the dressing with $\Gamma_{2,2}$ shifted lattices.
We display below the refined partition function in case $a)$, for the $\N_C=0$ sector $E$, 
\begin{align}
{1\over 2}\!\left.\Big(Z\big[{}^{1,\, 0}_{0, \, 1} \big]+Z\big[{}^{1,\, 1}_{1, \, 0} \big]+Z\big[{}^{0,\, 1}_{1, \, 1} \big]\Big)\!\right\abs_{2v,2\bar w}=&-8\, {\theta_4^2(v)\,\theta^2_3(v) \over \eta^6\,\theta^2_2} 
\, \Gamma_{2,2}\!\left[^{1,\, 0}_{\zeta,\, 1}\right] {\bar Z}\!\left[^1_0\big\abs{^0_1}\right]\!(2\bar w) \nonumber\\
&-8\, {\theta_3^2(v)\,\theta^2_2(v) \over \eta^6\,\theta^2_4} 
\, \Gamma_{2,2}\!\left[^{1-\zeta,\, 1}_{\,\;\,1\;\;\,,\, 0}\right] {\bar Z}\!\left[^1_1\big\abs{^1_0}\right]\!(2\bar w) \nonumber\\
& +8\, {\theta_2^2(v)\,\theta^2_4(v) \over \eta^6\,\theta^2_3} 
\, \Gamma_{2,2}\!\left[^{\,\;\,\zeta\;\;,\; 1}_{1-\zeta,\, 1}\right] {\bar Z}\!\left[^0_1\big\abs{^1_1}\right] \!(2\bar w)\, ,
\end{align}
and for the $\N_D=0$ sector $F$,
\begin{align}
{1\over 2}\!\left.\Big(Z\big[{}^{1,\, 0}_{1, \, 1} \big]+Z\big[{}^{0,\, 1}_{1, \, 0} \big]+Z\big[{}^{1,\, 1}_{0, \, 1} \big]\Big)\!\right\abs_{2v,2\bar w}=&+8\, {\theta_4^2(v)\,\theta^2_3(v) \over \eta^6\,\theta^2_2} 
\, \Gamma_{2,2}\!\left[^{\,\;\,1\;\;\,,\, 0}_{1-\zeta,\, 1}\right] {\bar Z}\!\left[^1_1\big\abs{^0_1}\right]\!(2\bar w) \nonumber\\
&+8\, {\theta_3^2(v)\,\theta^2_2(v) \over \eta^6\,\theta^2_4} 
\, \Gamma_{2,2}\!\left[^{\zeta,\, 1}_{1,\, 0}\right] {\bar Z}\!\left[^0_1\big\abs{^1_0}\right] \!(2\bar w)\nonumber\\
& -8\, {\theta_2^2(v)\,\theta^2_4(v) \over \eta^6\,\theta^2_3} 
\, \Gamma_{2,2}\!\left[^{1-\zeta,\, 1}_{\,\;\,\zeta\;\;,\; 1}\right] {\bar Z}\!\left[^1_0\big\abs{^1_1}\right] \!(2\bar w)\, .
\end{align}
In these expressions,  the $\bar Z$-factors are purely antiholomorphic.  The partition functions in case $b)$ are obtained from the above ones by exchanging the columns of the $\Gamma_{2,2}$-lattices. In case $c)$, the first columns of the  $\Gamma_{2,2}$-lattices are as above, while the second columns are obtained by changing $\zeta\to 1-\zeta$ in the first ones.

The key point here is that once  $\big\abs{}^{h\;H}_{g\; \,G}\big\abs\neq 0$, it is forbidden to have $h=H=0$ in the sectors $E$ and $F$. Therefore,  all individual terms in the associated partition functions are coupled with exponentially suppressed shifted lattices (see Eq.~(\ref{lattt})), when $\!\Im T$ is large and $U$ finite.  This shows explicitly  
that  in sectors $E$ and $F$, the contributions to the cosmological term and coupling constants can be neglected, 
\be
 \Delta_{E,F}={\cal O}\!\left(e^{-c\,\sqrt{\Im T}}\right) ,\quad 
 Y_{E,F}={\cal O}\!\left(e^{-c\,\sqrt{\Im T}}\right) , \quad {V_{\rm eff}}_{E,F}={\cal O}\!\left(e^{-c\,\sqrt{\Im T}}\right).
 \ee 
 
 %%%%%%%%%%%%%%%%%%%%%%%%%%%%%%%%%%%%%%%%%
\section{Analysis of  the $\boldsymbol{\N=4}$ and $\boldsymbol{1^{\rm st}}$  plane contributions}
\label{5}

Before investigating the $2^{\rm nd}$ and $3^{\rm rd}$ planes contributions   in  the $\Z_2\times \Z_2$ models  where only the $1^{\rm st}$ $\Z_2$ action is freely acting,  we would like to comment further on the structure of the corrections coming  
from the $\N=4\to \N=2\to \N=0$ susy breaking associated to the sectors $A$ to $F$  in these models.  Some explicit examples will also be given. Let us start by collecting the results found in the previous section :
\begin{description}

\item[\footnotesize $\bullet$]  In sector $A$, the contributions  to the effective potential, ${V_{\rm eff}}_A$, and  to the gauge thresholds, $\Delta^i_A$,  are always zero due to the ``mother'' $\N=4$ theory.

\item[\footnotesize $\bullet$]  There are two non-aligned $\N_C=2$ and $\N_D=2$ ``daughter'' supersymmetries  
in the sectors $C$ and $D$. In the former, the first two gravitini are massless, while in the latter the third and fourth gravitini are massless. Gauge coupling corrections $\Delta^i_{C,D}$ occur, while there are no contributions to the effective potential,   ${V_{\rm eff}}_{C,D}=0$.

\item[\footnotesize $\bullet$]  The sectors $E$ and $F$ are not supersymmetric and correspond to 
the breaking $\N_{C,D}=2\to \N_{C,D}=0$. 
However, their contributions ${V_{\rm eff}}_{E,F}$ and $\Delta^i_{E,F}$ are exponentially suppressed, when $\!\Im T$ is large and $U$  finite. 

\item[\footnotesize $\bullet$]  The contributions ${V_{\rm eff}}_{B}$ and $\Delta^i_{B}$ of the sector $B$ are the only ones arising from a non-supersymmetric sector. The latter realizes a spontaneous breaking of $\N=4$ to  $\N=0$.
Moreover, the sector $B$ is the only one that gives a non-vanishing (or non-negligible) cosmological term, which is  proportional to  $m_{3\over 2}^4 \equiv m_B^4 \propto 1/(\Im T)^2$.

\item[\footnotesize $\bullet$]  The non-trivial contributions to the gauge thresholds arise from the sectors $B$, $C$ and $D$. For any model $a)$, $b)$ or $c)$, with $\zeta=0$ or 1, $(\alpha_B,\beta_B)$, $(\alpha_C,\beta_C)$ and $(\alpha_D,\beta_D)$ take distinct values among the set $\{(1,0), (0,1), (1,1)\}$.  In fact, the 6 models realize the $3!$ allowed permutations of these parameters. 

\end{description} 

The contributions of  the sectors $A$ to $F$ are what is required to write the corrections to the gauge coupling constants and to the cosmological term in the $\Z_2$ non-symmetric orbifold models, where shifts along the untwisted plane realize an $\N=2\to \N=0$ spontaneous susy breaking \`a la SSS.
The running gauge couplings can be expressed in terms of the redefined  infrared regulator $Q^2 =\mu^2 {\pi^2\over 4}$, and are valid for $Q< m_B, m_C, m_D<c M_{\rm s}$, where $c$ is defined below Eq.~(\ref{DDelta}). They take the form
\begin{align}
{16\, \pi^2\over g_{i}^2(Q)} =  k^{i}{16\, \pi^2\over g_{\rm s}^2}  
&-{1\over 2}\big( b^i_{B}+b^i_{C}+b^i_{D}\big)\log  {Q^2\over M_{\rm s}^2} \nonumber \\
&- {1\over 2}b^i_{B}\log\!\left( \big|\theta\big[{}^{1-\beta_B}_{1-\alpha_B}\big](U) \big|^4
\Im T\Im U \right)\nonumber \\
& -{1\over 2}b^i_{C}\log\!\left( \big|\theta\big[{}^{1-\beta_C}_{1-\alpha_C}\big](U) \big|^4 \Im T\Im U \right)\nonumber \\
&-{1\over 2}b^i_{D}\log\!\left( \big|\theta\big[{}^{1-\beta_D}_{1-\alpha_D}\big](U) \big|^4 \Im T\Im U\right)+\O\!\left({1\over \Im T}\right),
\label{n=duo}
\end{align}
while the effective potential is
\be
\label{n=two}
V_{\rm eff}={1\over 2}{V_{\rm eff}}_B+{\cal O}\!\left(e^{-c\,\sqrt{\Im T}}\right) =  -{1\over 2}\, {2+d_{G_B}-n_{{\rm F}_B}\over 2\pi^7}\, {1\over (\Im T)^2}\, E_{(\alpha_B,\beta_B)}(U,3)+{\cal O}\!\left(e^{-c\,\sqrt{\Im T}}\right).
\ee
The  factors  ${1\over 2}$ in front of the $\beta$-function coefficients and in the expression of the  potential come from the normalization arising from  the
$\Z_2$ orbifold projection. The gravitino mass $m_{3\over 2}$ of the $\N=2\to \N=0$ model being equal to that of sector $B$, 
\be
\label{m}
m_{3\over 2}\equiv m_B={\sqrt{(\alpha_B \Im U)^2+(\alpha_B\abs\!\Re U\abs-\beta_B)^2} \over \sqrt{\Im T\Im U}}\, M_{\rm s}\, ,
\ee
the cosmological term is proportional to $m_{3\over 2}^4$. Note that no correction of order $M_{\rm s}^2m_{3\over 2}^2$ occurs.  
In order to make the physical interpretation of the gauge coupling threshold corrections
more transparent, it is convenient to introduce moduli-dependent mass scales,  
\begin{align}
\label{Mbcd}
&{1\over M_B^2} = {1\over M_{\rm s}^2}\, \big|\theta\big[{}^{1-\beta_B}_{1-\alpha_B}\big](U) \big|^4 \Im T\Im U\, , \nonumber \\
 &{1\over M_C^2} = {1\over M_{\rm s}^2}\, \big|\theta\big[{}^{1-\beta_C}_{1-\alpha_C}\big](U) \big|^4 \Im T\Im U\, ,  \nonumber\\
 &{1\over M_D^2} = {1\over M_{\rm s}^2}\, \big|\theta\big[{}^{1-\beta_D}_{1-\alpha_D}\big](U) \big|^4 \Im T\Im U\, ,  
\end{align}
in terms of which the coupling constant corrections for  $Q<M_B, M_C, M_D$ take the form  
\be
{16\, \pi^2\over g_{i}^2(Q)} = k^{i}{16\, \pi^2\over g_{\rm s}^2} 
-{1\over 2}b^i_{B}\log{Q^2\over M^2_B}
 -{1\over 2}b^i_{C}\log{Q^2\over M^2_C} 
 -{1\over 2}b^i_{D}\log {Q^2\over M^2_D}+\O\!\left({1\over \Im T}\right).
\label{n=deux}
\ee 

As we are going to see, the behavior of these thresholds depends crucially on the complex structure $U$.  In particular, the hierarchy between the moduli-dependent scales $M_B, M_C, M_D$ depends only on $U$.  To further investigate  the qualitative features of the $U$-dependence, we can focus on the particular susy breaking pattern of model $a)$, with $\zeta=0$, keeping in mind that the gauge coupling thresholds  in all six cases $a)$, $b)$, $c)$, with $\zeta=0,1$, are obtained by permutation of the defining expressions of the threshold scales $M_{B,C,D}$. In this  case, the shifted lattice involved in the threshold corrections is $\Gamma_{2,2}\big[{}^{0,\; 0}_{g,\, G}\big]$ and the susy breaking scales in sectors $B,C,D$ are  
\be
m_B={\abs U\abs\over \sqrt{\Im T\Im U}}\, M_{\rm s}\, , \quad m_C={1\over \sqrt{\Im T\Im U}}\, M_{\rm s}\, , \quad m_D={\sqrt{(\Im U)^2+(1-\abs\!\Re U\abs)^2} \over \sqrt{\Im T\Im U}}\, M_{\rm s}\, .
\ee
This shows that the scale  at which $\N=4$ is spontaneously broken to $\N=2$ is $m_C$, since $(g,G)=(0,1)$ is the value taken by $(\alpha_C,\beta_C)$. Similarly, the scale at which supersymmetry is spontaneously broken to $\N=0$ is $m_B$, since $(g,G)=(1,0)$ is the value taken by $(\alpha_B,\beta_B)$. These two scales  are relatively small compared to $M_{\rm s}$, as is also the third one, $m_D$, which  emerges for $(g,G)=(1,1)=(\alpha_D,\beta_D)$. 

To proceed, we specialize further to the situation where $\!\Re U=0$ and define 
\be
t=\!\Im T=R_1R_2\, , \qquad u=\!\Im U={R_2\over R_1}\, , 
\ee
where $R_1$ and $R_2$ are the radii of the shifted squared untwisted internal 2-torus. The susy breaking scales become 
\be
m^2_B= {u\over t}\, M_{\rm s}^2\, ,\qquad m^2_C={1\over tu}\, M_{\rm s}^2\, ,\qquad m^2_D=m_B^2+m_C^2\, ,
\ee
which implies $m_D$ is the largest one. The moduli-dependent scales $M_B$ and $M_C$  become
\begin{align}
&{1\over M_B^2} = {1\over M_{\rm s}^2}\, \big|\theta_2(iu) \big|^4 \, tu  ={1\over m^2_B} \,  \big|\theta_4({i/u}) \big|^4\, ,\nonumber \\
&{1\over M_C^2} = {1\over M_{\rm s}^2}\, \big|\theta_4(iu) \big|^4 \, tu  ={1\over m^2_C} \,  \big|\theta_4({iu}) \big|^4\, .
\end{align}
Utilizing the identity $\big|\theta_3(iu) \big|^4=\big|\theta_2(iu) \big|^4+\big|\theta_4(iu) \big|^4$, which is valid for pure imaginary arguments, we obtain the moduli-dependent threshold scale related to the  $\N_D=2$  supersymmetric sector $D$ as a function of $M_B$ and $M_C$, 
\be
{1\over M_D^2} = {1\over M_{\rm s}^2}\,\big|\theta_3(iu) \big|^4\, tu={1\over M_B^2}+{1\over M_C^2}\,.
\ee
This shows that in the present case, $M_D$ is the lowest threshold scale. This example is illuminating. It shows that the scales at which supersymmetry is restored in the sectors $B,C,D$ {\em are not} the associated gravitini masses $m_{B,C,D}$.  Instead, the relevant scales for supersymmetry restoration are the full threshold scales $M_{B,C,D}$, whose  hierarchy  differs from that of the scales $m_{B,C,D}$. For instance, since 
\be
M_B^2\sim m_B^2\, , \qquad 
M_C^2 \sim {m_B^2\over 16}\, e^{\pi/u}\, , \qquad 
M_D^2 \sim m_B^2\, ,\qquad \mbox{when }u\ll 1\, , 
\ee
 the full hierarchy of the threshold scales for small enough $u$  is 
$Q <M_D\le M_B \le M_C$, while we have $m_B<m_C\le m_D$. Moreover, in the limit where   $u$ is very  small, the scale $M_C$ grows exponentially, which gives  large corrections to the gauge couplings in Eq.~(\ref{n=deux}), proportional to $1/u=R_1/R_2$. On the contrary, since 
\be
M_B^2\sim {m_C^2\over 16}\, e^{\pi u}\, , \qquad 
M_C^2 \sim m_C^2\, , \qquad 
M_D^2 \sim m_C^2\, ,\qquad \mbox{when }u\gg1\, , 
\ee
the hierarchy of the threshold scales for large enough $u$  is $Q <M_D\le M_C \le M_B$, while $m_C<m_B\le m_D$. Furthermore, when $u$ is very large, the scale $M_B$, which grows exponentially with $u$, gives rise to large corrections to the couplings in Eq.~(\ref{n=deux}), proportional to  $u=R_2/R_1$. In the end, in both extreme limits summarized by the condition $u+1/u\gg1$, large linear corrections can  destroy the string perturbative  expansion, when  dressing $\beta$-function coefficients are negative. In such cases, one must assume that $u$ is not  too small or  large. 

In our low energy description, the range of permitted ratios $u={R_2 / R_1}$ can be derived by the requirement that the higher threshold scale must be smaller than the scale of the massive states we neglected \ie  $cM_{\rm s}$.   In general, the lowest threshold scale among $M_B$, $M_C$ and $M_D$ in Eq.~(\ref{Mbcd}) is the one that contains $\theta_3(U)$ in its definition. As we have just shown, this scale has a simple relation with the highest threshold scale in the extreme limits $u\gg1$ or $u\ll 1$.  The validity constraint in these two limits becomes
\be
{1\over 16}\, e^{\pi (u+1/u)}={M^2_{\rm high} \over M^2_{\rm low}}< 
c^2{M^2_{\rm s}\over M^2_{\rm low}}=
c^2{M^2_{\rm Planck}\over M_{\rm low}^{({\rm E})2}}\, ,
\ee
where $M_{\rm low}^{({\rm E})}$ is the lowest scale measured in the Einstein frame.
Notice that the ratio $M_{\rm high} /M_{\rm low}$ is independent of the frame. 
This gives the condition :
\be
u+{1\over u}< {2\over \pi}\, \log \!\left(4\, c \,{M_{\rm Planck} \over M_{\rm low}^{({\rm E})}}\right).
\ee
Assuming the lowest supersymmetry breaking scale measured in Einstein frame to be in the  1--10
TeV region, we take $M_{\rm low}^{({\rm E})} = {\cal O}(10^4)$ GeV, and given the gravity scale  
$M_{\rm Planck}=2.4\cdot10^{18}$ GeV, one finds for $c=\O(1)$ the permitted values of $u$ :
\be
u+{1\over u} < 22\, .
\ee

Once $u$ is in this region,  we can write the following  interpolating expression  
for the running gauge couplings, in terms of the physical energy scale measured in string frame, $Q< cM_{\rm s}$ (or $Q^{(\rm E)}\equiv Q/g_{\rm s}<cM_{\rm Planck}$ in the Einstein frame). It is valid for all supersymmetry breaking patterns \ie models $a)$, $b)$ or $c)$, with $\zeta=0$ or 1, and independently of the $U$-dependent hierarchy among the threshold scales $M_B$, $M_C$  and $M_D$ :
\be
{16\, \pi^2\over g_{i}^2(Q)} = k^{i}{16\, \pi^2\over g_{\rm s}^2} 
-{1\over 2}b^i_{B}\log\!\left({Q^2\over Q^2+M^2_B}\right)
\! -{1\over 2}b^i_{C}\log\!\left({Q^2\over Q^2+M^2_C}\right)
\! -{1\over 2}b^i_{D}\log\!\left( {Q^2\over Q^2+M^2_D}\right).
\label{n=2}
\ee
The above expression implements the successive decouplings of  
the  effective threshold mass scales $M_{B,C,D}$, which occur when the infrared cut-off scale $Q$ crosses them. $Q$ plays the role of a scattering  energy scale.  
For $Q$ smaller than the  three threshold scales, it can be neglected compared to them  and one recovers the threshold formula for small $Q$, Eq.~(\ref{n=deux}). Once $Q$  
becomes larger than one of the threshold scales, the latter can be neglected compared to $Q$, which is consistent with the fact that the whole tower of associated thresholds give negligible contribution. In particular : 
\begin{description}
\item[\footnotesize $\bullet$]  In the cases where the susy breaking pattern and the complex structure $U$  imply $M_B$ to be the lowest threshold scale, when the physical scale satisfies $M_B< Q< M_C, M_D$,  the two  non-aligned $\N_C=2$ and $\N_D=2$ supersymmetries  are restored. The full $\N=4$ supersymmetry is recovered when $Q$ is above $M_C$ and $M_D$.  
\item[\footnotesize $\bullet$]  In the cases where the susy breaking pattern and the complex structure $U$  imply $M_B$ to be the highest threshold scale, then the model describes a total $\N=4\to \N=0$ spontaneous susy breaking, when the physical scale satisfies $M_C, M_D<Q<M_B$. When $Q>M_B$, the full $\N=4$ supersymmetry is restored. 
\end{description}

\subsection{Example~1 : Gauge group factor $\boldsymbol{E_8}$ with $\boldsymbol{n_F=0}$}

Before analyzing the contributions of the $2^{\rm nd}$ and $3^{\rm rd}$ planes in the $\Z_2\times \Z_2$ models we consider, we would like to present typical examples in the $\Z_2$  case \ie where  $\N=4$ supersymmetry is spontaneously broken to $\N=2$ and further broken to $\N=0$  using the shifts and $T,U$ moduli of the untwisted $\Gamma_{2,2}$-lattice.  In fact, the $\beta$-function coefficients we are going to focus on can either be deduced by computing those associated to the sectors $B$, $C$ and $D$, or directly by considering the massless spectrum of the $\N=0$ theory. 

In our first example, we consider the models whose gauge groups contain  a factor $G^i=E_8$. The associated affine character in the adjoint representation, $\bar E_8(\bar \tau)$,  is realized  by 16  right-moving Majorana-Weyl worldsheet fermions,
\be
\bar E_8({\bar \tau})={1\over 2}\sum_{\gamma,\delta}{\bar \theta[^{\gamma}_{\delta} ]^8\over\bar \eta^{8}} \, .
\ee
The latter is factorized in the right-moving part of the partition function, whose relevant conformal block takes the form
\be
Z_{4,20}\big[{}^h_g\big\abs{^H_G} \big] \!= 
Z_{4,12}\big[{}^h_g\big\abs{^H_G} \big]  \bar E_8\, .
\ee 
The  adjoint character $\bar E_8$ can be written in terms of those associated to the adjoint and spinorial representations of $SO(16)$,
\be
\bar E_8=\bar O_{16} +\bar S_{16}\,,
\ee 
where our conventions for the holomorphic $SO(2N)$ characters are 
\begin{align}
&O_{2N}={\theta[^0_0]^N+\theta[^0_1]^N\over 2\eta^N}\, , \qquad \quad \,\,  V_{2N}={\theta[^0_0]^N-\theta[^0_1]^N\over 2\eta^N}\, ,\nonumber \\
&S_{2N}\,={\theta[^1_0]^N+(-i)^N\theta[^1_1]^N\over 2\eta^N}\, , \quad \!\!C_{2N}={\theta[^1_0]^N-(-i)^N\theta[^1_1]^N\over 2\eta^N}\, .
\label{charac}
\end{align}

Since the  character $\bar E_8$ is factorized, the gauge groups realized in the sectors $B,C,D$ contain a common factor, $G_B^i=G_C^i=G_D^i=E_8$. In sector $B$, the $\beta$-function coefficient arises from the bosonic part of an $\N=4$ vector multiplet (1 gauge boson + 6 real scalars) in the adjoint of $G^i_B$.  In the sectors $C$ and $D$, the $\beta$-function coefficients correspond to $\N_C=2$ and $\N_D=2$ vector multiplets in the adjoint of $G^i_C$ and $G^i_D$. Thus, we have
\be
b^i_B=-{8\over3} \, C(E_8)\, ,\qquad b^i_C=-2\, C(E_8)\, ,\qquad b^i_D=-2\, C(E_8)\, , 
\ee
where $C(E_8)=14+16=30$. The contribution 14 in $C(E_8)$ comes from the adjoint of $SO(16)$, $C(O_{16})=14$, while the contribution 16 comes from the spinorial of $SO(16)$, $C(S_{16})=16$. 
Thus, the sector by sector analysis leads to a $\beta$-function coefficient in the $\N=0$ theory given by 
\be
b^i={1\over 2} \big(b^i_B+b^i_C+b^i_D\big)=-{10\over3}\, C(E_8)=-100\, ,
\ee
which shows that the gauge theory is asymptotically free.

To cross check this value, we can directly compute $b^i$ from the point of view of an $\N=2\to \N=0$ spontaneously broken theory.  The massless spectrum contains the bosonic part of an $\N=2$ vector multiplet in the adjoint representation of $G^i$, namely 1 gauge boson for 2 real scalars, while the gauginos have become massive :
 \be
 b^i=\left(\!-{11\over 3} +{2\over 6}\right)\!  C(E_8)=-{10\over 3}\, C(E_8)\, .
  \ee
 
 \subsection{Example~2 : Gauge group factor $\boldsymbol{SO(16)}$ with $\boldsymbol{n_F\ne 0}$}
 \label{eg2}

The second models we would like to present have a gauge group factor $G^i=SO(16)$. The latter is obtained by coupling non-trivially the lattice shift $(h,g)$, with the $SO(16)$ spinorial representation initially present in the character $\bar E_8$. 
The coupling is implemented by a phase as follows :
\be
Z_{4,20}\big[{}^h_g\big\abs{^H_G} \big]\!= 
Z_{4,12}\big[{}^h_g\big\abs{^H_G} \big] Z_{0,8}\big[{}^h_g \big] \quad \mbox{where}\quad Z_{0,8}\big[{}^h_g \big]\!={1\over 2}
\sum_{\gamma, \delta} {\bar \theta[^{\gamma}_{\delta} ]^8\over\bar \eta^{8}} \,e^{i\pi(g\gamma +h\delta +hg)}\,,
\ee 
which breaks simultaneously $E_8\to SO(16)$ and supersymmetry to $\N=0$. 

The SSS phase changes effectively to
\be
S=e^{i\pi [g(a +\gamma)+h(b+\delta)]} \, .
\ee
This shows clearly that in the sector $B$, the fermions of the initially massless $\N=4$ vector multiplets in the $\bar O_{16}$ representation (\ie for $\gamma=0$) become massive, while the bosons remain massless. However, compared to Example~1, the new thing is that the situation is reversed for the states in the $\bar S_{16}$ representation (\ie for $\gamma=1$) : The bosons of the originally massless $\N=4$ vector multiplets become massive, while the fermions remain massless. In total, the gauge group factor in the non-supersymmetric sector $B$ is $G_B^i=SO(16)$ and the $\beta$-function coefficient is
\be
b^i_B=-{8\over3}\left\{C(O_{16})-C(S_{16})\right\} .
\ee
Notice that since $(h,g)=(0,0)$ in sector $C$, the gauge group factor $G_C^i=E_8$ is unbroken and the associated $\N_C=2$ supersymmetric $\beta$-function coefficient is identical to that of Example~1,
\be
b^i_C=-2\, C(E_8)\equiv -2\left\{C(O_{16})+C(S_{16})\right\} .
\ee
However, in sector $D$, where $(h,g)\neq (0,0)$, the $E_8$ gauge group is broken to $G_D^i=SO(16)$, with massless hypermultiplets in the spinorial representation $\bar S_{16}$. The $\N_D=2$ supersymmetric $\beta$-function coefficient is thus
\be
b^i_D=-2\left\{C(O_{16})-C(S_{16})\right\}.
\ee  
Taking into account the above sector by sector contributions, the  $\beta$-function coefficient of the $G^i=SO(16)$ non-supersymmetric gauge theory is 
\be
b^i={1\over 2}\big(b_B+b_C+b_D\big)=-{10\over3}\, C(O_{16})+{4\over3}\, C(S_{16})={76\over 3}\, .
\ee
Even if in this example the gauge theory  is non-asymptotically free, it remains a good exercize that illustrates the sector by sector analysis of the gauge threshold corrections. 

Here also, the agreement with the direct evaluation of the $\beta$-function coefficient of the $\N=2\to \N=0$ theory can be checked. This can be done in two steps. At the $\N=2$ level obtained by applying the $\Z_2$ action that breaks spontaneously $\N=4\to \N=2$, the massless spectrum contains an $\N=2$ vector multiplet in the adjoint representation of $G^i=SO(16)$, coupled to a hypermultiplet in the spinorial representation. Applying the final $\Z_2^{\rm shift}$ responsible for the $\N=2\to\N=0$ spontaneous breaking, the massless spectrum charged under the  $G^i=SO(16)$ gauge group factor are the bosons of the $\N=2$ vector multiplet in the adjoint representation of $SO(16)$, together with the fermions of  the hypermultiplet in the spinorial representation.  Consistently, one finds 
\be
b^i=\left(\!-{11\over 3}+{2\over 6}\right) \!C(O_{16})+{4\over 3}\, C(S_{16})=
-{10\over3}\, C(O_{16})+{4\over3}\, C(S_{16})\, .
\ee

\subsection{Example~3 : Gauge group factor $\boldsymbol{SO(8)\times SO(8)'}$ with $\boldsymbol{n_F\ne 0}$}
\label{eg3}
The third example we would like to present has a
$G^i=SO(8)\times SO(8)'$ gauge subgroup. It is obtained by coupling non-trivially both 
$(g,h)$ and $(G,H)$, with the vectorial and spinorial 
representations of  $SO(8)\times SO(8)'$ initially present in the $\bar E_8$ character :    
\be
Z_{4,20}\big[{}^h_g\big\abs{}^H_G \big]\!= 
Z_{4,12}\big[{}^h_g\big\abs{}^H_G \big] Z_{0,8}\big[{}^h_g\big\abs{}^H_G \big]\quad \mbox{where}\quad Z_{0,8}\big[{}^h_g\big\abs{}^H_G \big]\!={1\over 2}
\sum_{\gamma, \delta}  {\bar \theta^4[^{\gamma}_{\delta}]\, 
\bar \theta^4[^{\gamma+H}_{\delta+G}] \over\bar \eta^8} 
\, e^{i\pi(g\gamma +h\delta +hg +GH)} \, .
\ee 
As in Example~2, the coupling to $(h,g)$ breaks $E_8\to SO(16)$ and supersymmetry to $\N=0$, while the coupling to $(H,G)$ breaks further $SO(16)\to SO(8)\times SO(8)'$.  Here also, the SSS phase is effectively
\be
S=e^{i\pi [g(a +\gamma)+h(b+\delta)]}\, .
\ee

Since $(H,G)=(0,0)$ in sector $B$, the latter is identical to that of Example~2. Therefore, we have $G_B^i=SO(16)$, with $\beta$-function coefficient 
\be
b^i_B=-{8\over3}\left\{C(O_{16})-C(S_{16})\right\} .
\ee
However, since the overall gauge group factor of the model is $G^i=SO(8)\times SO(8)'$,  it is instructive to express the characters of $G_B^i=SO(16)$ in terms of  those of $SO(8)\times SO(8)'$ :
\be
\bar O_{16}=\bar O_{8} \bar O_{8}'+\bar V_8\bar V_8'\, , \qquad \bar S_{16}=\bar S_{8}\bar S_{8}'+  \bar C_{8}\bar C_{8}'\, .
\ee
Thus, the bosons of the initially massless $\N=4$ vector multiplets in the $\bar O_{16}$ representation (\ie for $\gamma=0$) are in the adjoint representation $\boldsymbol{(28, 1)\oplus (1, 28)}$ as well as in the bi-vectorial $\boldsymbol{(8_v, 8_v)}$ of $SO(8)\times SO(8)'$.  Moreover, the  fermions of the initially massless $\N=4$ vector multiplets in the $\bar S_{16}$ representation (\ie for $\gamma=1$) are in the $\boldsymbol{(8_s,8_s)}$ and $\boldsymbol{(8_c,8_c)}$ bi-spinorial representations of $SO(8)\times SO(8)'$. 

As said before, the model can be constructed by successive breaking,  
\be
\label{brea}
E_8\to SO(16)\to SO(8)\times SO(8)'\, , 
\ee
by first coupling the $SO(8)\times SO(8)'$ characters initially present in $\bar E_8$,
\be
\label{e8}
\bar E_8=\bar O_{16} +\bar S_{16}=\bar O_{8} \bar O_{8}'+\bar V_8\bar V_8'+\bar S_{8}\bar S_{8}'+  \bar C_{8}\bar C_{8}'\, ,
\ee
with $(h,g)$, and then with $(H,G)$. In the intermediate step, which is nothing but the sector $B$, the $G^i_B=SO(16)$ gauge theory is non-supersymmetric.  However, the analysis of the sectors $C$ and $D$ is more conveniently done by considering the model from two other viewpoints :
\begin{description}
\item[\footnotesize $\bullet$]  The breaking (\ref{brea}) can be realized by first coupling  the $SO(8)\times SO(8)'$ characters with $(H,G)$, and then with $(h,g)$. In the intermediate step, which is nothing but the sector $C$, we have an $\N_C=2$ supersymmetric $G_C^i=SO(16)$ gauge theory.
\item[\footnotesize $\bullet$]  The breaking (\ref{brea}) can also be realized by first coupling  the $SO(8)\times SO(8)'$ characters with $(h,g)=(H,G)$, and then with $(h-H,g-G)$. In the intermediate step, which is nothing but the sector $D$, we have an $\N_D=2$ supersymmetric $G^i_D=SO(16)$ gauge theory. 
\end{description}
Actually, the three intermediate gauge group factors $G^i_{B,C,D}=SO(16)$ are not aligned, so that the resulting unbroken gauge group of the combined final theory is $G^i=SO(8)\times SO(8)'$. Correspondingly, thanks to the triality symmetry of the three $SO(8)$ representations $\boldsymbol{8_v},\boldsymbol{8_s}, \boldsymbol{8_c}$, there are three alternative decompositions of the $SO(16)$ characters in terms of $SO(8)\times SO(8)'$ ones. If desired, these decompositions can be used to describe the spectra in sectors $B,C,D$ in terms of $SO(8)\times SO(8)'$ representations. They are  
\begin{align}
\mbox{in sector $B$ : }   \quad &\bar O_{16}=\bar O_{8} \bar O_{8}'+\bar V_8\bar V_8'\, , \quad \;\bar S_{16}=\bar S_{8}\bar S_{8}'+  \bar C_{8}\bar C_{8}'\, ,\nonumber\\
\mbox{in sector $C$ : }  \quad  & \bar O_{16}=\bar O_{8} \bar O_{8}'+\bar S_8\bar S_8'\;\, , \quad \bar S_{16}=  \bar C_{8}\bar C_{8}'+\bar V_{8}\bar V_{8}' ,\nonumber\\
\mbox{in sector $D$ : } \quad   & \bar O_{16}=\bar O_{8} \bar O_{8}'+\bar C_8\bar C_8'\; , \quad \bar S_{16}= \bar V_{8}\bar V_{8}'+\bar S_{8}\bar S_{8}'\, .
\end{align}
In any case, what we are interested in is  the massless spectrum in sector $C$, charged under the gauge group factor $G^i_C=SO(16)$. To find it, we start from the parent $\N=4$ theory, where the massless spectrum contains an $\N=4$ vector multiplet in the adjoint representation of $E_8$. Implementing the $(H,G)$-projection and using the fact that $\bar E_8=\bar O_{16}+\bar S_{16}$, we obtain the sector $C$, whose massless spectrum lies schematically in the representation   
\be
\label{sC}
\big(\N_C=2 \mbox{ vector multiplet}\big)\cdot \bar O_{16}\oplus \big(\N_C=2 \mbox{ hypermultiplet}\big)\cdot \bar S_{16}\, .
\ee
We have an $\N_C=2$ vector multiplet in the adjoint representation and a hypermultiplet in the spinorial representation, so that 
\be
b^i_C=-2\left\{C(O_{16})-C(S_{16})\right\}.
\ee  
By symmetry between the sectors $C$ and $D$, we also have in sector $D$ for the gauge group factor $G^i_D=SO(16)$,
\be
b^i_D=-2\left\{C(O_{16})-C(S_{16})\right\}.
\ee  
Combining the above results, the  $\beta$-function coefficient of the $G^i=SO(8)\times SO(8)'$ non-supersymmetric gauge theory is 
\be
b^i={1\over 2}\big(b_B+b_C+b_D\big)=-{10\over3}\left\{C(O_{16})- C(S_{16})\right\}={20\over 3}\, .
\ee
Here also, the gauge theory is non-asymptotically free. 

To check the above value of $b^i$, we can derive the massless spectrum of the theory that is charged under $G^i=SO(8)\times SO(8)'$. We have just seen that the implementation of the $(H,G)$-projection on the parent $\N=4$ theory leads to the massless spectrum of sector $C$, given in 
Eq.~(\ref{sC}). Using the decomposition of the $SO(16)$ characters in terms of $SO(8)\times SO(8)'$ ones valid in sector $C$, this spectrum can be written as 
\be
\big(\N_C=2 \mbox{ vector multiplet}\big)\cdot \big(\bar O_8\bar O_8'+\bar S_8\bar S_8'\big)\oplus \big(\N_C=2 \mbox{ hypermultiplet}\big)\cdot \big(\bar C_8\bar C_8'+\bar V_8\bar V_8'\big) .
\ee
We can now implement the final $(h,g)$-projection, which let us with  massless states schematically as follows :
\begin{align}
\big(\mbox{bosons of the vector multiplet}\big)\cdot \bar O_8\bar O_8' &\oplus\big(\mbox{fermions of the vector multiplet}\big)\cdot \bar S_8\bar S_8'\oplus \nonumber\\
\big(\mbox{bosons of the hypermultiplet}\big)\cdot \bar V_8\bar V_8'&\oplus 
\big(\mbox{fermions of the hypermultiplet}\big)\cdot \bar C_8\bar C_8'\, .
\end{align}
We have 1 gauge boson and 2 real scalars in the adjoint representation  
of $SO(8)\times SO(8)'$, $\boldsymbol{(28, 1)\oplus (1, 28)}$, together with 4 real  scalars 
in the $\boldsymbol{(8_v, 8_v)}$, and 4 Majorana fermions in the $\boldsymbol{(8_s, 8_s)\oplus (8_c, 8_c)}$. Since the gauge coupling of $G^i$ is equal to that of each of its $SO(8)$ subgroups, it is sufficient to calculate the $\beta$-function coefficient associated to one of them :
\be
b^i=\left(\!-{11\over 3}+{2\over 6}\right)\!C(O_8)+{4n(V_8')\over 6}\, C(V_8)+{4n(S_8')\over 3}\, 
C(S_8)+{4n(C_8')\over 3}\, C(C_{8})\, ,
\ee
where $C(O_8)=6$, $C(V_8)=C(S_8)=C(C_8):= C_{(8)}=1$ and the   
multiplicities arising from the second $SO(8)'$ factor are all equal, $n(V_8')=n(S_8')= n(C_8')=8$. In total, one has
\be
b^i=-{10\over 3}\, C(O_8)+ {80\over 3}\, C_{(8)} ={20\over 3}\, ,
\ee
which is in agreement with the sector by sector contributions. 

\subsection{The generic case}

The above examples illustrate  the  {\it universal  structure} of the running effective gauge couplings valid in the $\Z_2$ non-symmetric orbifold models that realize a spontaneous $\N=4\to \N=2\to \N=0$ supersymmetry breaking \`a la SSS, when  shifts (but no dual shifts)   are introduced along the untwisted torus. In these models, no dangerous linear dependence on the internal volume appears in the threshold corrections. The  result is given in Eq.~(\ref{n=deux}) for $Q<M_{B,C,D}$ (or Eq.~(\ref{n=2}) for $Q < cM_{\rm s}$), with the sector by sector $\beta$-function coefficients given by : 
\be
b^i_B=-{8\over3}\left\{C(O_B)-C({\cal R}_B) \right\} , ~~b^i_C=-2\left\{ C(O_C)-C({\cal R}_C) \right\} ,
~~b^i_D=-2\left\{C(O_D)-C({\cal R}_D)\right\}  .
\ee
The structures of the sectors $C$ and $D$  are simple to understand, since both of them describe $\N=2$ supersymmetric gauge theories. The associated gauge groups contain  factors $G^i_C$ and $G^i_D$, which may be different. The individual $\beta$-function coefficients are given in terms of vector multiplets contributions in the adjoint representations of $G^i_{C,D}$, denoted by  $-2C(O_{C,D}$), together with hypermultiplets contributions in the representations ${\cal R}_{C,D}$, denoted by $2C({\cal R}_{C,D})$. 

On the contrary, the structure of sector $B$, which describes a non-supersymmetric gauge theory with a gauge group factor $G^i_B$, is something new.  
The $-{8\over 3}C(O_B)$ contribution to $b_B^i$  comes from the  
bosons of  initially massless $\N=4$ vector multiplets in the parent $\N=4$ model, that remain massless. These bosons (1 vector and 2 real scalars) are in the adjoint representation of $G^i_B$. The second contribution, ${8\over 3}C({\cal R}_B)$, arises from the fermions  of  initially massless $\N=4$ vector multiplets in the parent theory, that remain massless. They are 4 Majorana  fermions in a spinorial representation ${\cal R}_B$. If as in Examples 2 and 3, ${\cal R}_B$ is  a spinorial representation of $SO(16)$, it is in general the spinorial representation of a subgroup of $E_8$, such as   $SO(16)$, $SO(8)\times SO(8)'$, $E_7\times SU(2)$, $SO(12)\times SO(4)$ or even $SO(4)^4$. All these cases can be easily realized by fermionic constructions.

%%%%%%%%%%%%%%%%%%%%%%%%%%%%%%%%%%%%%%%%%%
 \section{$\boldsymbol{2^{\rm nd}}$ plane, $\boldsymbol{3^{\rm rd}}$ plane and $\boldsymbol{\N=1}$ sector contributions\,: $\boldsymbol{(H_2 , G_2)\neq (0,0)}$}
 \label{6}
 
In Sects~\ref{twistshift} and \ref{5}, we have extensively analyzed the threshold corrections in $\Z_2$ non-symmetric orbifold models, where an $\N=4\to\N=2\to \N=0$ spontaneous breaking of supersymmetry is implemented with shifts such that the  running gauge couplings develop only logarithmic dependencies on the volume of the untwisted internal 2-torus. Up to an additional overall factor of ${1\over 2}$, these results are the contributions of the $\N=4$ and $1^{\rm st}$ complex plane in $\Z_2\times \Z_2$ non-symmetric orbifold models. In the present section, we proceed with the evaluations of the contributions arising from the remaining sectors, namely the $2^{\rm nd}$  and $3^{\rm rd}$ complex planes,  and the $\N=1$ sector.  All of them are twisted, with $(H_2,G_2)\neq(0,0)$. Moreover, the $2^{\rm nd}$   plane has $(H_1,G_1)=(0,0)$, the $3^{\rm rd}$ plane has $(H_3,G_3)=(0,0)$ and the $\N=1$ sector has $(H_1,G_1)\neq (0,0)$, $(H_1,G_1)\neq (H_2,G_2)$. 

 Our concern in the present paper is the decompactification problem only. In particular, we do not address  the issue of chirality and the models presented here are actually incompatible with the physical requirement that the $\N=1$ spectrum (further spontaneously broken to $\N=0$ \`a la SSS) be chiral. Forgetting for the moment the final breaking to $\N=0$, we remind that  at the level of $\N=1$ supersymmetric models constructed {\em via}  $\Z_2\times \Z_2$ non-symmetric moduli-deformed fermionic construction, the chiral  families always come from the $\N=2$ twisted sectors that have non-trivial fixed points. Thus, the  $\N=1$ untwisted spectrum ($H_1=H_2=0$ projected by $G_1,G_2$) is always non-chiral, while the spectrum arising in the $1^{\rm st}$ plane ($H_1=1$, $H_2=0$ projected by $G_1,G_2$) is massive, when the $1^{\rm st}$ $\Z_2$ action acts freely so that  no fixed point arises in this plane. To understand why the spectrum realized in the $2^{\rm nd}$ and $3^{\rm rd}$ planes is also non-chiral in this case, we reverse the role of the two $\Z_2$ actions. The freely acting $\Z_2$ can be viewed as responsible of  an  $\N=2\to \N=1$ spontaneous breaking of supersymmetry on a $T^6/\Z_2$ parent model. This means that in the large volume limit  of the $1^{\rm st}$ internal 2-torus, one recovers an $\N=2$ spectrum.  
However, in the $2^{\rm nd}$ and $3^{\rm rd}$ complex planes, the $1^{\rm st}$ torus, which is shifted, is also twisted since $H_2=1$. The spectrum arising from these sectors  is thus independent of the moduli $T_1, U_1$ and is {\em identical} to the non-chiral one present in the large volume limit, where $\N=2$ is recovered. In the case of $(2,2)$ compactifications, which correspond to Calabi-Yau internal spaces at fermionic points, the Euler characteristic vanishes~\cite{Ferrara:1995yx}.

Taking into account the final breaking of $\N=1\to \N=0$,  we have  in the partition function (see Appendix~\ref{deformed} and Ref.~\cite{RcorrectionsDuals})
\be
Z_{2,2}\Big[{}^{h_1^1,\, h_1^2} _{g_1^1,\, g_1^2} \Big \abs {}^{H_2}_{G_2} \Big]\!=\left\{
\begin{array}{ll}
\displaystyle {\Gamma_{2,2}\Big[{}^{h_1^1,\, h_1^2} _{g_1^1,\, g_1^2} \Big](T_1,U_1)\over (\eta\bar\eta)^2}\, , \phantom{\underset{\underset{a}{\cdot}}{\abs}}&\mbox{when }  (H_2, G_2)=(0,0)\, ,\\ 
\displaystyle {4\eta \bar \eta\over\ \theta[^{1-H_2}_{1-G_2} ]\, \bar\theta[^{1-H_2}_{1-G_2}]}\; 
\delta_{\big\abs{}^{h_1^1\;\;H_2}_{g_1^1\;\; G_2}\big\abs,0 \,\mbox{\scriptsize mod} \,2}\;\delta_{\big\abs{}^{h_2^1\;\;H_2}_{g_2^1\;\; G_2}\big\abs,0 \,\mbox{\scriptsize mod} \,2}\, , &\mbox{when }(H_2, G_2)\neq(0,0)\, ,
\end{array}
\right.
\label{toto}
\ee
where the  shifts $(h^i_1,g^i_1)$ are defined in Eq.~(\ref{lattt}) (we remind that $H,G$ denote $H_1,G_1$).  Therefore, the twisted spectrum of the  $2^{\rm nd}$ and $3^{\rm rd}$ complex planes ($H_2=1$ projected by $G_1,G_2$) 
is independent of the gravitino mass $m_{3\over 2}^{({\rm E})}\equiv m_B/g_{\rm s}$, which is in the desired 1--10 TeV region. The contributions of these modes to the partition function are identical to those evaluated at the fermionic point. We proceed by arguing that the SSS phase in the sectors $(H_2,G_2)\neq (0,0)$ must not break supersymmetry to $\N=0$. The reason for this comes in three steps. First, in the $2^{\rm nd}$ and  $3^{\rm rd}$ planes, the sub-sectors with $(h,g)=(0,0)$ always preserve $\N=2$ supersymmetry, and since  we choose to restrict to the case where they arise from non-free $\Z_2$ actions,  the order of magnitude of the moduli $T_I, U_I$, $I=2,3$, must be close enough to 1 for the decompactification problem not to arise. Second, if the $(h,g)\neq (0,0)$ sub-sectors of the $2^{\rm nd}$ and  $3^{\rm rd}$ planes were non-supersymmetric, the respective gravitini mass scales would be determined by $T_I, U_I$, $I=2,3$,  and thus of order $M_{\rm Planck}$, when measured in Einstein frame, which is something we want to exclude. Third, the $(h,g)\neq (0,0)$ sub-sector of the $\N=1$ sector must preserve supersymmetry as well, in order to not lead to an extremely large gravitino mass. To summarize, in our solution to the decompactification problem, the SSS phase $S$ in the sectors $(H_2,G_2)\neq (0,0)$ must not contain the factor $e^{i\pi (ag+bh+hg)}$ introduced in Eq.~(\ref{S}), which would otherwise break susy to $\N=0$ at tree level  at a high scale. The breaking of supersymmetry is transmitted to the $2^{\rm nd}$ and  $3^{\rm rd}$ planes twisted spectra ($H_2=1$ projected by $G_1,G_2$) by quantum corrections that involve states with broken supersymmetry ($H_2=0$ projected by $G_1,G_2$). To summarize,  the spectrum arising from the  $2^{\rm nd}$ and  $3^{\rm rd}$ planes presents at tree level an $\N=2$ extended supersymmetry and is non-chiral.  

Note that since the sectors $(H_2,G_2)=(0,0)$ and  $(H_2,G_2)\neq(0,0)$ are independent orbits of the worldsheet modular group, the associated choices of SSS phases do not need to be correlated to guaranty the consistency of the whole $\Z_2\times \Z_2$ model. In the sectors  $(H_2,G_2)\neq(0,0)$, a certainly valid susy preserving choice is $S\equiv 1$. However, playing with the quantum numbers $(H_1,G_1)$ and $(H_2,G_2)$, we can have
\be
\mbox{In the sectors $(H_2,G_2)\neq(0,0)$, } \quad S=e^{i\pi[\zeta_1 (aG_1+bH_1+H_1G_1)+\zeta_2 (aG_2+bH_2+H_2G_2)]}\, ,
\ee
where $\zeta_1$ and $\zeta_2$ can be fixed to 0 or 1. As we just noticed, $\zeta_1$ may not be equal to $\zeta'$ we introduced in Eq.~(\ref{S}). To see that $(\zeta_1,\zeta_2)=(0,0)$ is not the only allowed choice, we consider the conformal block associated to the left-moving fermionic degrees of freedom, 
\begin{align}
&{1\over 2}
\sum_{a,b}e^{i\pi (a+b+ab)}\, e^{i\pi[\zeta_1 (aG_1+bH_1+H_1G_1)+\zeta_2 (aG_2+bH_2+H_2G_2)]}\, \theta[^a_b ] (2v)\, \theta [^{a+H_2}_{b+G_2}]\, \theta[^{a+H_1}_{b+G_1}]\, \theta[^{a-H_1-H_2}_{b-G_1-G_2} ] \, \nonumber\\
& =e^{i\pi (\zeta_1+\zeta_2)(H_1G_2 - G_1H_2)}\,  e^{i\pi (G_1+G_2)(1+H_1+H_2)}\, \theta[^{1}_{1}](v) \, 
 \theta[^{1-H_2}_{1-G_2} ](v)\, \theta[^{1-H_1}_{1-G_1}](v) \, 
 \theta[^{1+H_1+H_2}_{1+G_1+G_2} ](v)  \, .
\end{align}
To show this equality, one can redefine  $a=A-\zeta_1 H_1-\zeta_2H_2$, $b=B-\zeta_1 G_1-\zeta_2G_2$ and sum over $A,B$ equal to 0 or 1. Given that  $(H_2,G_2)\neq (0,0)$, we see that $\N=2$ supersymmetry is preserved in the $2^{\rm nd}$ plane, $(H_1,G_1)=(0,0)$,  and  in the $3^{\rm rd}$ plane, $(H_1,G_1)=(H_2,G_2)$, (or $(H_3,G_3)=(0,0)$). Supersymmetry is also preserved in the $\N=1$ sector, $\big\abs{}^{H_1\;H_2}_{G_1\; G_2}\big\abs\neq 0$. Two distinct cases arise however, $\zeta_1=\zeta_2$ or $\zeta_1=1-\zeta_2$, corresponding to different choices of discrete torsions that yield opposite contributions of the $\N=1$ sector to the partition function.  

The $\N_I=2$, $I=2,3$, unbroken supersymmetries of the  $2^{\rm nd}$ and $3^{\rm rd}$  planes are  {\it not aligned} to one another, as well as 
{\em non-aligned}  with the $\N_C=2$ and $\N_D=2$ supersymmetries appearing in the sectors $C$ and $D$ of the  $1^{\rm st}$ complex plane. 
Being supersymmetric, the  $2^{\rm nd}$ plane, $3^{\rm rd}$ plane and $\N=1$ sector do not contribute to the effective potential. Moreover, their contributions to the gauge coupling thresholds are  identical to those present in the $\N=1$ supersymmetric $\Z_2\times\Z_2$ moduli-deformed fermionic models.  In this class of theories, the $\N=1$ sectors do not contribute. The reason for this is that the helicity operator ${\cal Q}[^a_b]$ acting on an $\N=1$ sector involves
\be
\partial_v^2\Big(\theta[^1_1](v)\, \theta[^{1-H_2}_{1-G_2}](v)\, \theta[^{1-H_1}_{1-G_1}](v)\,  \theta[^{1+H_1+H_2}_{1+G_1+G_2}](v)\Big)\!\Big\abs_{v=0}\propto \partial_v^2\Big(\theta_1(v)\, \theta_2(v)\, \theta_3(v)\,  \theta_4(v)\Big)\!\Big\abs_{v=0}=0\, ,
\ee
thanks to the fact that $\theta_1(v)$ is odd and $\theta_{2,3,4}(v)$ are even.
Therefore, corrections to the gauge couplings occur only from the $\N=2$ planes. The case of $\N=2$ planes in symmetric orbifolds, which are characterized by $(2,2)$ superconformal symmetry, have been analyzed extensively in the literature~\cite{solving}. However, even if the analysis for  non-symmetric orbifolds that posses $(2,0)$ superconformal symmetry has not yet been fully completed, our conclusions will remain valid in this case, as mentioned  later in   this section. 
 
 Let us start by considering the $2^{\rm nd}$ and $3^{\rm rd}$ planes in the $(2,2)$ case.  As was shown in Refs~\cite{thresholds, solving,universality,pr},  the gauge coupling  corrections are given in terms of two threshold  functions,  
\be
\Delta^i_I=b^i_I\Delta(T_I,U_I)-k^i Y(T_I,U_I)\, ,\quad I=2,3\, ,
\label{66a}
\ee
where $b^i_I$ are the $\N=2$ $\beta$-function coefficients in each  planes\footnote{\label{foot}In our conventions, $b^i_I$, $I=2,3$, are  $\beta$-function coefficients in the parent theories obtained by acting with a single $\Z_2$. In the $\Z_2\times \Z_2$ models we are interested in, overall factors ${1\over 2}$ must be included in the r.h.s. of Eq.~(\ref{66a}), for the thresholds to the correctly normalized.},  
\begin{align}
\Delta(T_I,U_I)&=-\log\!\left(4\pi^2 \big\abs\eta(T_I)\big\abs^4 \,\big\abs\eta(U_I)\big\abs^4 \Im T_I  \Im U_I\right),\phantom{\underset{\underset{a}{\cdot}}{\abs}}\nonumber\\
Y(T_I,U_I)&=-{\xi \over 12}\int_{\cal F}{d^2\tau\over \tau_2}\,
\Gamma_{2,2}(T_I,U_I) \!\left[ \Big(\bar E_2-{3\over \pi\tau_2}\Big){\bar E_4 \bar E_6\over \bar\eta^{24}}-\bar \jmath+1008 \right] .
\label{67}
\end{align}
In these expressions, $E_{2,4,6}$ are holomorphic Eisenstein series, with modular weights 2,4,6,
\begin{align}
E_2&= {12\over i \pi}\, \partial_\tau\log \eta =1-24\sum_{n=1}^{\infty}{n\, q^n\over 1-q^n}\, ,\nonumber\\
E_4&= {1 \over 2}(\theta_2^8+\theta_3^8+\theta_4^8)=1+240\sum_{n=1}^{\infty}{n^3q^n\over 1-q^n}\, ,\nonumber \\
E_6&={1\over 2}({\theta}_2^4 + \theta_3^4)(\theta_3^4 + \theta_4^4)(\theta_4^4 - \theta_2^4)
=1-504\sum_{n=1}^{\infty}{n^5q^n\over 1-q^n}\, ,
\label{61b}
\end{align}
while $j={1\over q}+744+\O(q)$ is holomorphic and modular invariant. 
$\xi$ is a constant that can be expressed in terms of the numbers
of massless vector multiplets and hypermultiplets per plane.  
Using the relation between gauge and $\R^2$-term 
renormalizations~\cite{RcorrectionsDuals}, it is fixed to $\xi=-1$,  thanks to the anomaly cancellation conditions~\cite{sixDimAnom} valid in the six dimensional decompactification limits~\cite{solving, RcorrectionsDuals}. This property being general in all $\N=2$ theories with underlying $(2,2)$ superconformal symmetries, the threshold corrections are universal in this case~\cite{ek,universality}, modulo the $\beta$-function coefficients and Kac-Moody levels.

As anticipated, what is relevant to note is that these threshold corrections scale linearly with the volume of the untwisted 2-tori. For $\!\Im T_I\gg 1$ and $U_I$ finite, one has
\be
\Delta(T_I,U_I)={\pi\over 3}\Im T_I-\log(\!\Im T_I)+\O(1)\; ,\quad Y(T_I,U_I)=4\pi \Im T_I+\O\!\left({1\over\Im T_I}\right),
\ee
which invalidates the string perturbative expansion (when the dressing $\beta$-function coefficient is negative). As follows from target space duality, similar dangerous behaviors occur in all limits, where the  K\"ahler and/or complex structures of the untwisted 2-tori are large or small : $T_I\to \infty$ or 0, and/or $U_I\to \infty$ or 0. This is not a surprise, since we have seen in the previous sections (and also in Ref.~\cite{solving})  that for the linear terms not to arise, $\N=4$ supersymmetry must be restored on the moduli space boundary. However, this cannot be the case in our $2^{\rm nd}$ and $3^{\rm rd}$ complex planes, since the breaking from $\N=4$ to $\N=2$ in these sectors is not spontaneous.  As announced before in this section, 
these considerations force us to assume that the order of magnitude of the moduli of the $2^{\rm nd}$ and $3^{\rm rd}$  planes, $T_I$ and $U_I$, $I=2,3$, are not too far from 1. This justifies that we took the order of magnitude of the  coefficient $c$ introduced in Eq.~(\ref{DDelta}) to be not far from 1. Moreover, the moduli-dependent scales $M_I$'s that control the threshold corrections are   
\begin{align}
{1\over M_I^2}&= {16 \over M^2_{\rm s}}\, \big\abs\eta(T_I)\abs^4 \, \big\abs\eta(U_I)\abs^4\Im T_I \Im U_I\, ,\qquad I=2,3\phantom{\underset{\underset{\cdot}{\cdot}}{\cdot}}\nonumber\\
&= {16\over M^2_{\rm Planck}}\, \big\abs\eta(T_I)\big\abs^4 \, \big\abs\eta(U_I)\big\abs^4\Im S_{\rm dil}\Im T_I \Im U_I\, ,
\end{align}
and are close to the string scale $M_{\rm s}$. In the above expression, we introduce the string coupling constant, which is related to the dilaton field,  $g_{\rm s}^2 =1/\Im S_{\rm dil}$, in order to display the threshold masses in  units of gravitational scale.  

The contributions $b_I^i\Delta(T_I,U_I)$ controlled by the $M_I$'s have to be completed by the universal contribution $-k_iY(T_I,U_I)$, whose order of magnitude is close to 1. Being infrared finite, these corrections are continuous functions that remain finite even at special values of $(T_I,U_I)$, where additional massless states arise. Thus, we are free to absorb them in a redefinition of the string coupling~\cite{pr} :
\be
{16\, \pi^2\over g_{\rm renor}^2}={16\, \pi^2\over g_{\rm s}^2}-{1\over 2}Y(T_2,U_2)-{1\over 2}Y(T_3,U_3)\, ,
\ee  
where the factors ${1\over 2}$ arise from  the action of the second $\Z_2$ (see Footnote \ref{foot}) and the ``renormalized'' string coupling is
\be
 g_{\rm renor}^2={g_{\rm s}^2\over  1-{1\over 32\pi^2}\big(Y(T_2,U_2)+Y(T_3,U_3)\big)\, g_{\rm s}^2}\, .
\ee

When the $2^{\rm nd}$ and $3^{\rm rd}$ complex planes are realized as $(2,0)$ non-symmetric compactifications {\em via} fermionic constructions, the natural values for 
$\!\Im T_I$ and $\!\Im U_I$ are of order 1. 
Moreover, the target space dualities $SL(2, \Z)_{T_I}\times SL(2, \Z)_{U_I}$ of the $(2,2)$ case are broken to some sub-groups. Consequently, $\big\abs\eta(T_I)\big\abs^4$ and $\big\abs\eta(U_I)\big\abs^4$ are replaced  by products of other modular functions, with however identical weights.  In all cases, $(2,2)$ and $(2,0)$, the orders of magnitude of the dressed threshold scales $M_I$, $I=2,3$, remain close to the string scale.

We are now ready to collect all our previous results and present the 1-loop effective potential and running gauge couplings arising in $\Z_2\times\Z_2$ moduli-deformed fermionic construction. We consider models where $\N=1$ supersymmetry is further spontaneously broken to $\N=0$ at a scale in the TeV regime, $m_{3\over 2}^{({\rm E})}= {\cal O}$(1--10) TeV, while  the validity of perturbation theory is preserved. Our work is restricted to the case where only the $1^{\rm st}$ $\Z_2$ action is free. The second one and the product of the two have fixed points. Under these conditions, only one internal 2-torus, the $1^{\rm st}$ in our conventions, is large and involved in the $\N=4\to \N=2$ breaking and $\N=1\to \N=0$ breaking, which are both spontaneous. This was done by introducing suitable shifts along this torus but dual shifts may have been considered.

In these models, we find remarkable  that the $\N=4$ sector spontaneously broken to $\N=0$, which is referred as sector $B$, is the only one leading to a substantial contribution
to the effective potential (the cosmological term), when $m_B\equiv m_{3\over 2}$ is small  compared to $cM_{\rm s}$, 
\be
\label{n=two'}
V_{\rm eff}={1\over 4}{V_{\rm eff}}_B+{\cal O}\!\left(e^{-c\,\sqrt{\Im T_1}}\right) =  -{1\over 4}\, {2+d_{G_B}-n_{{\rm F}_B}\over 2\pi^7}\, {1\over (\Im T_1)^2}\, E_{\alpha_B,\beta_B}(U_1\abs \,3)+{\cal O}\!\left(e^{-c\,\sqrt{\Im T_1}}\right),
\ee
which is proportional to $m_{3\over 2}^4$. 
Moreover, the relevant threshold corrections to the gauge couplings arise from the sector $B$, as well as from four sectors exhibiting  exact $\N=2$ supersymmetries : The sectors $C$ and $D$, which are actually sub-sectors of the  ``massive" $1^{\rm st}$ complex plane,  and the $2^{\rm nd}$ and $3^{\rm rd}$ complex planes.  The associated $\N_C$, $\N_D$, $\N_{2}$, $\N_{3}=2$ supersymmetries are all non-aligned.  
These five contributions   to the gauge coupling thresholds are characterized by effective mass scales : $M_{B,C,D}$ depend on the  ``massive"  $1^{\rm st}$  plane moduli $T_1, U_1$, while $M_I$, $I=2,3$, depends on the $I^{\rm th}$ plane moduli $T_I, U_I$ and is modular invariant, with respect to some target space duality sub-group of $SL(2,\Z)_{T_I}\times SL(2,\Z)_{U_I}$.   

The running  effective coupling constants in the $\N=1\to \N=0$ models take a very simple form, once expressed in terms of the dressed mass scales and coupling $g_{\rm renor}$, 
\begin{align}
\label{thfinal}
{16\, \pi^2\over g_{i}^2(Q)} = k^{i}{16\, \pi^2\over g_{\rm renor}^2} 
&-{1\over 4}b^i_{B}\log\!\left({Q^2\over Q^2+M^2_B}\right)
\! -{1\over 4}b^i_{C}\log\!\left({Q^2\over Q^2+M^2_C}\right)
\! -{1\over 4}b^i_{D}\log\!\left( {Q^2\over Q^2+M^2_D}\right)\phantom{\underset{\underset{\cdot}{\cdot}}{\abs}}\nonumber \\
 & -{1\over 2} b^i_{2} \log\!\left({Q^2\over M^2_{2}} \right) 
 \!-{1\over 2}b^i_{3}\log\!\left( {Q^2\over M^2_{3}}\right) ,
\end{align}
where $Q<cM_{\rm s}$ is the energy scale measured in string frame ($Q^{({\rm E})}<cM_{\rm Planck}$ in the Einstein frame) and the sector by sector $\beta$-coefficients are
\begin{align}
b^i_{B}=-{8\over3}\left\{C(O_B)-C({\cal R}_B) \right\}, \; b^i_{C}&=-2\left\{C(O_C)-C({\cal R}_C)\right\}, \; b^i_{D}=-2\left\{C(O_D)-C({\cal R}_D)\right\},\nonumber \\
b^i_{2}&=-2\left\{C(O_2)-C({\cal R}_2)\right\},\; b^i_{3}=-2\left\{C(O_3)-C({\cal R}_3)\right\} .
\label{bi}
\end{align}
The $O_{B,C,D,2,3}$ and ${\cal R}_{B,C,D,2,3}$ symbols refer to adjoint and matter representations of gauge group factors  $G^i_{B,C,D,2,3}$ that are realized in the sectors $B,C,D$ and $I=2,3$, respectively. In total,  the $\beta$-function coefficient of the $\N=1\to \N=0$ model, for $Q$ smaller than all threshold scales, is given by
\be
b^i={1\over 4}\big (b^i_{B}+b^i_{C}+b^i_{D}\big)+{1\over 2}\big(b^i_{2}+b^i_{3}\big) .
\label{btot}
\ee
When $\!\Im U_1=\O(1)$, the  dressed masses measured in Einstein frame, $M_{B,C,D}^{({\rm E})}=M_{B,C,D}/g_{\rm s}$, are all in the TeV region. Thus, they decouple in Eq.~(\ref{thfinal}), when $Q^{({\rm E})}=Q/g_{\rm s}$ reaches larger energy scales, thanks to the restoration of $\N=4$ supersymmetry in the sector $B$ and  $1^{\rm st}$ plane.  When $\!\Im U_1$ or $1/\Im U_1$ is larger, say up to 20 or so, only two scales among $M_{B,C,D}^{({\rm E})}$ are in the TeV region, while the remaining one can be up to $cM_{\rm Planck}$. In this case, the full restoration of $\N=4$ supersymmetry in the sector $B$ and $1^{\rm st}$ plane occurs only at energies above this highest threshold scale. 
In Eq.~(\ref{thfinal}), the reason why we do not add $Q^2$-terms in front of the $M_I^2$'s, $I=2,3$, is that the order of magnitude of these two threshold masses is close to the string scale $M_{\rm s}$, and that in our effective description, the physical energy $Q$ must not exceed $cM_{\rm s}$.

From the effective field theory viewpoint, the SSS susy breaking gives rise to a specific $\N=1$ supergravity no-scale model, with so-called 
``$S_{\rm dil}T_1U_1$''-breaking mechanism~\cite{FKZ}. We remind that $S_{\rm dil}$ is the four dimensional dilaton, while $T_1,U_1$ are the moduli of the  ``massive" $1^{\rm st}$  complex plane.   The moduli of the $2^{\rm nd}$ and $3^{\rm rd}$ planes do not participate in the supersymmetry breaking. As explained in Ref.~\cite{FKZ}, the 
determination  {\em via} radiative corrections of the vacuum expectation value of the   ``no-scale modulus" and thus of the $\N=1$  gravitino mass  $m^{({\rm E})}_{3\over 2}$~\cite{noscale}, at relatively low scale of order 1--10 TeV, requires  that  the genus-1 effective potential is free from terms  that scale like $\big(m^{({\rm E})}_{3\over 2}\big)^2\, \Lambda^2$. In such  terms, $\Lambda$ is the cut-off of the effective field theory, which in principle can be as large as $M_{\rm Planck}$ or  $M_{\rm s}$.  Thus, it is remarkable that in the setup we consider  in this work to break spontaneously $\N=1\to \N=0$, such terms are absent, thanks to the underlying $\N=4\to \N=0$ supersymmetry breaking structure of the sector $B$, which imposes the genus-1 effective potential to scale like~$\big(m^{({\rm E})}_{3\over 2}\big)^4$.
     
%%%%%%%%%%%%%%%%%%%%%%%%%%%%%%%%%%%%%%%%%%
\section{Conclusions} 
\label{8}

In this paper, our concern is to implement a low scale spontaneous breaking of supersymmetry in  $\N=1$ models,  while maintaining the validity of gauge coupling perturbation theory. We address this question within the context of $\N=1$ $\Z_2\times \Z_2$ non-symmetric orbifolds, realized by moduli-deformed fermionic constructions. At the $\N=1$ supersymmetric level, it is known that an $\N=2$ complex plane realized as an $\N=4\to \N=2$  spontaneous breaking of supersymmetry yields threshold corrections to the gauge couplings, with a mild logarithmic  dependence on the  complex plane volume~\cite{solving}.  This contrasts with the case where the $\Z_2$ action responsible of the $\N=4$ breaking to $\N=2$ is not freely acting. Indeed, a linear dependence of the thresholds on the complex plane volume arises in this case, invalidating perturbation theory once the volume is large.  What we have shown in the present work is that the above solution to the ``decompactification problem'' can be extended to the case where $\N=1$ supersymmetry is further spontaneously broken to $\N=0$ at a low scale, by implementing an additional  $\Z_2^{\rm shift}$ orbifold shift acting along the large internal dimensions and coupled with the helicity charges $(a,b)$. 

To arrive at this conclusion, we develop a sector by sector analysis of the models and analyze systematically the associated induced threshold corrections. We find that one of the $\Z_2$ twists, which for instance preserves the $1^{\rm st}$ complex plane, must act freely. Restricting to the case where no ``dual shifts" are implemented along this plane, the $\Z_2$ twist acts on it as a shift. Allowing the volume of the $1^{\rm st}$ plane to be large, we can further implement the $\Z_2^{\rm shift}$ shift responsible for the susy breaking to $\N=0$ along this plane. As desired, the gravitino mass $m_{3\over 2}^{({\rm E})}$ generated this way is low. We find that taking into account the first $\Z_2$ (which has a free action) and the additional $\Z_2^{\rm shift}$ only,  three sub-sectors denoted as  $B$, $C$ and $D$ contribute substantially to the thresholds. What is meant by ``substantially"  is that other sub-sectors that are non-supersymmetric contribute in the $1^{\rm st}$ complex plane, but their  effects are however exponentially suppressed when the gravitino  mass  is small, $m_{3\over 2}^{({\rm E})}\ll M_{\rm Planck}$.  Moreover, this hierarchy allows another great simplification, since it implies the contributions of the massive excitations of the string are also exponentially suppressed, compared to those arising from the Kaluza-Klein towers of states above the charged massless states.

 The above discussion is general if the $2^{\rm nd}$ $\Z_2$ twist and the diagonal product of both $\Z_2$'s have fixed points. 
In this case, the $2^{\rm nd}$ and $3^{\rm rd}$ planes do not arise from a spontaneous breaking of $\N=4$ supersymmetry and their volume (in Planck units) and shape moduli must be close to 1, in order not to introduce the decompactification problem back.  In addition, supersymmetry has to be preserved at tree level in these sectors, since otherwise an extremely large gravitino mass would be generated. These  two planes are the remaining sectors that contribute to the thresholds. Of course, other models where both $\Z_2$ actions (and eventually their diagonal product as well) are freely acting could be analyzed. In these cases, both the $1^{\rm st}$ and $2^{\rm nd}$ (and eventually the $3^{\rm rd}$) internal 2-tori are allowed to be large and involved in the spontaneous breaking of the supersymmetries.   

In total, the five relevant sectors   in the $\Z_2\times \Z_2$  models we consider in the present paper, which have $\N=1$ supersymmetry spontaneously broken to $\N=0$ at low scale \`a  la SSS, are as follows : 
\begin{description}
\item[\footnotesize $\bullet$] The sector $B$, describes the $\N=0$ spontaneously broken phase of the $\N=4$ spectrum of the initial parent theory. Surprisingly, {\em this sector is the only non-supersymmetric one that is relevant for the gauge coupling thresholds and effective potential.} In fact, the other sectors relevant for the gauge couplings being supersymmetric, the sector $B$ is solely responsible for the generation of the cosmological term. The latter  is proportional to $\big(m_{3\over 2}^{({\rm E})}\big)^4$ and no $M_{\rm Planck}^2 \big(m_{3\over 2}^{({\rm E})}\big)^2$ term is induced.

\item[\footnotesize $\bullet$] The sectors $C$ and $D$, which are both sub-sectors of the non-chiral $1^{\rm st}$ complex plane, preserve  $\N_C=2$ and $\N_D=2$ supersymmetries, respectively. 

\item[\footnotesize $\bullet$] The $2^{\rm nd}$ and $3^{\rm nd}$ chiral complex planes preserve $\N_2=2$ and $\N_3=2$ supersymmetries, respectively.
\end{description}

The gauge coupling thresholds arising from the above sectors are controlled by associated mass scales, which are functions of the K\"alher and complex structures $T_I, U_I$ of the corresponding planes, $I=1,2,3$. In the $1^{\rm st}$ plane, the smallest of the masses $M_{B}^{({\rm E})}$,  $M_{C}^{({\rm E})}$ and $M_{D}^{({\rm E})}$ is about  1--10 TeV (as is the case for all of them if $U_1\simeq i$). However, any hierarchy among these scales can be achieved by permuting the formal expressions of  $M_{B,C,D}^{({\rm E})}$, which can be done by changing the pattern of shifts along the $1^{\rm st}$ complex plane. On the contrary,  in the $2^{\rm nd}$ and $3^{\rm rd}$ planes, $M_{2}^{({\rm E})}$ and $M_{3}^{({\rm E})}$ are close to the Planck scale. Finally, additional universal contributions of order 1 arising from these $2^{\rm nd}$ and $3^{\rm rd}$ planes correct slightly the large inverse bare coupling, $k^i/g_{\rm s}^2$.  

What we have found is the complete dependence of the running effective gauge couplings on the physical scale $Q^{({\rm E})}$, up to $cM_{\rm Planck}$, including when $Q^{({\rm E})}$ crosses the thresholds scales $M_{B,C,D}^{({\rm E})}$ and that the associated Kaluza-Klein towers of states decouple from the thresholds.  The upper bound $cM_{\rm Planck}$, where $c$ is not far from 1, is the order of magnitude of the massive string modes in Planck units, whose exponentially suppressed contributions have been neglected. The result, displayed in Eq.~({\ref{thfinal}), takes a universal form that depends only on the $\beta$-function coefficients associated to the above listed five relevant sectors. Moreover, the form itself of the $\beta$-function coefficients is universal, Eqs (\ref{bi}). The factors $\mp 2$ in the coefficients $b_C^i, b_D^i, b_2^i, b_3^i$ arise from the massless  vector multiplets and hypermultiplets charged under the gauge group factors $G^i_{C,D,2,3}$, which are realized in each sectors. The factors $\mp{8\over 3}$ in $b_B^i$ follow from specific truncations to $\N=0$ of the massless $\N=4$ vector multiplets in the parent models : 1 vector boson plus 6 real scalars contribute  $-{8\over 3}$, while 4 Majorana fermions contribute ${8\over 3}$. All these states are charged under a gauge group factor $G_B^i$, realized in the sector $B$.

  While the early examples of realistic 
free fermionic models consisted in isolated examples \cite{earlyffmodels},
in more recent years, systematic classification methods have been developed that enable scanning large classes of three generations models, with viable phenomenological  properties \cite{classi}.  
However, in all these vacua \cite{earlyffmodels, classi}, as well as in other quasi-realistic heterotic string models \cite{others}, $\N=1$ supersymmetry is unbroken and its spontaneous breaking to $\N=0$ needs to be implemented. When this is done {\em via}  Stringy Scherk-Schwarz mechanism in $\Z_2\times \Z_2$ fermionic construction, the conditions  for the present solution to the decompactification problem to be valid are however incompatible with the physical requirement that the spectrum be chiral (the large volume limit of the $1^{\rm st}$ internal 2-torus leads to an $\N=2$ spectrum and the twisted spectra of the $2^{\rm nd}$ and $3^{\rm rd}$ planes are independent of this volume). Thus, implementing an $\N=1\to \N=0$ spontaneous breaking of supersymmetry in a realistic, chiral model, while preserving perturbation theory remains a challenge.  We also note the recent work of Ref.~\cite{ad} on the partition functions of non-supersymmetric heterotic string vacua.

%%%%%%%%%%%%%%%%%%%%%%%%%%%%%%%%%%%%%
%%%%%%%%%%%%%%%%%%%%%%%%%%%%
  
 \section*{Acknowledgement}
 
We are grateful to  
C. Angelantonj, I. Antoniadis,  D. Luest, I. Florakis, J. Rizos and N. Toumbas 
for fruitful discussions. 
A.F. acknowledges the 
Laboratoire de Physique Th\'eorique of Ecole Normale Sup\'erieure, 
the Centre de Physique Th\'eorique of Ecole Polytechnique, 
the CERN theory division, the theoretical physics department of Oxford University
and the Mainz Institute for Theoretical Physics
for hospitality. 
C.K. and H.P. would like to thank the University of Liverpool for hospitality, as well as the CERN theory division and the university of Cyprus where a part of this work was done.  
C. K. acknowledges
the theoretical physics group of Ludwig Maximilians University and 
Max-Planck-Institute in Munich for hospitality.\\
The work of A.F.  is partially supported by the UK
Science and Technology Facilities Council (STFC) under grant number ST/G00062X/1. The work of C.K. and H.P is partially supported by  the ANR 05-BLAN-NT09-573739 and EU Program IRSES UNIFY No 269217, the CEFIPRA/IFCPAR 4104-2 contract  and a PICS France/Cyprus. The work of C.K. is partially supported by his  Gay Lussac-Humboldt Research Award 2014, in the  Ludwig Maximilians University and Max-Planck-Institute for Physics.  The work of H.P. is partially supported by the EU contracts PITN GA-2009-237920, ERC-AG-226371 and PICS  France/Greece, France/USA.
%\newpage

%%%%%%%%%%%%%%%%%%%%%%%%%%%%%%%%%%%%%%%%%%%%%%%%
%%%%%%%%%%%%%%%%%%%%%%%%%%%%%%%%%%%%%%%%%%%%%%%%

\appendix

\section{The moduli-deformed  fermionic construction}
\label{deformed}
\renewcommand{\theequation}{A.\arabic{equation}}
\setcounter{equation}{0}

The context of our study is within the framework of fermionic constructions, where marginal $(1,1)$-current-current deformations are implemented. We restrict to the introduction of the moduli $T_I$ and $U_I$, $I=1,2,3$, associated to the three internal 2-tori involved in the $\Z_2\times \Z_2$ models in bosonic language. The goal of this appendix is to review the procedure to achieve these deformations. Throughout this paper, our definition for the $\theta$-functions is, for $\alpha, \beta\in\Real$, 
\be
\label{th}
\theta[^\alpha_\beta](v\abs \tau)=\sum_m q^{{1\over 2}(m-{\alpha\over 2})^2}e^{2i\pi(v-{\beta\over 2})(m-{\alpha\over 2})}\,  , \quad \mbox{where } \quad q=e^{2i\pi \tau}\, .
\ee

%%%%%%%%%%%%%%%%%%%%%%%%%%%%%%%%%%%%%%%%%%%%%%%

\subsection{One coordinate compactification}

In the fermionic construction, one deals with two holomorphic and two antiholomorphic worldsheet Majorana-Weyl fermions $\omega, y,  \bar \omega, \bar y$, rather than an internal compactified coordinate $X(z,\bar z)=X_L(z)+X_R(\bar z)$. The well known fermion-boson equivalence in two dimensions is established {\em via} the definitions and identifications
\be
\psi= {\omega+iy\over \sqrt{2}}\equiv \; :\!e^{i\sqrt{2} X_L}\!:\, , \qquad \bar \psi= {\bar\omega-i\bar y\over \sqrt{2}}\equiv \; :\!e^{-i\sqrt{2}X_R}\!: \,,
\ee
where the periodicity of $X$ is $2\pi R_0$, with $R_0=1/\sqrt{2}$.
These systems lead to the same $U(1)$ left- and right-moving current algebras generated by
\be
J_L=\, :\!\psi\psi^*\!:\, \equiv i\sqrt{2}\, \partial X\, ,\qquad    J_R=\, :\!\bar\psi\bar \psi^*\!:\,\equiv -i \sqrt{2}\, \bar\partial X\, .
\ee

In general, a non-left/right-symmetric model involves sectors characterized by specific boundary conditions of the complex fermions $\psi$, $\bar\psi$ on the genus one Riemann surface,
\begin{align}
&\psi(z+1)=e^{i\pi(\gamma+h_L)}\,\psi(z)\, , \qquad \bar\psi(\bar z+1)=e^{-i\pi(\gamma+h_R)}\,\bar\psi(\bar z)\, , \nonumber\\
&\psi(z+\tau)=e^{i\pi(\delta+g_L)}\,\psi(z)\; , \qquad \bar\psi(\bar z+\bar\tau)=e^{-i\pi(\delta+g_R)}\,\bar\psi(\bar z)\; , 
\label{BC}
\end{align}
whose bosonic counterpart for the chiral bosons $X_{L,R}$ leads
\begin{align}
&X_L(z+1)=X_L(z)+\pi R_0 (\gamma+h_L)\, ,\qquad X_R(\bar z+1)=X_R(\bar z)+\pi R_0 (\gamma+h_R)\, , \nonumber\\
&X_L(z+\tau)=X_L(z)+\pi R_0 (\delta+g_L) \; , \qquad X_R(\bar z+\bar \tau)=X_R(\bar z)+\pi R_0 (\delta+g_R)\; .
\label{shi}
\end{align}
In the above notations, which are chosen for later convenience, $\gamma$ and $\delta$ are integers, while $h_L, g_L$ and $h_R, g_R$ are real constants referred as left-moving and right-moving shifts.  

The model can also involve a $\Z_2$ twist action on the bosonic coordinate, $X\to -X$, whose translation in fermionic language is $(\omega, y,\bar \omega ,\bar y)\to (\omega, -y,\bar \omega, -\bar y)$, \ie
\be
\psi\to \psi^*\, , \quad \bar\psi\to \bar \psi^*\qquad \Longleftrightarrow \qquad X\to -X\, .
\ee 
In this case, four sectors labeled by pairs $(H,G)$ of integers arise, as dictated by the  boundary conditions
\begin{align}
& J_L(z+1)=(-1)^H J_L(z)\, , \qquad J_R(\bar z+1)=(-1)^H J_R(\bar z)\, ,\nonumber \\
& J_L(z+\tau)=(-1)^G J_L(z)\, , \qquad J_R(\bar z+\bar \tau)=(-1)^G J_R(\bar z)\, .
\label{HG}
\end{align}

The marginal deformation we want to consider is implemented by the addition of the operator $\lambda J_LJ_R$, known as Thirring interaction in fermionic language,  in the worldsheet Lagrangian density. Its effect in the bosonic picture is clear, since  $\lambda J_LJ_R=2\lambda \partial X\bar\partial X$, which corresponds  to a change of circle squared radius, $R^2_0\to R^2=R^2_0(1+2\lambda)$. In the fermionic picture, the Thirring interactions can be totally absorbed by changing the boundary conditions of the worldsheet complex fermions $\psi, \bar \psi$. 

%%%%%%%%%%%%%%%%%%%%%%%%%%%%%%%%%%%%%%%%%%%%%%%
\vspace{0.2cm}
\noindent{\em \large Untwisted sector} 

\noindent In the present context, we refer as ``untwisted" the sector where $J_L$ and $J_R$ are periodic along both directions of the worldsheet torus, $(H,G)=(0,0)$.
The contribution to the one-loop partition function of the complex left-moving and right-moving fermions $\psi$, $\bar\psi$ is expressed in terms of $\theta$-functions according to the boundary conditions (\ref{BC}),
\be
\label{fermio}
{\theta[^{\gamma +h_L}_{\delta+g_L}]\, \bar\theta[^{\gamma +h_R}_{\delta+g_R}]\over\eta  \bar \eta}\equiv e^{i\pi{h^{\prime}\over 2}(\delta+{\hat g}')} \, Z_{1,1}\Big[{}^{\gamma; \, h_L, \, h_R }_{\delta; \; g_L, \; g_R }\Big \abs{}^0_0 \Big](R_0)\, ,
\ee
where the r.h.s. expresses the result in the bosonic picture, which is valid at the fermionic point $R_0$. The bosonic side involves naturally 
\be
(h^{\prime}, g^{\prime})=( h_L-h_R,  g_L- g_R)\, , \qquad ({\hat h}',{\hat g}')=\Big({h_L+h_R\over 2}, {g_L+g_R\over 2}\Big)  , 
\label{hh}
\ee
in terms of which  we have for arbitrary radius $R$,
\be
Z_{1,1}\Big[{}^{\gamma; \, h_L, \, h_R }_{\delta; \; g_L, \; g_R }\Big \abs{}^0_0 \Big](R)
={R\over\sqrt{\tau_2}\eta\bar\eta}\sum_{\tilde m,n}
e^{-{\pi R^2 \over \tau_2}  \left\abs(\tilde m-{g^{\prime}\over 2})+(n-{h^{\prime} \over 2})\tau \right\abs^2 + i\pi\left(\tilde m {{\hat h}'}- n{{\hat g}'}\right)}
 \, e^{i\pi\left(\tilde m n+\tilde m {\gamma}- n{\delta}\right)}\, .
\ee
The identity (\ref{fermio})  can be  derived by writing the powers of $q$ and $\bar q$ in the $\theta$-functions as $q^{{1\over 2}(m-{\gamma+h_L\over 2})^2}\bar q^{{1\over 2}(m-n-{\gamma+h_R\over 2})^2}$ and performing a Poisson resummation on the momentum charge $m$~\cite{Kounnas:1990ww}.
The phase $ e^{i\pi{h^{\prime}\over 2}(\delta+{\hat g}')}$  expresses the non-trivial behavior of the $(1,1)$-conformal block under modular transformation, while $Z_{1,1}$ is modular covariant. Actually,  $Z_{1,1}$ couples the modular covariant $\Gamma_{1,1}$-lattice shifted by $(h^{\prime},g^{\prime})$ and $({\hat h}',{\hat g}')$,
\be
\Gamma_{1,1}\Big[{}^{h^{\prime}, \, {\hat h}' }_{g^{\prime}, \, {\hat g}' } \Big](R)=
{R\over\sqrt{\tau_2}}\sum_{\tilde m,n}
e^{-{\pi R^2 \over \tau_2}  \left\abs(\tilde m-{g^{\prime}\over 2})+(n-{h^{\prime} \over 2})\tau \right\abs^2 + i\pi\left(\tilde m {{\hat h}'}- n{{\hat g}'}\right)}\, ,
\ee
to the characters $(\gamma,\delta)$ {\it via} the modular invariant phase $e^{i\pi\left(\tilde m n+\tilde m {\gamma}- n{\delta}\right)}$. The modular transformations act as
\begin{align}
\label{motrans}
\tau\to -{1\over \tau} \quad &\Longleftrightarrow \quad (h^{\prime},g^{\prime})\to (h^{\prime},g^{\prime}) \S\,, \quad ({\hat h}',{\hat g}')\to ({\hat h}',{\hat g}')\S\, ,\quad  (\gamma,\delta)\to (\gamma,\delta)\S\, , \nonumber\\
\tau\to \tau+1 \quad &\Longleftrightarrow \quad (h^{\prime},g^{\prime})\to (h^{\prime},g^{\prime})\T\,, \quad({\hat h}',{\hat g}')\to ({\hat h}',{\hat g}')\T\, ,\quad  \!(\gamma,\delta)\to (\gamma,\delta+\gamma-1)\, ,\nonumber\\
&\!\mbox{where} \qquad \S=\left(\!\!\begin{array}{cc} 0 &-1\\ 1 & 0\end{array}\!\!\right)\, , \qquad \T=\left(\!\!\begin{array}{cc} 1 &1\\ 0 & 1\end{array}\!\!\right) .
\end{align}

Given the fact that the marginal deformation by $J_LJ_R$ in the bosonic picture amounts  to changing the argument $R$ of $Z_{1,1}$, whose modular properties are $R$-independent, the contribution to the partition function of {\em the untwisted sector of the $R$ modulus-deformed fermionic construction is obtained by replacing} 
\be
{\theta[^{\gamma +h_L}_{\delta+g_L}]\, \bar\theta[^{\gamma +h_R}_{\delta+g_R}]\over \eta \bar \eta}\, 
\quad \longrightarrow\quad e^{i\pi{h^{\prime}\over 2}(\delta+{\hat g}')} \, Z_{1,1}\Big[{}^{\gamma; \, h_L, \, h_R }_{\delta; \; g_L, \; g_R }\Big \abs{}^0_0 \Big](R)\, .
\label{RRR}
\ee
Note that the particular values $R= {p\over q}R_0$ for ${p\over q}$ rational can be realized in fermionic language by implementing a $\Z_p\times \Z_q$ quotient on the theory, were the orbifold generators act as phases similar to Eq.~(\ref{BC}), or shifts similar to  Eq.~(\ref{shi}) in bosonic language. A well known example of this procedure is that the left/right-symmetric compactification on $S^1(R)/\Z^{\rm shift}_2$ is equivalent to that on $S^1(R/2)$. 

%%%%%%%%%%%%%%%%%%%%%%%%%%%%%%%%%%%%%%%%%%%%%%%
\vspace{0.2cm}
\noindent{\em \large Twisted sectors} 

\noindent The twisted sectors, which have $H, G$ not both even, can be considered in the bosonic language for arbitrary radius $R$. The boundary conditions (\ref{HG}) imply $\partial X$ and $\bar \partial X$ have vanishing constant modes, so that no $R$-dependent zero mode lattice arises in these sectors and the $J_LJ_R$ marginal deformation is trivial. The alternative point of view, where the switch from $R_0$ to $R= {p\over q}R_0$ is implemented in the fermionic construction by a $\Z_p\times \Z_q$ orbifold action, leads to the same conclusion. For instance, when $H=1$, the key point is that the boundary conditions for some phases $\varphi_L,\varphi_R$ are
\be
\psi(z+1)=(e^{i\varphi_L}\psi)^*(z)\, , \qquad \bar\psi(\bar z+1)=(e^{i\varphi_R}\bar \psi)^*(\bar z)\, ,
 \ee
and become trivial under the redefinitions
\be
\tilde \psi(z)\equiv e^{{i\over 2}\varphi_L}\, \psi(z)\, , \qquad \tilde{\bar\psi}(\bar z)\equiv e^{{i\over 2}\varphi_R}\, \bar\psi(\bar z)\, .
\ee
In other words, {\em the twisted sectors of the $R$ modulus-deformed fermionic construction are those of the undeformed one}. 

In a twisted sector, the boundary conditions of $\omega,y,\bar \omega,\bar y$ along the cycles of the genus one Riemann surface are either periodic or antiperiodic. In other words, when a $\Z_2$ twist is implemented, $h_{L,R}$ and $g_{L,R}$ are restricted to be integer.  The contribution of $\omega,y,\bar \omega,\bar y$ to the one-loop partition function is
 \be
 \label{tz}
{1 \over \eta \bar \eta}\, \theta^{1\over 2}[^{\gamma +h_L}_{ \delta +g_L}]\, \theta^{1\over 2}[^{\gamma+h_L+H}_{\delta + g_L + G}] \,\bar\theta^{1\over 2}[^{\gamma+h_R}_{\delta+g_R}]\, \bar\theta^{1\over 2}[^{\gamma+h_R+ H}_{\delta+ g_R+ G}]\equiv e^{i\varphi\left[^{\gamma;\, h_L,\, h_R}_{\delta;\; g_L,\; g_R}\big\abs{}^{H}_{G}\right]}\,  Z_{1,1}\Big[{}^{\gamma; \, h_L, \, h_R }_{\delta; \; g_L, \; g_R }\Big \abs{}^H_G \Big],
\ee
where the r.h.s. shows the result in non-left/right-symmetric orbifold language. In fact, the bosons yield 
  \be
Z_{1,1}\Big[{}^{\gamma; \, h_L, \, h_R }_{\delta; \; g_L, \; g_R }\Big \abs{}^H_G \Big]\!=\left \abs {2\eta \over \theta[^{1-H}_{1-G}] } \right\abs   {\cal P}\left[^{\gamma;\, h_L, \, h_R}_{\delta;\;  g_L, \;g_R} \Big\abs{}^{H}_{G}\right] ,
\ee 
where  ${\cal P}$ is a modular invariant projector that picks up the only non-trivial contributions, which arise from the fixed points of the non-symmetric $\Z_2$ orbifold,
\begin{align}
 {\cal P}\left[^{\gamma;\, h_L, \,h_R}_{\delta;\;  g_L, \;g_R} \Big\abs{}^{H}_{G}\right] =\; &
 {1\over 2}\!\left(1+e^{i\pi(\gamma+h_L)(\delta +  g_L)}\right) \,{1\over 2}\! \left(1+e^{i\pi(\gamma+h_L+H)(\delta +  g_L+G)}\right)\, \times\nonumber \\
& {1\over 2}\!\left(1+e^{i\pi(\gamma+h_R)(\delta +  g_R)}\right)\, {1\over 2}\!\left(1+e^{i\pi(\gamma+h_R+H)(\delta +  g_R+G)}\right).
\label{deltass}
\end{align}
Beside Eq.~(\ref{motrans}), the modular transformations act on $(H,G)$ as,
\be
\tau\to -{1\over \tau} \; \Longleftrightarrow \; (H,G)\to (H,G) \S\;, \quad \tau\to \tau+1 \;\Longleftrightarrow \; (H,G)\to (H,G)\T\, .
\ee

The relation (\ref{tz}) is obtained {\em via} the  $\theta$-function identities 
 \be
\theta[^1_0] \theta[^0_0] \theta[^0_1] = 2\eta^3\, , \quad \theta[^1_1] =0\qquad \mbox{\em i.e.} 
\qquad \theta_2\theta_3\theta_4=2\eta^3 \, ,\quad \theta_1=0\, ,
\ee
while from the fermionic point of view, the projector ${\cal P}$ captures the fact that the sectors that involve $\theta_1$ are vanishing. The phase in the r.h.s. of Eq.~(\ref{tz}) is
\be
\varphi\left[^{\gamma;\, h_L,\, h_R}_{\delta;\; g_L,\; g_R}\Big\abs{}^{H}_{G}\right]={\pi\over 2}(g_L-g_R) (1-H-G)  \quad \mbox{for} \quad \delta+g_L, \delta+g_R, H,G\in\{0,1\}\, ,
\ee
but varies accordingly, when some of the above arguments take other integer values. 

%%%%%%%%%%%%%%%%%%%%%%%%%%%%%%%%%%%%%%%%%%%%%%%
\vspace{0.2cm}
\noindent{\em \large Left/right-symmetric case} 

\noindent At this stage, the left- and right-moving shifts we have described are the most general ones. In the following, we concentrate on a case of particular interest that corresponds to the left/right-symmetric bosonic compactification.

In sectors where $\hat h'$ and $\hat g'$ vanish, we define
\be
\label{hgsh}
(h,g)\!:=(h_R,g_R)=(-h_L,-g_L) \quad \bigg(\mbox{\ie } (h,g)=\Big(\!\!-\!{h'\over 2},-{g'\over 2}\Big) \mbox{ and }(\hat h',\hat g')=(0,0)\bigg),
\ee
and consider the fermionic block 
\be
\label{deffer}
e^{i\pi(k-{1\over 2})(hG-gH)}\!\left(e^{i\pi h\delta} \, {\theta[^{\gamma -h}_{\delta-g}]\, \bar\theta[^{\gamma +h}_{\delta+g}]\over \eta \bar \eta}\right)^{1\over 2}
\left(e^{i\pi h(\delta+G)} \, {\theta[^{\gamma+H -h}_{\delta+G-g}]\, \bar\theta[^{\gamma+H +h}_{\delta+G+g}]\over \eta \bar \eta}\right)^{1\over 2}.
\ee
Since the quantity $hG-gH$ is modular invariant, the phase $e^{i\pi(k-{1\over 2})(hG-gH)}$ can be introduced for any real $k$. Moreover, we see from 
Eq.~(\ref{fermio}) that the specific insertion of phase $e^{i\pi h(\delta+{G\over 2})}$ makes the fermionic block modular covariant and allows $\gamma,\delta$ to be defined modulo 2. Summing over $\gamma,\delta$ equal to $0,1$, we obtain when $h,g$ are restricted to be integer, 
\begin{align}
Z_{1,1}^{{\rm fer}, k}\Big[{}^{h}_{g}\Big \abs{}^H_G \Big]\! :\! &= e^{i\pi(k-{1\over 2})(hG-gH)}\, {1\over 2}\sum_{\gamma,\delta}
e^{i\pi [h(\delta+{G\over 2})-g(\gamma+h+{H\over 2})]}\left\abs{\theta[^{\gamma+h}_{\delta+g}]\over \eta}\right\abs \left\abs{\theta[^{\gamma+h+H}_{\delta+g+G}]\over \eta}\right\abs\nonumber \\
&=e^{ik\pi (hG-gH)} \, {1\over 2}\sum_{\tilde \gamma,\tilde \delta}
e^{i\pi (-g\tilde \gamma+h\tilde \delta-hg)}\left\abs{\theta[^{\tilde \gamma}_{\tilde \delta}]\over \eta}\right\abs \left\abs{\theta[^{\tilde \gamma+H}_{\tilde \delta+G}]\over \eta}\right\abs\, ,
\label{zz}
\end{align}
where we have defined  $\tilde \gamma = \gamma+h$ and $\tilde \delta =\delta+g$ in the second line. From now on, we restrict $k$ to be integer modulo 2, so that $h,g$ and $H,G$ are defined modulo 2 in the above expression. In this case, we also have
\be
Z_{1,1}^{{\rm fer}, k}\Big[{}^{h}_{g}\Big \abs{}^H_G \Big]\! =e^{i(1-k)\pi (hG-gH)} \, {1\over 2}\sum_{\hat \gamma,\hat \delta}
e^{i\pi (-g\hat \gamma+h\hat \delta-hg)}\left\abs{\theta[^{\hat \gamma+H}_{\hat \delta+G}]\over \eta}\right\abs \left\abs{\theta[^{\hat \gamma}_{\hat \delta}]\over \eta}\right\abs\, ,
\ee
which shows that changing $k\to 1-k$ corresponds to imposing the twist to act on $\omega, \bar\omega$ instead of $y, \bar y$, which leads to an equivalent model. 

For $(H,G)=(0,0)$, we obtain from the definition (\ref{deffer})
\begin{align}
Z_{1,1}^{{\rm fer}, k}\Big[{}^{h}_{g}\Big \abs{}^0_0 \Big]\! &={1\over 2} \sum_{\gamma,\delta}e^{i\pi h\delta} \, {\theta[^{\gamma -h}_{\delta-g}]\, \bar\theta[^{\gamma +h}_{\delta+g}]\over \eta \bar \eta}\nonumber \\
&={1\over 2} \sum_{\gamma,\delta}{R_0\over \sqrt{\tau_2}\eta\bar\eta}\sum_{\tilde m,n}e^{-{\pi R_0^2\over \tau_2}\left\abs(\tilde m+g)+(n+h)\tau\right\abs^2}\, e^{i\pi(\tilde m n+\tilde m \gamma-n\delta)}\nonumber\\
&={2R_0\over \sqrt{\tau_2}\eta\bar\eta}\sum_{\tilde m',n'}e^{-{\pi (2R_0)^2\over \tau_2}\left\abs(\tilde m'+{g\over 2})+(n'+{h\over 2})\tau\right\abs^2}\!:= {\Gamma_{1,1}[^{h}_{g}](2R_0)\over \eta\bar\eta}\, ,
\end{align}
where the sum over $\gamma,\delta$ projects out the odd values of $\tilde m$ and $n$. Thus, we recover the well know bosonic $\Gamma_{1,1}$-lattice considered in  Eq.~(\ref{Gshift}), with shifts $(h,g)$ and radius $R_1=2R_0$. 

For $(H,G)\neq (0,0)$ modulo 2,  we use Eqs (\ref{tz})--(\ref{deltass}) applied for $h_{L,R}=g_{L,R}=0$  to write
\begin{align}
Z_{1,1}^{{\rm fer}, k}\Big[{}^{h}_{g}\Big \abs{}^H_G \Big]\! &=e^{ik\pi (hG-gH)} \, {1\over 2}\sum_{ \gamma, \delta}
e^{i\pi (-g\gamma+h \delta-hg)}\left \abs {2\eta \over \theta[^{1-H}_{1-G}] } \right\abs \delta_{\gamma\delta,0\,  \mbox{\scriptsize mod}\, 2}\; \delta_{(\gamma+H)(\delta+G),0  \,\mbox{\scriptsize mod}\, 2}\nonumber\\
&= e^{ik\pi (hG-gH)} \left \abs {2\eta \over \theta[^{1-H}_{1-G}] } \right\abs \Big(\delta_{(h,g),(0,0) \, \mbox{\scriptsize mod}\, 2}+ \delta_{(h,g),(H,G)  \,\mbox{\scriptsize mod}\, 2}\Big)\nonumber\\
&= \left \abs {2\eta \over \theta[^{1-H}_{1-G}] } \right\abs \delta_{\big\abs{}^{h\;\;H}_{g\;\; \,G}\big\abs,0 \, \mbox{\scriptsize mod}\, 2}\, ,
\end{align}
which is nothing but the $(H,G)$-twisted and $(h,g)$-shifted sector of a circle compactification~\cite{RcorrectionsDuals}.  

Using the rule shown in Eq.~(\ref{RRR}), the $(1,1)$-block of the $R$-modulus deformed fermionic construction that realizes the left/right-symmetric case in bosonic language is obtained by substituting
\begin{align}
\label{subs}
Z_{1,1}^{{\rm fer}, k}\Big[{}^{h}_{g}\Big \abs{}^H_G \Big]\! &=e^{ik\pi (hG-gH)} \, {1\over 2}\sum_{\gamma, \delta}
e^{i\pi (-g \gamma+h\delta-hg)}\, \Bigg\abs{\theta[^{ \gamma}_{\delta}]\over \eta}\Bigg\abs \left\abs{\theta[^{ \gamma+H}_{ \delta+G}]\over \eta}\right\abs\quad \longrightarrow \quad Z_{1,1}\Big[{}^{h}_{g}\Big \abs{}^H_G \Big](2R)\, ,
\end{align}
where the r.h.s. is the block associated to  a twisted and shifted circle compactification at arbitrary radius $2R$,
\be
Z_{1,1}\Big[{}^{h}_{g}\Big \abs{}^H_G \Big](2R)= \left\{
\begin{array}{ll}
\displaystyle {\Gamma_{1,1}[^{h}_{g}](2R)\over \eta\bar\eta}\, , \phantom{\underset{\underset{a}{\cdot}}{\abs}}&\mbox{when }(H, G)=(0,0) \mbox{ mod 2}\\ 
\displaystyle  \left \abs {2\eta \over \theta[^{1-H}_{1-G}] } \right\abs \delta_{\big\abs{}^{h\;\;H}_{g\;\; \,G}\big\abs,0 \,\mbox{\scriptsize mod} \,2}\, , &\mbox{when }(H, G)\neq(0,0)\mbox{ mod 2}\, .
\end{array}
\right.
\label{totoi}
\ee

Before considering the two coordinates compactification, we would like to make some remarks. Summing over the  shifts $h,g$, we obtain
\be
{1\over 2}\sum_{h,g} Z_{1,1}^{{\rm fer}, k}\Big[{}^{h}_{g}\Big \abs{}^H_G \Big]\!={1\over 2}\sum_{\gamma, \delta}e^{-i\pi (\gamma+kH)(\delta+kG)}\, \Bigg\abs{\theta[^{ \gamma}_{\delta}]\over \eta}\Bigg\abs \left\abs{\theta[^{ \gamma+H}_{ \delta+G}]\over \eta}\right\abs=Z_{1,1}[^H_G](R_0)\, ,
\label{/2}
\ee
where 
\be
Z_{1,1}[^H_G](R)= \left\{
\begin{array}{ll}
\displaystyle {\Gamma_{1,1}(R)\over \eta\bar\eta}\, , \phantom{\underset{\underset{a}{\cdot}}{\abs}}&\mbox{when } (H, G)=(0,0) \mbox{ mod 2}\\ 
\displaystyle  \left \abs {2\eta \over \theta[^{1-H}_{1-G}] } \right\abs  , &\mbox{when }  (H, G)\neq(0,0)\mbox{ mod 2}
\end{array}
\right.
\ee
and $\Gamma_{1,1}(R)\equiv\Gamma_{1,1}[^0_0](R)$ is the circle compactification lattice. Eq.~(\ref{/2}) expresses the geometrical fact that 
\be
{S^1(2R_0)\over \Z^{\rm shift}_2\times \Z_2} ={S^1(R_0)\over \Z_2}\,,
\ee
\ie that the shift divides the radius of compactification by a factor of 2, even when the circle is twisted. However, from the fermionic point of view, the natural definition of the twisted $(1,1)$-conformal block is without  the phase $e^{-i\pi (\gamma+kH)(\delta+kG)}$ present in Eq.~(\ref{/2}). Thus, we take
\be
\label{defi}
Z_{1,1}^{\rm fer}[^H_G]\!:={1\over 2}\sum_{\gamma, \delta}\Bigg\abs{\theta[^{ \gamma}_{\delta}]\over \eta}\Bigg\abs \left\abs{\theta[^{ \gamma+H}_{ \delta+G}]\over \eta}\right\abs=\left\{
\begin{array}{ll}
\displaystyle {1\over 2}\sum_{\gamma, \delta}{\theta[^{ \gamma}_{\delta}]\, \bar \theta[^{ \gamma}_{ \delta}]\over \eta\bar \eta}\, , \phantom{\underset{\underset{a}{\cdot}}{\abs}}\mbox{when } (H, G)=(0,0) \mbox{ mod 2}\\ 
\displaystyle  \left \abs {2\eta \over \theta[^{1-H}_{1-G}] } \right\abs  , \;\; \;\qquad \mbox{when } (H, G)\neq(0,0)\mbox{ mod 2}\, ,
\end{array}
\right.
\ee
where we have used Eqs (\ref{tz})--(\ref{deltass}) for $h_{L,R}=g_{L,R}=0$ in the second line. Since Eq.~(\ref{fermio}) gives 
\begin{align}
{1\over 2}\sum_{\gamma, \delta}{\theta[^{ \gamma}_{\delta}]\, \bar \theta[^{ \gamma}_{ \delta}]\over \eta\bar \eta}&={R_0\over\sqrt{\tau_2}\eta\bar\eta}\sum_{\tilde m,n}
e^{-{\pi R_0^2 \over \tau_2}  \left\abs \tilde m+n\tau \right\abs^2}\,
 {1\over 2}\sum_{\gamma, \delta}e^{i\pi\left(\tilde m n+\tilde m {\gamma}- n{\delta}\right)}\nonumber\\
 &= {2R_0\over\sqrt{\tau_2}\eta\bar\eta}\sum_{\tilde m',n'}
e^{-{\pi (2R_0)^2 \over \tau_2}  \left\abs \tilde m'+n'\tau \right\abs^2}={\Gamma(2R_0)\over \eta\bar\eta}\, , 
\end{align}
we finally conclude as expected that 
\be
\label{x2}
Z_{1,1}^{\rm fer}[^H_G]=Z_{1,1}[^H_G](2R_0)\, .
\ee
Comparing Eqs (\ref{/2}), (\ref{defi}) and (\ref{x2}), we see that if the  shift divides the radius by 2 in bosonic language, it flips the signs in front of $\theta_1$-functions in fermionic language. We thus have $Z_{1,1}[^H_G](2R_0)=Z_{1,1}[^H_G](R_0)$, a fact that can be understood as a T-duality. Actually, since $2R_0=1/R_0$, the shift operation  that changes  $2R_0\to R_0$ is equivalent to the operation $1/R_0\to R_0$. 

Before concluding this subsection, we would like to mention that in order to  simplify formulas in the core of our paper, we have used the convention to take $Z_{1,1}\Big[{}^{h}_{g}\Big \abs{}^H_G \Big](R)$ rather than $Z_{1,1}\Big[{}^{h}_{g}\Big \abs{}^H_G \Big](2R)$ in the r.h.s. of the substitution (\ref{subs}).

%%%%%%%%%%%%%%%%%%%%%%%%%%%%%%%%%%%%%%%%%%%%%%%

\subsection{Two coordinates compactification}

Proceeding in a similar way for a second coordinate,  we can  deform even further an initial fermionic model by switching on the full metric $G_{ij}$ and antisymmetric tensor $B_{ij}$ moduli, $i=1,2$. This is done without changing the modular properties of the initial model constructed at the fermionic point. As before,  we introduce integers $\gamma^i,\delta^i$, together with real left- and right-moving shifts $h_{L,R}^i,g_{L,R}^i$. In case a $\Z_2$ twist is implemented, we suppose it acts simultaneously on the two coordinates. 

%%%%%%%%%%%%%%%%%%%%%%%%%%%%%%%%%%%%%%%%%%%%%%%
\vspace{0.2cm}
\noindent{\em \large Untwisted sector} 

\noindent We start with the sector $(H,G)=(0,0)$. Defining linear combinations $h^{\prime i}, g^{\prime i}, \hat h^{\prime i}, \hat g^{\prime i}$ as in Eq.~(\ref{hh}), the undeformed $(2,2)$-conformal block takes the form
\be
{\theta[^{\gamma^1 +h^1_L}_{\delta^1+g^1_L}]\, \bar\theta[^{\gamma^1 +h^1_R}_{\delta^1+g^1_R}]\over \eta \bar \eta}\, {\theta[^{\gamma^2 +h^{2}_L}_{\delta^2+g^{2}_L}]\, \bar\theta[^{\gamma^2 +h^{2}_R}_{\delta^2+g^{2}_R}]\over \eta \bar \eta}
\equiv e^{i\pi\left[{h^{\prime 1}\over 2}(\delta^1+{\hat g}^{\prime 1})+{h^{\prime 2}\over 2}(\delta^2+{\hat g}^{\prime 2})\right]}
\, Z_{2,2}\Big[{}^{\gamma^i; \, h_L^i, \, h_R^i }_{\delta^i; \; g_L^i, \; g_R^i}\Big \abs{}^0_0 \Big](T_0,U_0)\, ,
\label{pp}
\ee
where $(T_0,U_0)=({i\over 2},i)$ and, for arbitrary $T$ and $U$, 
\begin{align}
Z_{2,2}\Big[{}^{\gamma^i; \, h_L^i, \, h_R^i }_{\delta^i; \;g_L^i, \; g_R^i}\Big \abs{}^0_0 \Big](T,U)=
{\sqrt{\det G}\over \tau_2(\eta\bar \eta)^2} \sum_{\tilde m^i,n^i}&\;e^{-{\pi\over \tau_2}\big[\tilde m^i-{g^{\prime i}\over 2}+\left(n^i-{h^{\prime i}\over 2}\right)\tau\big](G_{ij}+B_{ij})\big[\tilde m^j-{g^{\prime j}\over 2}+\left(n^j-{h^{\prime j}\over 2}\right)\bar\tau\big]}\, \times \nonumber \\
&\,\,e^{i\pi( \tilde m_i \hat h^{\prime i} - n_i\hat g^{\prime i})}\; e^{i\pi(\tilde m_in^i+\tilde m_i\gamma^i-n_i\delta^i)} \, ,
\end{align}
with $T, U$ related to the metric and antisymmetric tensor as
\be
G_{ij}={\Im T\over \Im U}\left(\!\!\begin{array}{cc} 1 &\Re U\\ \Re U & \abs U\abs^2\end{array}\!\!\right)\; , \qquad B_{ij}=\Re T \left(\!\!\begin{array}{cc} 0 &1\\ -1 & 0\end{array}\!\!\right).
\ee
Here also, $Z_{2,2}$ couples non-trivially the $\Gamma_{2,2}$-lattice shifted by $(h^i,g^i)$ and  $(\hat h^i, \hat g^i)$,
\be
{\sqrt{\det G}\over \tau_2} \sum_{\tilde m^i,n^i}e^{-{\pi\over \tau_2}\big[\tilde m^i-{g^{\prime i}\over 2}+\left(n^i-{h^{\prime i}\over 2}\right)\tau\big](G_{ij}+B_{ij})\big[\tilde m^j-{g^{\prime j}\over 2}+\left(n^j-{h^{\prime j}\over 2}\right)\bar\tau\big]}\, e^{i\pi( \tilde m_i \hat h^{\prime i} - n_i\hat g^{\prime i})}
\ee
to the characters $(\gamma^i, \delta^i)$, {\em via} the modular invariant phase  $e^{i\pi\left(\tilde m_i n^i+\tilde m_i {\gamma}^i- n_i{\delta}^i\right)}$. 

The contribution to the partition function of {\em the untwisted sector of the $T,U$ moduli-deformed fermionic construction is obtained by replacing $T_0,U_0$ by arbitrary $T$ and $U$ :}
\be
{\theta[^{\gamma^1 +h^1_L}_{\delta^1+g^1_L}]\,\bar\theta[^{\gamma^1 +h^1_R}_{\delta^1+g^1_R}]\over \eta \bar \eta}\, {\theta[^{\gamma^2 +h^{2}_L}_{\delta^2+g^{2}_L}]\,\bar\theta[^{\gamma^2 +h^{2}_R}_{\delta^2+g^{2}_R}]\over \eta \bar \eta}\quad \longrightarrow \quad e^{i\pi\left[{h^{\prime 1}\over 2}(\delta^1+{\hat g}^{\prime 1})+{h^{\prime 2}\over 2}(\delta^2+{\hat g}^{\prime 2})\right]}
\, Z_{2,2}\Big[{}^{\gamma^i; \, h_L^i, \, h_R^i }_{\delta^i; \; g_L^i, \; g_R^i}\Big \abs{}^0_0 \Big](T,U)\, .
\label{titi}
\ee

%%%%%%%%%%%%%%%%%%%%%%%%%%%%%%%%%%%%%%%%%%%%%%%
\vspace{0.2cm}
\noindent{\em \large Twisted sectors} 

\noindent When $H$ and $G$ are not both even, the associated {\em twisted sectors of the $T,U$ moduli-deformed fermionic construction are those of the undeformed one}. This is again due to the fact that they are moduli-independent, which implies that the expressions of their conformal blocks are those given at the fermionic point $(T_0,U_0)$ :
 \be
 \label{tz2}
{1 \over (\eta \bar \eta)^2}\prod_{i}\left( \theta^{1\over 2}[^{\gamma^i +h_L^i}_{ \delta^i +g_L^i}]\, \theta^{1\over 2}[^{\gamma^i+h_L^i+H}_{\delta^i + g_L^i + G}] \,\bar\theta^{1\over 2}[^{\gamma^i+h_R^i}_{\delta^i+g_R^i}]\, \bar\theta^{1\over 2}[^{\gamma^i+h_R^i+ H}_{\delta^i+ g_R^i+ G}]\right)\equiv e^{i\sum_i \varphi\left[^{\gamma^i;\, h_L^i,\, h_R^i}_{\delta^i;\; g_L^i,\; g_R^i}\big\abs{}^{H}_{G}\right]}\,  Z_{2,2}\Big[{}^{\gamma^j; \, h_L^j, \, h_R^j }_{\delta^j; \; g_L^j, \; g_R^j}\Big \abs{}^H_G \Big],
\ee
where in bosonic  language we have
\be
Z_{2,2}\Big[{}^{\gamma^j; \, h_L^j, \, h_R^j }_{\delta^j; \; g_L^j, \; g_R^j}\Big \abs{}^H_G \Big]\!=\!\left \abs {2\eta \over \theta[^{1-H}_{1-G}] } \right\abs^2  \prod_i {\cal P}\left[^{\gamma^i;\, h_L^i, \, h_R^i}_{\delta^i;\;  g_L^i, \;g_R^i} \Big\abs{}^{H}_{G}\right] .
\ee
 
%%%%%%%%%%%%%%%%%%%%%%%%%%%%%%%%%%%%%%%%%%%%%%%
\vspace{0.2cm}
\noindent{\em \large Left/right-symmetric case} 

\noindent Defining  shifts $h^i,g^i$ as in Eq.~(\ref{hgsh}), we consider for integer $k^i$'s the fermionic conformal block 
\be
\prod_ie^{i\pi(k^i-{1\over 2})(h^iG-g^iH)}\!\left(e^{i\pi h^i\delta^i} \, {\theta[^{\gamma^i -h^i}_{\delta^i-g^i}]\, \bar\theta[^{\gamma^i +h^i}_{\delta^i+g^i}]\over \eta \bar \eta}\right)^{1\over 2}
\left(e^{i\pi h^i(\delta^i+G)} \, {\theta[^{\gamma^i+H -h^i}_{\delta^i+G-g^i}]\, \bar\theta[^{\gamma^i+H +h^i}_{\delta^i+G+g^i}]\over \eta \bar \eta}\right)^{1\over 2}\, , 
\ee
where $\gamma^i,\delta^i$ are integer modulo 2. Proceeding as in the one coordinate case, we sum over $\gamma^i,\delta^i$  and find, when $h^i,g^i$ are integer, 
\begin{align}
Z_{2,2}^{{\rm fer}, k^j}\Big[{}^{h^1,\, h^2}_{g^1,\;g^2}\Big \abs{}^H_G \Big]\! :\! &=\prod_ie^{ik^i\pi (h^iG-g^iH)} \, {1\over 2}\sum_{\gamma^i,\delta^i}
e^{i\pi (-g^i\gamma^i+h^i\delta^i -h^i g^i)}\left\abs{\theta[^{\gamma^i}_{\delta^i}]\over \eta}\right\abs \left\abs{\theta[^{\gamma^i+H}_{\delta^i+G}]\over \eta}\right\abs
\nonumber\\
&= \left\{
\begin{array}{ll}
\displaystyle {\Gamma_{2,2}[^{h^1,\, h^2}_{g^1,\; g^2}](4T_0,U_0)\over (\eta\bar\eta)^2}\, , \phantom{\underset{\underset{a}{\cdot}}{\abs}}&\!\!\!\!\mbox{when }(H, G)=(0,0) \mbox{ mod 2}\\ 
\displaystyle  {4\eta\bar\eta \over \theta[^{1-H}_{1-G}] \, \bar \theta[^{1-H}_{1-G}] } \; \delta_{\big\abs{}^{h^1\;\;H}_{g^1\;\; \,G}\big\abs,0 \,\mbox{\scriptsize mod} \,2}\; \delta_{\big\abs{}^{h^2\;\;H}_{g^2\;\; \,G}\big\abs,0 \,\mbox{\scriptsize mod} \,2}\, , &\!\!\!\!\mbox{when }(H, G)\neq(0,0)\mbox{ mod 2}\, ,
\end{array}
\right.
\end{align}
where the $\Gamma_{2,2}$ shifted lattice is defined in Eq.~(\ref{latt}). 

As said in Eq.~(\ref{titi}), the moduli deformation amounts to changing the argument of the lattice as $(4T_0,U_0)\to (4T,U)$, where $T,U$ are arbitrary. However, in the core of the paper, we found convenient to take the lattice argument at arbitrary point in moduli space to be  $(T,U)$, as indicated in 
Eq.~(\ref{toto}).

%%%%%%%%%%%%%%%%%%%%%%%%%%%%%%%%%%%%%%%%%%
%\vspace{.4cm}
%\newpage

\end{document}